%% file: paper.tex
\documentclass[letterpaper, twocolumn, 10pt]{article}
\usepackage{usenix2019_v3}

\usepackage[english]{babel}
\usepackage{blindtext}
\usepackage{xspace}
\usepackage{lipsum}
\usepackage{amsmath}
\usepackage{amsthm}

\usepackage[ruled,vlined,linesnumbered]{algorithm2e}

\usepackage{amssymb}
\usepackage{times}
\usepackage{graphicx}
\usepackage{caption}
\usepackage{url}
\usepackage{multirow}
\usepackage{stfloats}
\usepackage{float}
\usepackage{authblk}
\usepackage{cases}
\usepackage{color}
\usepackage{comment}
\usepackage{subcaption}
\usepackage{array}
\usepackage[compact]{titlesec}

\newsavebox{\imagebox}

\newcommand{\name}{{\textsc{Sensei}}\xspace}

\newcommand{\xu}[1]{\xspace}
\newcommand{\sid}[1]{\xspace}
\newcommand{\ignore}[1]{}

\newcounter{packednmbr}

\newenvironment{packeditemize}{\begin{list}{$\bullet$}{\setlength{\itemsep}{0.5pt}\addtolength{\labelwidth}{-4pt}\setlength{\leftmargin}{2ex}\setlength{\listparindent}{\parindent}\setlength{\parsep}{1pt}\setlength{\topsep}{2pt}}}{\end{list}}

\newcommand{\tightcaption}[1]{\vspace{-0.05cm}\caption{{\normalfont{\textit{{#1}}}}\vspace{-0.1cm}}}
\newcommand{\tightsection}[1]{\vspace{-0.02cm}\section{#1}\vspace{-0.02cm}}
\newcommand{\tightsubsection}[1]{\vspace{-0.02cm}\subsection{#1}\vspace{-0.02cm}}

\newcommand{\eg}{{\it e.g.,}\xspace}
\newcommand{\ie}{{\it i.e.,}\xspace}

\newcommand{\bigO}{\mathrm{O}}

\newcommand{\mypara}[1]{\vspace{0.07cm}\noindent{\bf {#1}:}~}
\newcommand{\myparaq}[1]{\vspace{0.07cm}\noindent{\bf {#1}?}~}

\newcommand{\myparaitalic}[1]{\vspace{0.05cm}\noindent{\em {#1}:}~}

\begin{document}
% \title{Aligning Adaptive Video Streaming to Dynamic User Sensitivity}
% \title{{\huge Improving Perceived Video Quality by Harnessing \\Dynamic User Sensitivity}}
%\title{{\Large \bf Improving Video Streaming Experience by Aligning\\ Quality Adaptation with Dynamic User Sensitivity}}

%\title{\vspace{-0.4in}{\Large \bf \name: Aligning Video Streaming Quality with Dynamic User Sensitivity\vspace{-0.2in}}}
\title{\vspace{-0.4in}{\Large \bf \name: Aligning Video Streaming Quality with Dynamic User Sensitivity}}

% \\\sid{Or remove ``Using the crowd'' and say ``Aligning Video\ldots''}\\\jc{how about ``Sensei: Using the Wisdom of the Crowd to Align Adaptive Video Streaming with Dynamic Quality Sensitivity''}}

\author{Xu Zhang, Yiyang Ou, Siddhartha Sen$^*$, Junchen Jiang\\
% \authornote{Note} \orcid{1234-5678-9012}
University of Chicago~~~  $^*$Microsoft Research\\
  %\streetaddress{}
%   \city{\{zhangxu, daniar, haryadi, junchenj\}@uchicago.edu, sidsen@microsoft.com} 
%   \state{State} 
%  \postcode{Zipcode}
}

\maketitle

\input{macro}

\input{abstract}

\input{intro-new}
% \input{intro-jj}
% \blindtext

% \input{motivation-outline}

\input{motivation}

\input{qoemodel}

\input{abr}
\input{system}

% \input{eval}
\input{eval_xu_modified}

\input{relatedwork}

\section{Conclusion}
We have descried \name, a video streaming system that optimizes video quality by exploiting dynamic quality sensitivity. 
\name scales out the profiling of quality sensitivity, which is unique to each video, through a combination of crowdsourcing and the insight that quality sensitivity is inherent to the video content.
We show that with minor modifications to state-of-the-art video adaptation algorithms, \name can improve their QoE by 15.1\% or save bandwidth by 26.8\% on average.

\bibliographystyle{plain}
\bibliography{reference}

\input{appendix}

\end{document}

%% file: macro.tex
\newcommand{\NumVideosPerSurvey}{\ensuremath{K}\xspace}
\newcommand{\NumVideos}{\ensuremath{V}\xspace}
\newcommand{\TotalLenVideos}{\ensuremath{L}\xspace}
\newcommand{\NumTurkers}{\ensuremath{M}\xspace}
\newcommand{\NumChunks}{\ensuremath{N}\xspace}
\newcommand{\NumParams}{\ensuremath{P}\xspace}
\newcommand{\QoE}{\ensuremath{Q}\xspace}
\newcommand{\Quality}{\ensuremath{q}\xspace}
\newcommand{\Weight}{\ensuremath{w}\xspace}

\newcommand{\NumBitrates}{\ensuremath{B}\xspace}
\newcommand{\NumBufferings}{\ensuremath{F}\xspace}
\newcommand{\ThreshDiff}{\ensuremath{\alpha}\xspace}

%% file: abstract.tex
\begin{abstract}

This paper aims to improve video streaming by leveraging a simple observation---users are more
sensitive to low quality in certain parts of a video than in others.  For instance, rebuffering
during key moments of a sports video (\eg before a goal is scored) is more annoying than rebuffering
during normal gameplay.
Such dynamic quality sensitivity, however, is rarely captured by current approaches, which predict
QoE (quality-of-experience) using one-size-fits-all heuristics that are too simplistic to understand
the nuances of video content, or that are biased towards the video content they are trained on (in
the case of learned heuristics). 

% Such dynamic quality sensitivity, however, is rarely captured by mainstream QoE
% (quality-of-experience) models; even when it {\em is} considered, people tend to use heuristics
% that are no match to the complexity of video content (\eg some believe users are more sensitive
% during more dynamic scenes, but garbage time can be dynamic too).
% These problems are symptomatic of an underlying issue: prior work seeks to predict QoE with
% one-size-fits-all {\em heuristics} (handcrafted/learned), which can be oversimplifying or biased
% to training data.

The problem is that none of these approaches know the true dynamic quality sensitivity of a video
they have never seen before. Therefore, instead of proposing yet another heuristic, we take a
different approach: we run a {\em separate crowdsourcing experiment for each video} to derive 
users' quality sensitivity at different parts of the video. Of course, the cost of doing this at
scale can be prohibitive, but we show that careful experiment design combined with a suite of
pruning techniques can make the cost negligible compared to how much content providers invest in
content generation and distribution. For example with a budget of just \$31.4 per min video, we can predict QoE
up to 37.1\% more accurately than state-of-the-art QoE models.

Our ability to accurately profile time-varying user sensitivity inspires a new approach to video
streaming---{\em dynamically aligning higher (lower) quality with higher (lower) sensitivity
periods}.
% We show that with a budget of \$\fillme, we can predict QoE \fillme\% more accurately than
% state-of-the-art models.
% The ability to accurately profile time-varying user sensitivity inspires a new approach to
% bitrate-adaptive streaming---{\em dynamically aligning low-quality incidents with low-sensitive
% periods}.
We present a new video streaming system called \name that incorporates dynamic quality sensitivity
into existing quality adaptation algorithms. We apply \name to two state-of-the-art adaptation
algorithms, one rule-based and one based on deep reinforcement learning.
% \name uses deep reinforcement learning to incorporate user sensitivity into its bitrate adaptation
% algorithm. It can \sid{Took out ``or the buffer is not empty'' as that's implied by sufficient
% bandwidth and also confusing if people forget how ABR works}
\name can take seemingly unusual actions: \eg lowering bitrate (or initiating a rebuffering event)
even when bandwidth is sufficient so that it can maintain a higher bitrate without rebuffering when
quality sensitivity becomes higher in the near future.
Compared to state-of-the-art approaches, \name improves QoE by 15.1\% or achieves the same QoE
with 26.8\% less bandwidth on average.

\end{abstract}

%% file: intro-new.tex
\tightsection{Introduction}

% video bandwidth consumption is exploding.
An inflection point in Internet video traffic is afoot, driven by more ultra-high resolution videos,
more large-screen devices, and ever-lower user patience for low quality~\cite{video-trends,Sandvine-trends}.
At the same time, the video streaming industry over its several decades of evolution has largely
saturated the room for improvement: recent adaptive bitrate (ABR) algorithms (e.g.,
\cite{mao2017neural,huang2014buffer,yan2020learning}) achieve near-optimal balance between
bitrate and rebuffering events, and recent video codecs (\eg \cite{h265,neuralencoding}) improve encoding efficiency but
require an order of magnitude more computing power than their predecessors.
% and video encoding~\cite{h265,neuralencoding} (\eg H.265 requires 10$\times$ more computing power
% than H.264~\cite{??}).
% Yet, online video traffic continues to grow rapidly.
The confluence of these trends means that the Internet may soon be overwhelmed by online video
traffic,\footnote{This is vividly illustrated by the recent actions taken by YouTube and Netflix
(and many others) to lower video quality in order to save ISPs from collapsing as more people stay
at home and binge watch online videos~\cite{quality-drop}.} and new ways are needed to attain fundamentally better
tradeoffs between bandwidth usage and user-perceived QoE (quality of experience).

% part of the contributing factor is the conventional wisdom to optimize quality everywhere, but we
% observe the dynamic sensitivity.
We argue that a key limiting factor is the conventional wisdom that users care about quality in the
same way throughout a video, so video quality should be optimized using the same standard {\em
everywhere} in a video. This means that lower quality---due to rebuffering, low visual quality, or
quality switches---should be avoided identically from the beginning to the end.
We argue that this assumption is not accurate. In sports videos, a rebuffering event that coincides
with scoring tends to inflict more negative impact on user experience than rebuffering during
normal gameplay, as shown in Figure~\ref{fig:soccer_content}.  Similarly, bitrate switches or
rebuffering during key moments of a movie, when tensions have built up, tend to be more noticeable to viewers.
In other words, user sensitivity to lower quality {\em varies with the video content dynamically
over time}.

Unfortunately, both the literature on ABR algorithms and the literature on QoE modeling adopt the
conventional wisdom.  Most ABR algorithms completely ignore the content of each video chunk:
they focus on balancing the bitrate with the available network bandwidth, and thus consider only the
size and download speed of the chunks. Traditional ways of modeling QoE are also agnostic to
the video content, although recent QoE models (\eg PSNR~\cite{gonzalez2004digital}, SSIM~\cite{wang2003multiscale}, VMAF~\cite{vmaf}, and
deep-learning models~\cite{eswara2019streaming,kim2018deep}) try to predict which content users will
be more sensitive to by studying pixel-level differences between frames. These models seek
heuristics that generalize across {\em all videos} and thus resort to generic measures (like
pixel-level differences), but it is unclear if any heuristic can capture the diverse and dynamic
influence a video's content can have on users' sensitivity to quality.

For example, models like LSTM-QoE~\cite{eswara2019streaming} assume that users are more
sensitive to rebuffering events in more ``dynamic'' scenes. In sports videos, however, non-essential
content like ads and quick scans of the players can be highly dynamic, but users may care less about
quality during those moments. In the video in Figure~\ref{fig:soccer_content}, LSTE-QoE considers
normal gameplay to be the most dynamic part, but the most quality sensitive
part of the video according to the user study is the goal.
% On the other hand, there are other videos in our study where the dynamism of a chunk is aligned
% with users' quality sensitivity.
This shows that quality sensitivity can vary depending on the content, and may be unique
for each video.

The dynamic nature of quality sensitivity suggests a new opportunity for QoE improvement and
bandwidth savings:
\begin{packeditemize}
    \item {\em Similar bandwidth, higher QoE:} If users are about to become more quality sensitive,
    we can carefully lower the current quality---\eg lower the bitrate or
    add a short rebuffering event---to save bandwidth and ensure higher quality
    when users become more sensitive. Similarly, we can also increase the current bitrate for a few chunks
    %beyond the available bandwidth
    if quality sensitivity is expected to drop in the near future.
    \item {\em Similar QoE, lower bandwidth:} When quality sensitivity is low, we can judiciously
    lower the bitrate to save bandwidth while minimally impacting the perceived QoE.
    perceived.
\end{packeditemize}
% \jc{quote some key findings from section 2}
In short, 
%, while prior work raises quality of video chunks to the same extent, 
we seek to {\em align higher (lower) quality of video chunks with higher (lower) quality
sensitivity of users}.

% and can't be utilized with conventional approaches
We present \name, a video streaming system that incorporates dynamic quality sensitivity into its
QoE model and video quality adaptation.
\name addresses two key challenges.
% This paper presents \name, a novel video streaming system to address these challenges. %\footnote{The name \name refers to {\bf ATTEN}tion-driven adap{\bf TIVE} streaming.}
% It makes two contributions.
% this makes two innovations

\myparaitalic{Challenge 1}
{\em How do we profile the unique dynamic quality sensitivity of each video in an accurate and
scalable manner?} 
As we observed, existing QoE models use general heuristics that are unable to predict the
differences in quality sensitivity between chunks of the same video, or across videos.

% Traditional ways of modeling QoE are agnostic to the substance of individual video.
% % crowdsourcing-based real user rating
% % We are not the first to model the impact of video content on quality sensitivity of viewers.
% Although previous QoE models (\eg PSNR~\cite{??}, SSIM~\cite{??}, VMAF~\cite{??} and recent deep-learning models~\cite{deepvqa,lstm-qoe}) aim to predict under what video content will user be more sensitive to video quality, they share an underlying issue---they seek heuristics that generalize to {\em all} videos, but it is unclear if any heuristics can capture the diverse and dynamic influence of content on user sensitivity of video quality.
% % There are also heuristics that seek to identify rebuffering/quality switches with higher impact. 
% %(\eg rebuffering during dynamic content~\cite{??} or the first rebuffering of a session).
% For instance, some heuristics assumes users are more sensitive to buffering at more dynamic scenes~\cite{??}, but in sports, non-essential content, like ads, can be highly dynamic too, but users will tend to care more about quality during more important moments.
% Recent computer vision techniques also struggle to infer when users pay more attention to the video content. (We will show more details in \S\ref{??}.)

\mypara{Crowdsourcing the true quality sensitivity per video}
Instead of proposing yet another heuristic, \name takes a different approach.
We run a {\em separate crowdsourcing experiment for each video} to derive the quality sensitivity of
users at different parts of the video. Specifically, we elicit quality ratings directly from real
users (obtaining a ``ground truth'' of their QoE) for multiple renderings of the same video,
where each rendering involves some degradation in quality. We use these ratings to estimate a
weight for each video chunk that encodes its quality sensivity.
\name automates and scales this process out using a public crowdsourcing platform (Amazon MTurk),
which provides a large pool of raters.
%\sid{Junchen, I think this fits better here rather than in the second challenge.}
Our empirical results show that the relative quality sensitivity of chunks is robust to any
particular low-quality incident, allowing us to encode each chunk's quality
sensitivity using a single weight, independent of the quality of other chunks. We combine
this with a suite of cost-cutting techniques to reduce the number of rendered videos that need to be
rated.
% cost and delay of profiling a new video---e.g., selecting representative chunks to profile and
% coarsening the quality levels to allow just enough accuracy to differentiate the chunks.
% \eg \name judiciously selects a small fraction of video chunks to be rated and automatically
% assigns raters among different video chunks to maximize profiling accuracy within a budget.
% \sid{This can go in the summary of results\ldots} Moreover, this cost is negligible compared to
% how much content providers invest in content generation and distribution.

\myparaitalic{Challenge 2} {\em How do we incorporate dynamic quality sensitivity into a video
streaming system to enable new decisions?}
% Today's video players are not design per video and do not optimize different parts of a video
% differently.
% new abr logic leveraging user attention across chunks
Today's video players are designed to be ``greedy'': they pick a bitrate that maximizes the quality
of the next chunk while avoiding rebuffering events. But in order to utilize dynamic
quality sensitivity, a player must ``schedule'' bitrate choices over {\em multiple} future chunks,
each having a potentially different quality sensitivity.
This means that some well-established behaviors of video players, \eg only rebuffer
when the buffer is empty,
% \sid{Didn't understand this, also not clear if we do this, and one example seems sufficient.} or
% the player should download a new chunk only after a chunk has been played (during the steady
% state)
may prohibit the envisioned optimizations.

% that ``borrows'' bandwidth from low quality-sensitivity chunks to high quality-sensitivity chunks
% support new quality adaptation actions, such as initiating rebuffering when the buffer is not empty to save bandwidth for near future chunks that have higher quality sensitivity.
% in today's players rebuffering occurs only when the buffer is empty, but with dynamic user sensitivity, it would be beneficial to move rebuffering to when users have low quality sensitivity.

\mypara{Refactoring video streaming to align quality adaptation with quality sensitivity}
Instead of proposing a new video streaming protocol, \name takes a pragmatic approach by working
within the popular DASH framework, but introduces two practical modifications.
First, it integrates the per-chunk weights mentioned above into the existing DASH protocol in a
natural way.
% our empirical results show that the temporal pattern of quality sensitivity is largely agnostic to
% particular low-quality incidents, so we encode the dynamic quality sensitivity by a score for each
% video chunk, independently to the quality assigned to each chunk, if users are sensitive to one
% type of quality incidents during some part of a video, they are likely to be sensitive to other
% types of quality incidents in that part of the video.
% the the importance of current video content, rather than of the type of the incidents (\eg
% rebuffering vs. quality switch).
% This empirical finding suggests one can This finding allows us to encode the dynamic quality
% sensitivity by a score for each video chunk, independently to the quality assigned to each chunk,
% and these per-chunk weights can fit in nicely in existing DASH protocol.
These weights are then incorporated into an existing ABR algorithm to leverage the dynamic quality
sensitivity of upcoming video chunks when making quality adaptation decisions. We apply \name
to two state-of-the-art ABR algorithms: Fugu~\cite{yan2020learning}, a more traditional
rule-based algorithm, and Pensieve~\cite{mao2017neural}, a deep reinforcement learning
algorithm. Second, \name enables new adaptation actions that ``borrow''
resources from low-sensitivity chunks and give them to high-sensitivity chunks.
For example, it can lower the bitrate (or initiate a rebuffering event) even when bandwidth is
sufficient (or the buffer is not empty), in order to maintain a higher bitrate when quality
sensitivity becomes higher.
% by leveraging interfaces commonly available in mainstream video player implementations. \jc{Xu,
% please add details here}

%\jc{Summary of evaluation results.}
%\sid{This paragraph may need tweaking after we fill in the \fillme}
Using its scalable crowdsourcing approach, \name can predict QoE more
accurately than state-of-the-art QoE models. For example, with a budget of just \$31.4 per min video, \name
achieves 55\% less QoE prediction errors than existing models.
% can predict QoE 11.8\% more accurately than existing models.  
Compared to state-of-the-art ABR
algorithms, \name improves QoE on average by 15.1\% or achieves the same QoE with 32\%
less bandwidth. 
%\sid{Should this say compared to state-of-the-art ABR algorithms or ``video
%streaming approaches''? I'm not sure if the latter specifically pertains to the ABR algorithm
%performance.}

\mypara{Contributions and roadmap} Our key contributions are:
\begin{packeditemize}
    \item A measurement study revealing substantial temporal variability in users' quality
    sensitivity and the potential of improving video streaming QoE and bandwidth efficiency by
    embracing this variable sensitivity (\S\ref{sec:moti}).
    \item The design and implementation of \name, including: 
    \begin{packeditemize}
	\item A scalable, low-cost crowdsourcing solution to profiling the true dynamic user
	sensitivity of each video (\S\ref{sec:qoe_model}),\footnote{Our study was IRB-approved (IRB18-1851).} and
	\item A new ABR algorithm that incorporates dynamic user sensitivity into
	existing algorithms and frameworks (\S\ref{sec:abr}).
\end{packeditemize}
\end{packeditemize}

%% file: motivation.tex
\tightsection{Motivation}
\label{sec:moti}

\begin{figure}[t]
    \centering
    \includegraphics[width=1.0\linewidth]{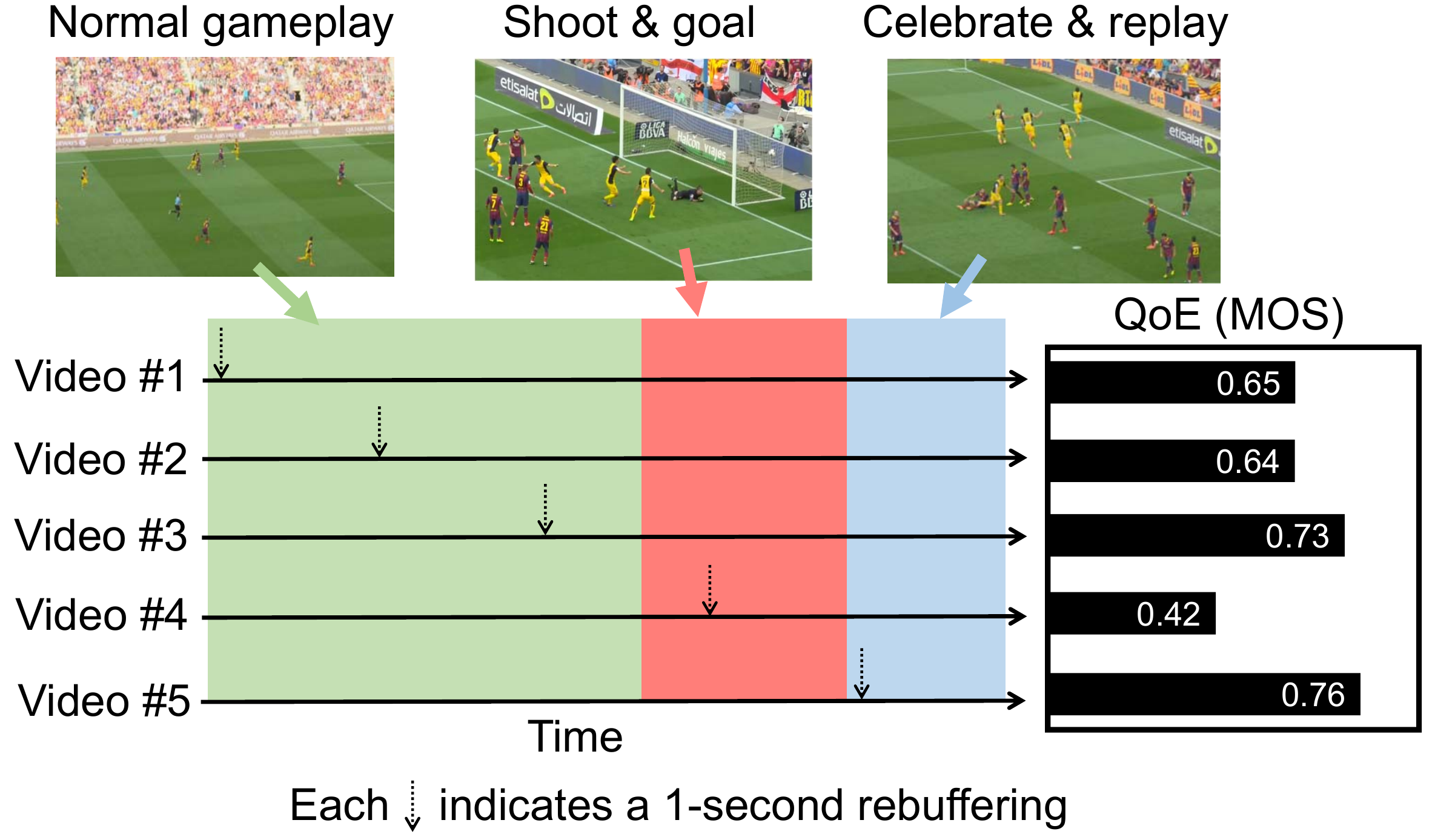}

    \tightcaption{Example of dynamic quality sensitivity. Users are asked to rate the quality (on
    a scale of 1 to 5) of different renderings of a source video (\texttt{Soccer1}), where a
    1-second rebuffering event occurs at a different place in each rendering. We observe
    substantial differences in the QoE impact (measured by mean opinion score, or MOS) across the
    renderings.}
    \label{fig:soccer_content}
\end{figure}

We begin by showing that existing approaches to modeling video streaming QoE
(\ref{sec:moti:modelling}) fail to accurately capture the true user-perceived QoE
(\ref{sec:moti:limit}). 
We then present user studies that reveal a missing piece in today's QoE
modeling---users' quality sensitivity varies dynamically throughout a video
(\ref{sec:moti:missing}).
% and that such dynamism of quality sensitivity is prevalent across videos and quality incident
% types (\ref{sec:moti:missing}).
% We illustrate the temporal variability of the video chunks' quality sensitivity
% (\ref{sec:moti:missing}), and the properties of this variability (\ref{sec:moti:prop}).
Finally, we show that by introducing dynamic quality sensitivity to ABR algorithms, we could
significantly improve QoE and save bandwidth (\ref{sec:moti:improve}).

\tightsubsection{Prior approaches to QoE modeling}
\label{sec:moti:modelling}

\mypara{Taxonomy of QoE models}
QoE models are a key component of modern video streaming systems.
A QoE model takes a streamed video as input and returns a predicted QoE as output.
% By the user ratings, we can train a \emph{QoE model} that takes videos as the input and outputs
% their predicted MOS.
When streaming a video, the video player optimizes the predicted QoE
(explicitly~\cite{yan2020learning,mao2017neural} or implicitly~\cite{huang2014buffer}) by adapting the bitrate of each video
chunk to the available bandwidth.
% The QoE models are set as the optimization objectives for the video streaming systems selecting
% the bitrate for every video chunk.
% QoE (quality of experience) is a central concept in video streaming.
% It captures how users are satisfied with the quality of a video.
QoE is often measured by mean opinion score (MOS), or the mean rating (\eg on a scale of 1 to 5)
assigned by a group of users to the quality of a video.\footnote{QoE is sometimes measured by other
metrics, such as user engagement (how long users watch a video). Our methodology extends to other
metrics.}

%\mypara{Taxonomy of QoE modeling}
Current QoE models focus on two aspects:
\begin{packeditemize}
\item {\em Pixel-based visual quality} tries to capture the impact of visual distortion on QoE,
based on the differences in pixel values between the encoded frames and the original
(uneconded) frames.
Metrics of pixel-based quality, sometimes called visual quality assessment (VQA), include
quantization parameter (QP)~\cite{rec2005h}, PSNR~\cite{gonzalez2004digital},
SSIM~\cite{wang2003multiscale,wang2004image}, STRRED~\cite{soundararajan2012video},
VMAF~\cite{vmaf} and DeepVQA~\cite{kim2018deep}.
\item {\em Streaming quality incidents} include events during the streaming process that negatively
impact user experience, such as rebuffering, low bitrate, and visual quality
switches.
%\sid{Replace ``low frame rate'' -> ``low bitrate''?} 
The impact of these incidents is modeled via summary metrics, such as rebuffering ratio, average
bitrate, frequency of bitrate switches during a video (\eg~\cite{dobrian2011understanding,balachandran2013developing}).
%\sid{Please check: I replaced ``buffering switches'' with ``bitrate switches''}
\end{packeditemize}
Some work also takes into account contextual factors (\eg viewer's emotion, acoustic conditions,
etc.), which are orthogonal to our focus on the impact of objective quality on QoE.
%\sid{Might explain this a bit if we have time/space, but low priority.}

% Existing QoE models have explored various ways of combining the pixel-based visual quality and quality incidents to predict the subjective QoE.

% use mathematical models to calculate the pixel-level difference between the source video and the user-viewed video; (2) Subjective VQA models, \eg VMAF \cite{vmaf}, and DeepVQA \cite{kim2018deep}, take the user feedback into account for detecting the user-perceivable visual distortion.

% try to calculate the visual distortion, and can be categorized into (1) Objective VQA models, such as quantization parameter (QP)\cite{rec2005h}, PSNR \cite{gonzalez2004digital}, SSIM \cite{wang2003multiscale,wang2004image} and STRREED \cite{soundararajan2012video}, use mathematical models to calculate the pixel-level difference between the source video and the user-viewed video; (2) Subjective VQA models, \eg VMAF \cite{vmaf}, and DeepVQA \cite{kim2018deep}, take the user feedback into account for detecting the user-perceivable visual distortion.

\mypara{State-of-the-art QoE models}
The latest QoE models combine pixel-based visual quality metrics and quality-incident metrics to
achieve better QoE prediction.
We focus on three such QoE models: KSQI, P.1203, and LSTM-QoE, which were proposed within the past
two years and have open-source implementations.
KSQI~\cite{duanmu2019knowledge} combines VMAF, rebuffering ratio, and quality switches in a linear
regression model.
% in a principle of not violating the prior knowledge on the human vision system.
P.1203 \cite{robitza2017modular} combines QP values and quality incident metrics in a 
random-forest model.
More recently, LSTM-QoE \cite{eswara2019streaming} takes STRRED and individual quality incidents as
input to a long short-term memory (LSTM) network designed to capture the ``memory effect'' of human
perception of past quality incidents. (We discuss related work in \S\ref{sec:relatedwork}.)

\tightsubsection{QoE prediction accuracy today}
\label{sec:moti:limit}

To evaluate the prediction accuracy of these QoE models, we created a video set of 16 source videos
randomly selected from four public
datasets~\cite{ghadiyaram2017subjective,bampis2018towards,duanmu2018quality,wang2019youtube} and 7
throughput traces randomly selected from real-world cellular networks~\cite{riiser2013commute,
hsdpa}.
The source videos (summarized in Table~\ref{tab:summary}) cover a wide range of content
genres (sports, scenic, movies, etc.), and the throughput traces exhibit bandwidths from 200{Kbps}
to 6{Mbps}.
Following recent work (\eg~\cite{duanmu2019knowledge,mao2017neural}), we replay each trace and
emulate the process of streaming each video using different ABR algorithms:
Fugu~\cite{yan2020learning}, Pensieve~\cite{mao2017neural}, and BBA~\cite{huang2014buffer}.
This creates 336 (16$\times$7$\times$3) rendered videos.
% To avoid biases caused by difference among the training data of these QoE models,

Using this dataset, we train KSQI and LSTM-QoE on a subset of 315 videos and test them on the
remaining 21 videos. (We ensure that our trained versions have higher accuracy on the test set than
the pre-trained models.) To obtain the ground truth, we elicit QoE ratings from crowdsourced workers
on Amazon MTurk~\cite{mturk} following the methodology described in \S\ref{sec:qoe_model}. We use
the MOS over a sufficient number of ratings (>30) as the true QoE for each of the 336 videos.  We
normalize the output range of the QoE models (and the true QoE) to $[0,1]$.

%\sid{What does this last sentence mean?}

% \jc{xu, say something about how these models are trained?}
% We randomly select 16 source videos from four open datasets: LIVE Mobile Stall Video Database-II (LIVE-MOBILE) \cite{ghadiyaram2017subjective}, LIVE-NFLX-II \cite{bampis2018towards}, WaterlooSQOE-III \cite{duanmu2018quality} and YouTube-UGC \cite{wang2019youtube}.
% The details of the source videos are in the Figure~\ref{fig:dataset_screenshot}.
% We also randomly select 7 throughput traces collected from real-world 3G networks with a wide range of network bandwidth\cite{riiser2013commute, hsdpa}.
% There are three state-of-the-art ABR algorithms for generating the video sequences: BBA \cite{huang2014buffer}, Pensieve \cite{mao2017neural}, Puffer \cite{yan2020learning} \footnote{We are not using Oboe\cite{akhtar2018oboe} which is for tuning the parameters in existing ABR algorithms, not suitable for this experiement}.

\begin{figure}[t]
	\centering
	\includegraphics[width=0.33\textwidth]{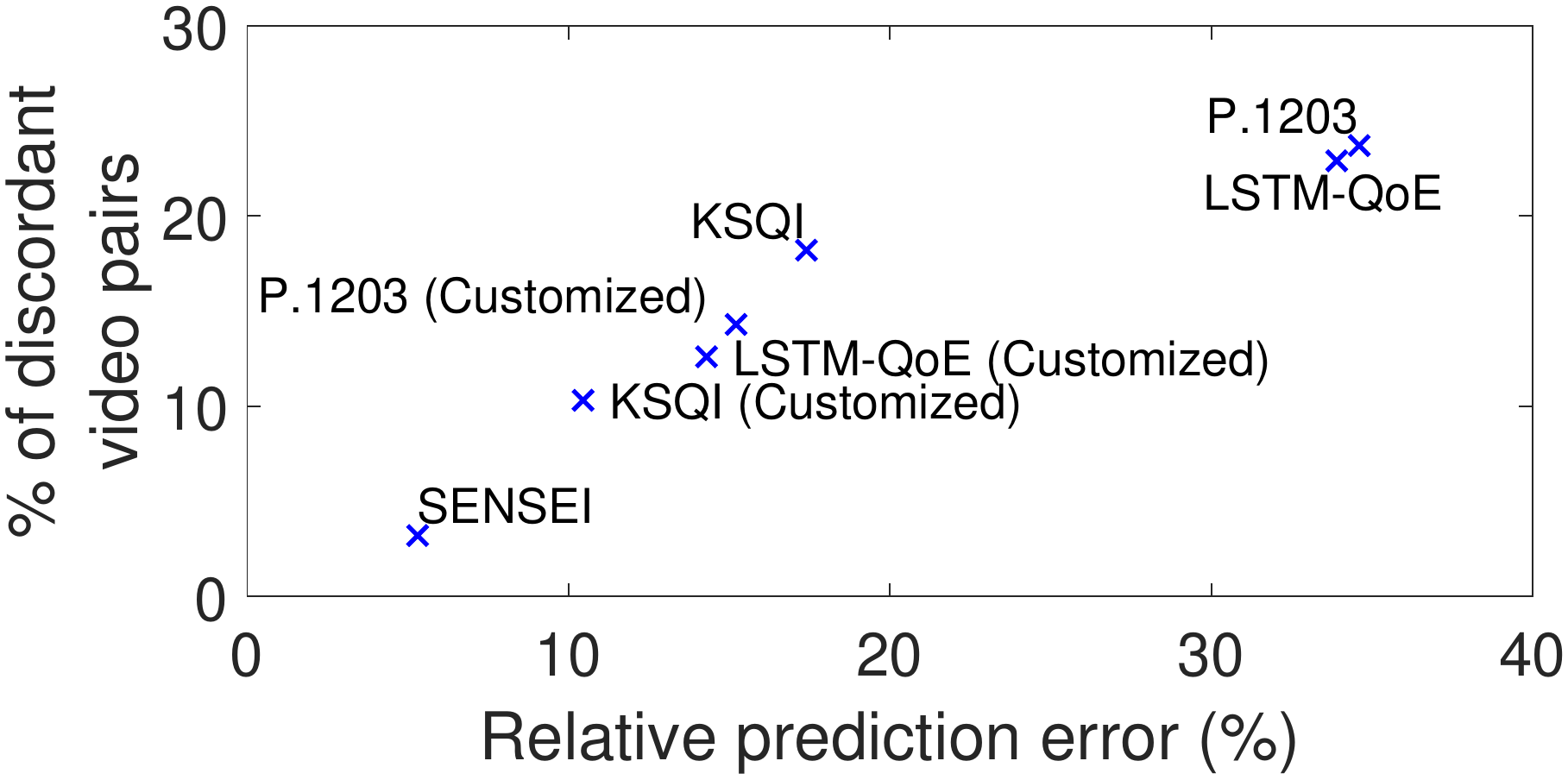}
	\tightcaption{Existing QoE models exhibit substantial QoE prediction errors (x-axis), which cause them
	to frequently mis-predict the relative QoE ranking between two ABR algorithms on the same video,
	\ie a discordant pair (y-axis).}
	\label{fig:moti:accurcy}
\end{figure}

% We measure their QoE prediction accuracy over the 336 videos in two ways.
The x-axis of Figure~\ref{fig:moti:accurcy} shows the mean relative prediction error of each QoE
model on our test set, where the relative prediction error is defined as
${\scriptstyle|Q_{predict}-Q_{true}|/{Q_{true}}}$ for each video; ${\scriptstyle
Q_{predict}}$ and ${\scriptstyle Q_{true}}$ are the predicted and true QoE of the video.
We see that these errors are nontrivial; even the most accurate QoE model has over 10.4\%
error on average.
% We see that they all have substantial prediction errors: even the relatively most accuracy model,
% KSQI, has over \fillme mean error.

We also examine whether these models can correctly rank the QoE achieved by two different ABR
algorithms.
% , despite their non-trivial QoE prediction error.
For each pair of source video and throughput trace, we first rank every two of the three ABR
algorithms using their true QoE and then rank them using the predicted QoE.
If the rank is different, this pair is called a discordant pair.
% We first group the videos by their source video and throughput trace so each group has three
% videos resulting from streaming the same source video and throughput trace by different ABR
% algorithms.
% Then for each pair of videos in a group, we call it a discordant pair if the QoE ranking predicted
% by the QoE models is opposite to their real QoE ranking.
The y-axis of Figure~\ref{fig:moti:accurcy} shows the fraction of discordant pairs among all
possible pairs (a common measure used in rank correlation metrics). We see that over 10.2\% of
pairs are discordant even for the most accurate QoE model.
% Figure~\ref{fig:moti:accurcy} shows that the state-of-the-art QoE models have low prediction
% accuracy and monotonicity.
% Such prediction errors could mislead the ABR algorithms to believe that one bitrate selection
% decision is better than the other, even when the real ranking is on the other way around.
% Unsurprisingly, all three QoE models will frequently rank the QoE of one ABR algorithm over
% another while the true QoE of the latter is better.
This has significant implications, because content providers and academic researchers heavily rely
on QoE models to design and compare different ABR algorithms~\cite{mao2017neural, yan2020learning,
huang2014buffer}.

\tightsubsection{Temporal variability of quality sensitivity}
\label{sec:moti:missing}

% \begin{figure}[t]
%     \centering
%     \includegraphics[width=0.24\linewidth]{eval_graphs/gap_std.pdf}
%     \caption{The CDF of QoE STD when we add a quality incident at different positions in the source video.}
%     \label{fig:ubiqui_gap_std}
% \end{figure}

% \begin{figure}[t]
%     \centering
%     \includegraphics[width=0.24\linewidth]{eval_graphs/biggest_gap.pdf}
%     \caption{The CDF of the biggest QoE gap when we add a quality incident at different positions in the source videos.}
%     \label{fig:ubiqui_biggest_gap}
% \end{figure}

Unlike the prior methods, Figure~\ref{fig:moti:accurcy} shows that our QoE model (which we present
in \S\ref{sec:qoe_model}) can predict QoE and rank ABRs significantly more accurately when evaluated
on the same train/test set.
We argue that the source of this gap is an underlying premise shared by all previous QoE models,
which is that all factors affecting QoE can be captured by a handful of
objective metrics.  This premise ignores the impact of high-level video content (rather than
low-level pixels and frames) on users' \emph{subjective sensitivity} to video quality at different
parts of the video.
We now demonstrate how this quality sensitivity varies as video content changes. 
%and why such % dynamism is hard for today's QoE models to capture. %the temporal variability of the
% user attention to the quality.

\mypara{Revealing dynamic quality sensitivity}
To measure the true quality sensitivity of users at different parts of a video, we create a {\bf
video series} for each source video and quality incident as follows.
Videos in a video series have the same source content and highest quality (highest bitrate
without rebuffering), except that a low-quality incident (a rebuffering event or a bitrate
drop) is deliberately added at different positions (\eg at the $4^{\textrm{th}}$ second,
$8^{\textrm{th}}$ second, and so forth).
Then, as before, we use Amazon MTurk to crowdsource the true QoE
of each video, measured by MOS (see \S\ref{sec:qoe_model}).
%, making sure each video QoE is based on at least 30 ratings.

%This process ensures that quality of videos in the same series would appear identical from most QoE
%metrics,\footnote{Although some VQA metric will return different QoE estimates if bitrate drops
%happen at different places, but as we will see their estimates are still weakly correlated with the
%real QoE rating.} and any QoE differences among them will likely result from the content of video
%content.

Figure~\ref{fig:soccer_content} shows an example video series created using a 25-second soccer video
as the source video and a one-second rebuffering event as the low-quality incident.
% multiple videos all of which have the same source video (a 25-second soccer video) encoded in the
% same bitrate and have a one-second rebuffering event, and the only difference among these videos
% is when the rebuffering occurs.
% Most prior QoE models would have predicted same QoE for these videos.
We observe significant differences between the QoE of these videos.
The gap between the highest QoE (rebuffering at the $10^{\textrm{th}}$ second) and lowest QoE
(rebuffering at the $15^{\textrm{th}}$ second) is over 40\%.
Notice that the quality sensitivity varies in just a few seconds, which suggests that the same
low-quality incident can have significantly higher/lower impact on user experience if it
occurs a few seconds earlier or later.

In contrast, most prior QoE models would predict the same QoE for all videos in the series. The few
QoE models that do give different ratings to the videos yield ratings that have little correlation
with the videos' true ratings.
For example, VMAF~\cite{vmaf} (the visual quality metric used by KSQI) gives lower QoE estimates if
a bitrate drop occurs when the frame pixels are more ``complex'', and
LSTM-QoE~\cite{eswara2019streaming} gives lower QoE estimates if rebuffering occurs at more
``dynamic'' scenes. In Figure~\ref{fig:soccer_content}, the true lowest QoE occurs when the
low-quality incident occurs during the goal, but both VMAF and LSTM-QoE predict the lowest QoE
when low-quality incidents occur during normal gameplay. We confirm this phenomenon in other videos
as well. The inaccuracy of these models is symptomatic of their assumption that the impact of
video content can be captured by pixels and motions between frames.

Another alternative is to use computer vision models~\cite{gao2019content} to identify temporal key moments in a video.
% Other efforts rely on users watching similar content in the past~\cite{gao2018optimizing} or use off-the-shelf vision model~\cite{gao2019content} to identify temporal key moments in a video. 
We show in Appendix~\ref{sec:cv}, however, that these models also fall short.

\begin{figure}[t]
	\centering
% 	\begin{subfigure}[t]{.45\linewidth}
%         \centering
%         \includegraphics[width=\linewidth]{eval_graphs/gap_std.pdf}  
%         \caption{Standard deviation of QoE}
%         \label{fig:ubiqui_gap_std}
%     \end{subfigure}
%     \hfill
%     \begin{subfigure}[t]{.45\linewidth}
%         \centering
%         \includegraphics[width=\linewidth]{eval_graphs/biggest_gap.pdf}
%         \caption{Difference between max and min QoE}
%         \label{fig:ubiqui_biggest_gap}
%     \end{subfigure}
     \includegraphics[width=0.3\textwidth]{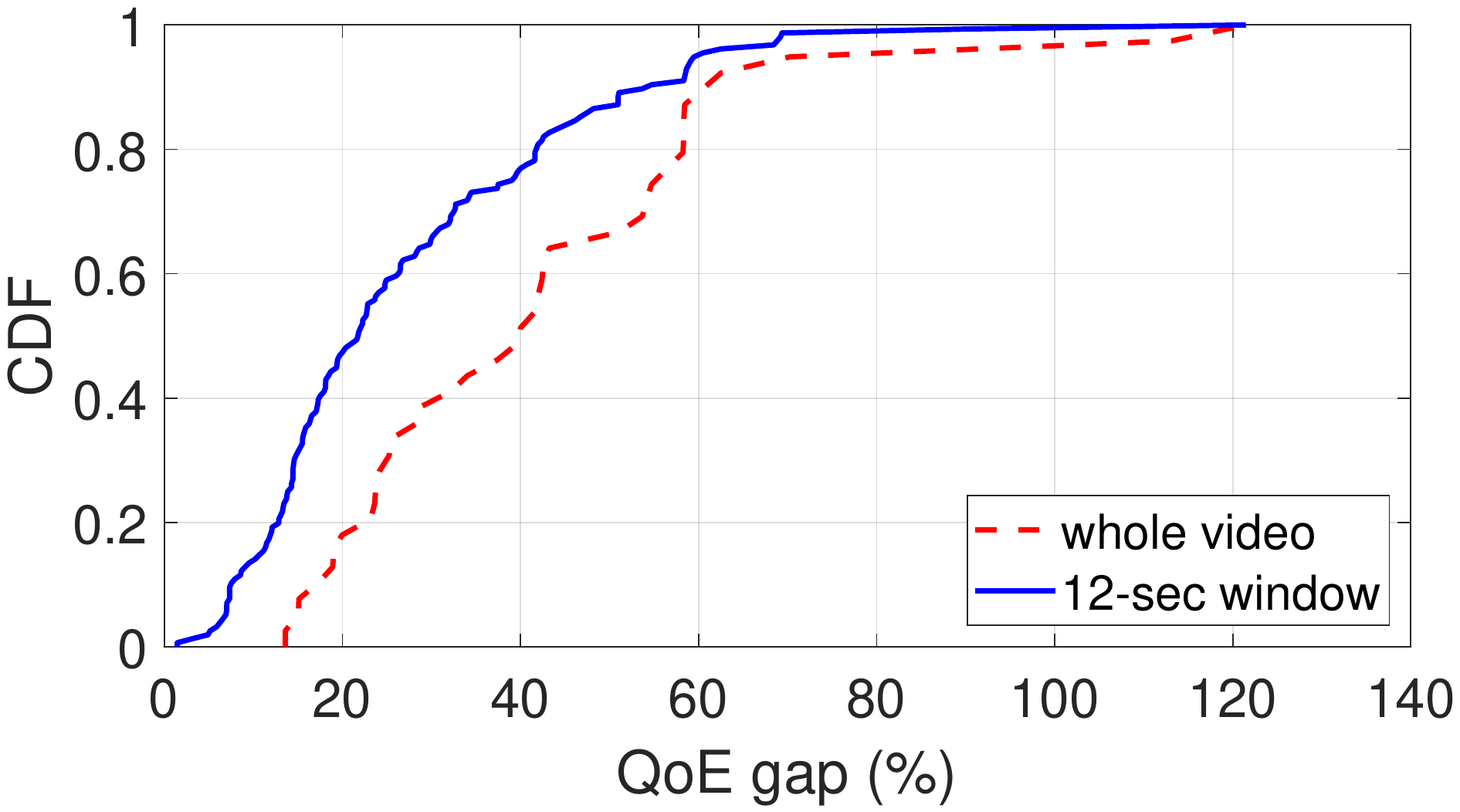}  
    \tightcaption{
    %Distribution of QoE variance when a quality incident (1-second rebuffering, 4-second rebuffering, or a bitrate drop for 4 seconds) is added at different places in the same video.
    Distribution of the max-min QoE gap when a low-quality incident (1-second rebuffering, 4-second
    rebuffering, or a bitrate drop for 4 seconds) is added at different places in the same video.
    %\sid{Xu: is this the right graph/labels? What does ``distribution of QoE variance'' mean?}
    %\xu{This is the right graph. QoE variance here means the QoE gap (defined in the text). The
    % lines here mean the QoE gap if we add a quality incident within the whole video, a 12-second window inside the video, a 16-second window of the video. May be we should use QoE gap instead of QoE variance here? }
    The trend is similar even if the low-quality incident and QoE gap is localized to a 12-second window.}
    \label{fig:variance}
\end{figure}

\mypara{Dynamic quality sensitivity is common}
We repeat the same experiment on all 16 source videos in our dataset and three low-quality
incidents: 1-second rebuffering, 4-second rebuffering, and a bitrate drop from 3Mbps to 0.3Mbps for
4 seconds. This creates 48 video series in total.
% we show the temporal variability of quality sensitivity is substantial across the genres.
As summarized in Figure~\ref{fig:variance}, we observe a similar temporal variability in quality
sensitivity.
% (though with different magnitudes of variability).
% Figure~\ref{fig:ubiqui_gap_std} shows that \fillme of the 48 video series have relative QoE
% standard deviation (\ie std/mean) of over \fillme\%.
The figure plots ${\scriptstyle (Q_{max}-Q_{min})/Q_{min}}$ for each video series, where $Q_{max}$
and $Q_{min}$ are the maximum and minimum QoE of the videos in a series.
We see that 21 of the 48 video series have a max-min QoE gap of over 40.1\%. A similar trend holds
even if we localize the low-quality incident and max-min QoE gap measurement to a 12-second window (repeated for all such
windows at 4-second boundaries). This shows that quality sensitivity varies substantially even among
very nearby chunks.

\begin{figure}[t]
	\centering
	\begin{subfigure}[t]{.36\linewidth}
        \centering
        \includegraphics[width=\linewidth]{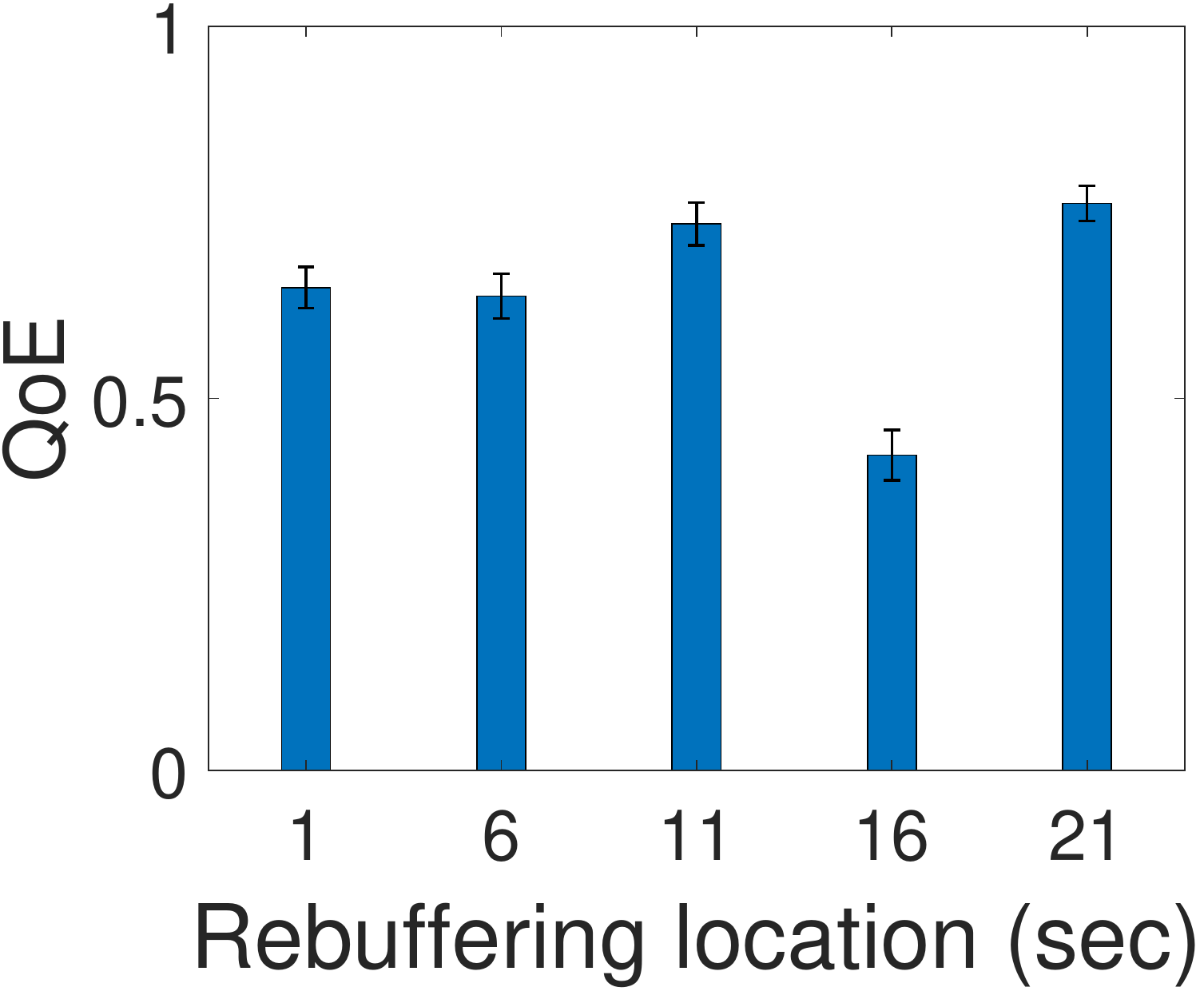}  
        \caption{1-sec rebuffering}
        \label{fig:moti:soccer:sub1}
    \end{subfigure}
    \begin{subfigure}[t]{.30\linewidth}
        \centering
        \includegraphics[width=\linewidth]{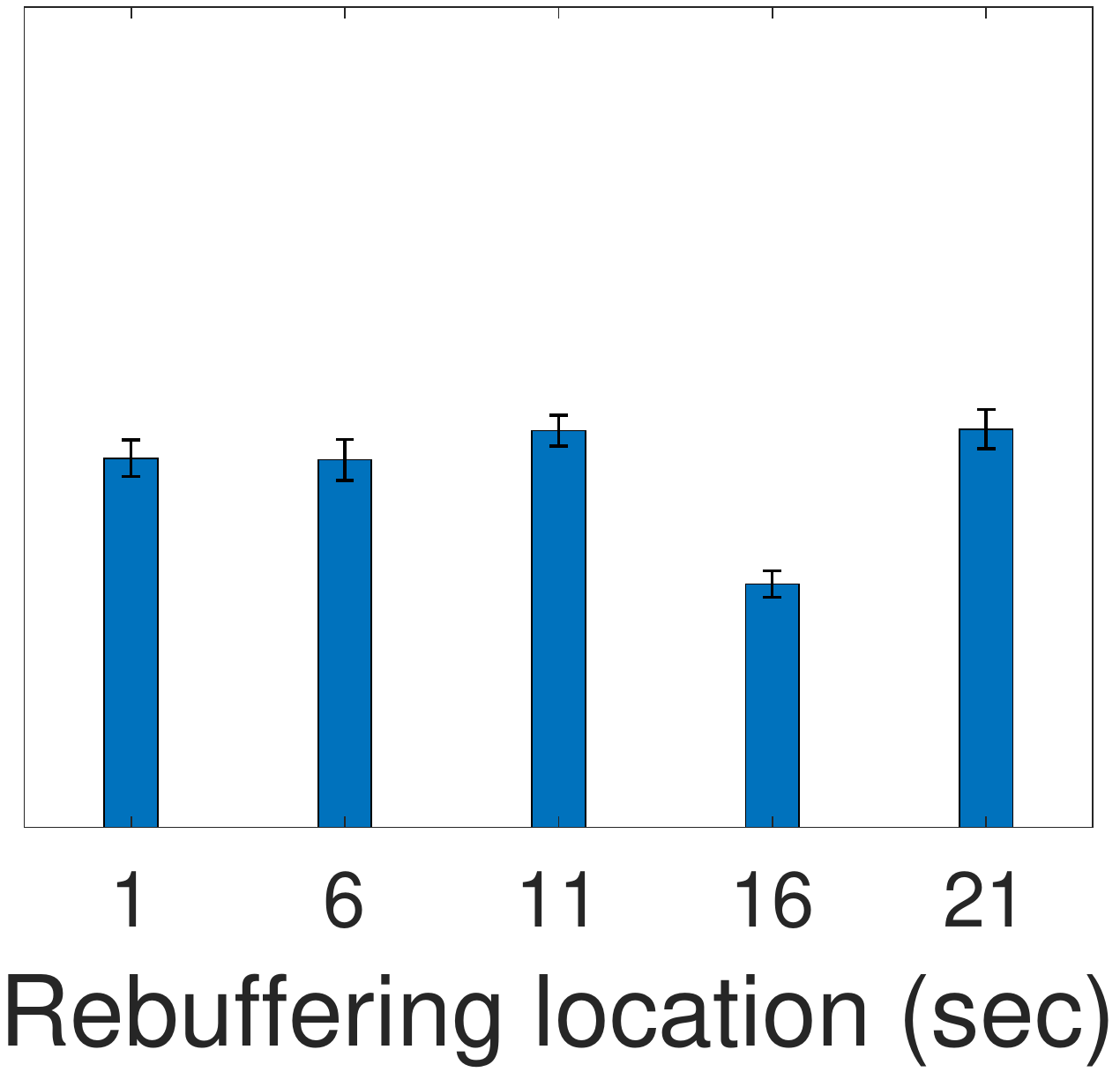}
        \caption{4-sec rebuffering}
        \label{fig:moti:soccer:sub2}
    \end{subfigure}
	\begin{subfigure}[t]{.298\linewidth}
        \centering
        \includegraphics[width=\linewidth]{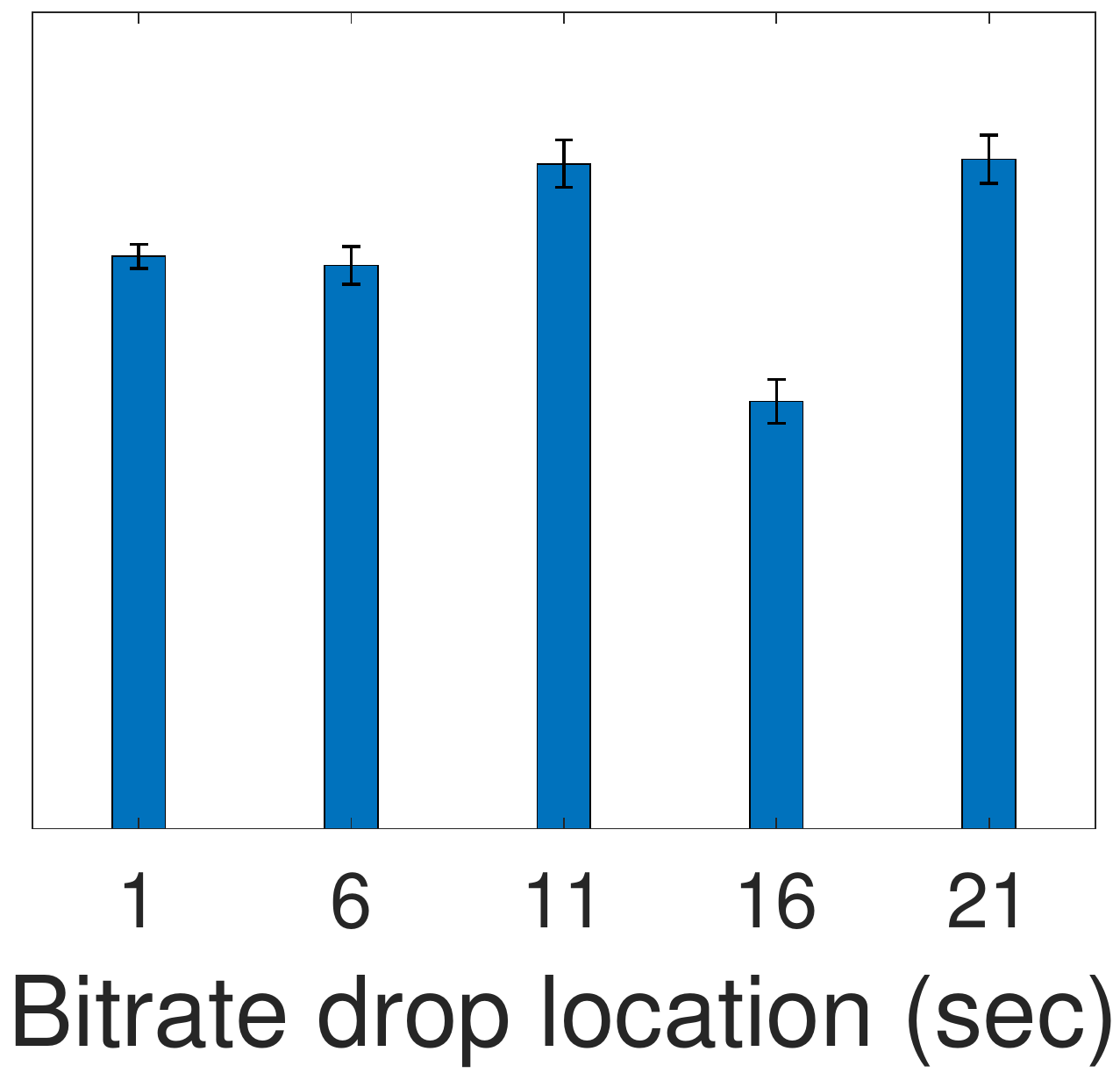}
        \caption{Bitrate drop}
        \label{fig:moti:soccer:sub3}
    \end{subfigure}
	\tightcaption{The QoE variability if we put the same quality incident at different places of the video in Figure~\ref{fig:soccer_content}.
	The pattern of variability remains the same under different quality incidents being added.}
	\label{fig:moti:soccer}
\end{figure}

\begin{figure}[t]
	\centering
	\begin{subfigure}[t]{.44\linewidth}
        \centering
        \includegraphics[width=0.95\linewidth]{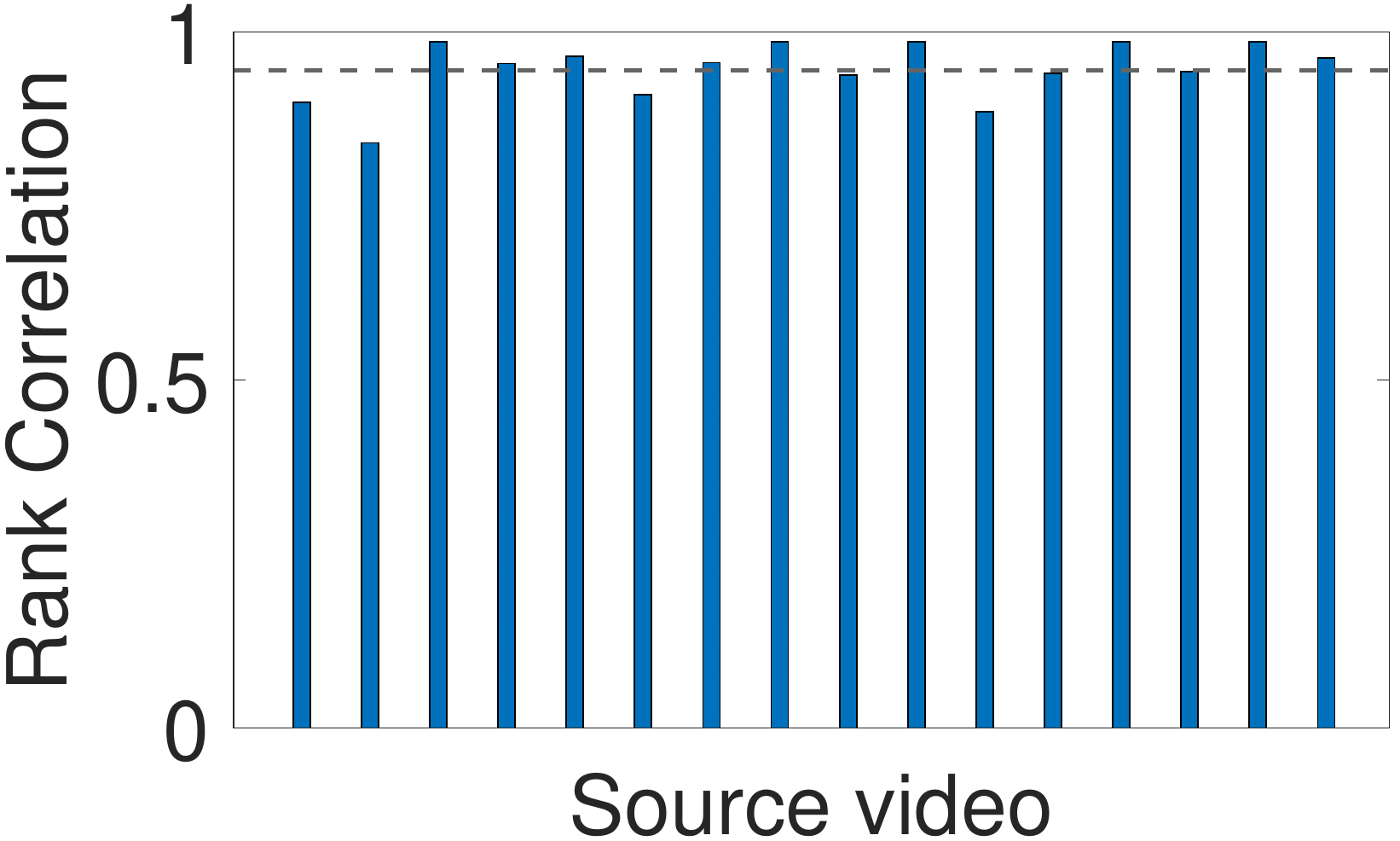}
        \caption{1-sec vs. 4-sec rebuffering event}
        \label{fig:moti:agnostic:sub1}
    \end{subfigure}
    \begin{subfigure}[t]{.44\linewidth}
        \centering
        \includegraphics[width=0.95\linewidth]{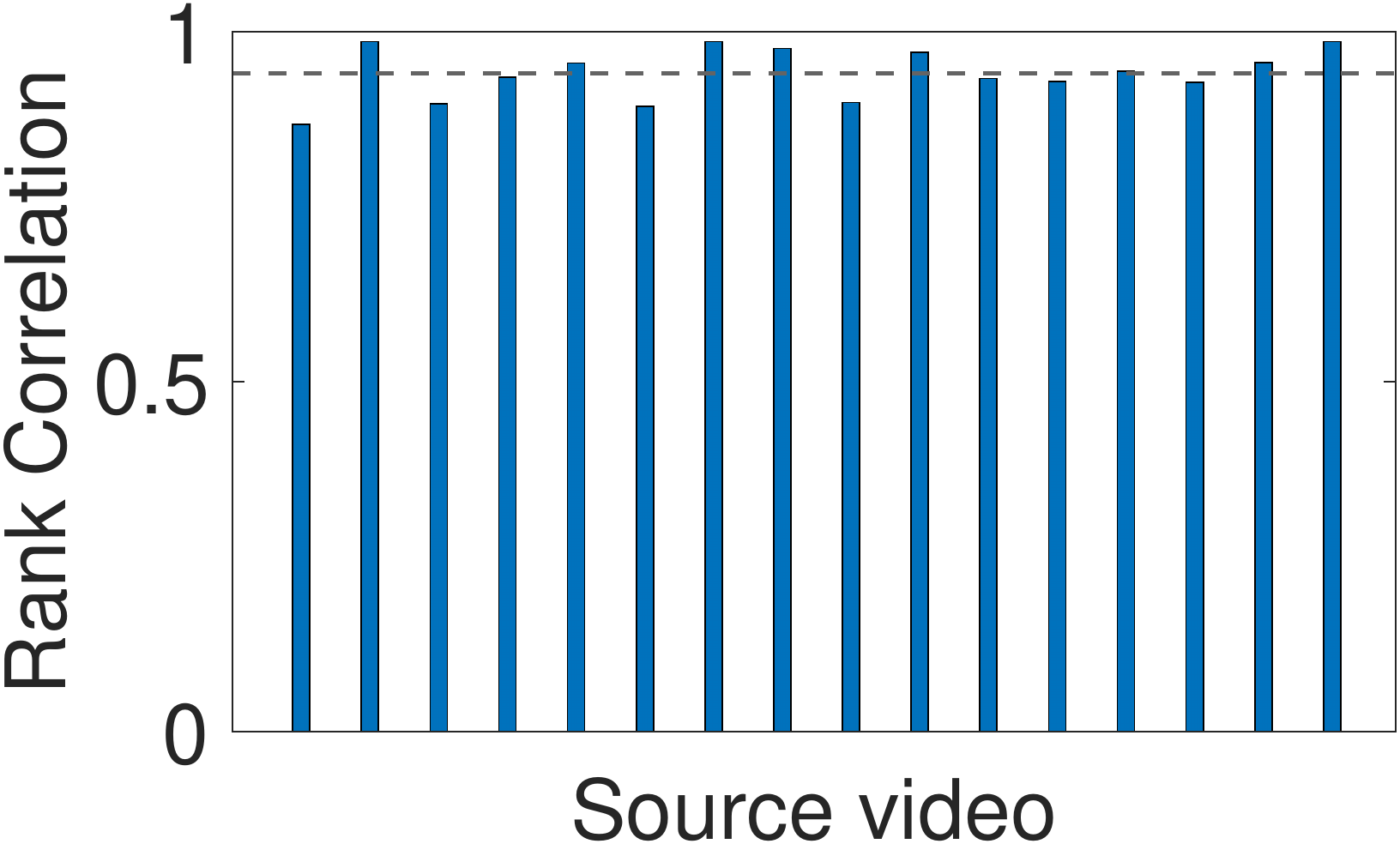}  
        \caption{1-sec rebuffering event vs. a 4-sec bitrate decrease}
        
        \label{fig:moti:agnostic:sub2}
    \end{subfigure}
    \tightcaption{QoE correlation between quality incidents}
    \label{fig:agnostic}
\end{figure}

\mypara{Quality sensitivity is inherent to video content}
Our user study also suggests that, although the type of low-quality incident does affect absolute
QoE, the QoE ranking within a video series is largely agnostic to the particular incident. Instead,
quality sensitivity seems to be inherent to different parts (contents) of the video. In
other words, if users are more sensitive to a 1-second rebuffering event at chunk $i$ than at chunk $j$, they
tend to be more sensitive to other low-quality incidents at chunk $i$ than at chunk $j$.
Figure~\ref{fig:moti:soccer} shows the dynamic user sensitivity of three low-quality incidents:
1-second rebuffering, 4-second rebuffering, and a bitrate drop.
Although the absolute values of the QoE depend on the quality incident, the relative
rankings are identical.
To generalize this finding, we also calculate the rank correlation (Spearman's rank coefficient)
between the QoE values of two video series from the same source video generated using different
low-quality incidents. Figure~\ref{fig:agnostic} shows the rank correlation between rebuffering for
different lengths of time, and between rebuffering and a bitrate drop. Both show strong rank
correlation.

%\sid{Moved to where Figure 1 is discussed, as this seems out of place here.}
% Now, we acknowledge that some QoE models do give different ratings to videos in a video series, but their ratings have little correlation with the videos' real ratings. 
% To use the soccer video in Figure~\ref{fig:soccer_content} as an example, VMAF~\cite{vmaf} (visual quality metric of KSQI) gives lower QoE estimates when a bitrate drop occurs when the frame pixels are more ``complex'' and LSTM-QoE~\cite{eswara2019streaming} also gives lower QoE predictions when rebuffering occurs at more ``dynamic'' scenes. 
% The true lowest QoE happens when the quality incidents occur at the goal, but VMAF rates bitrate drop at normal playtime the lowest and LSTM-QoE also rates rebuffering at normal playtime the lowest. 
% We confirm this phenomenon in other videos as well.
% 
% The inability these models are symptomatic of the common assumptions that the impact of video content (if any) can be captured by pixels and motions between frames, but as we have seen, they all fall short.
% So what really causes quality sensitivity to vary?

\mypara{Sources of dynamic quality sensitivity} 
We speculate that the source of dynamic quality sensitivity stems from users paying
different degrees of attention to different parts of a video.
In our dataset, we identify at least three types of moments when users tend to be more (or less)
attentive to video quality than usual.
The first are key moments in the storyline of a video when tensions have built up; \eg in one of the
animation videos (\texttt{BigBuckBunny}) when the tiny
bullies fall into a trap designed by the bunny, or when a goal is scored in the soccer video (\texttt{Soccer1}).
The second are moments when users must pay attention to get important
information; \eg change of the scoreboard in a sports video (\texttt{Soccer2}), or obtaining
supplies after killing an enemy (\texttt{FPS2}).
The third are transitional periods with scenic backgrounds, when users tend to be less
attentive to quality; \eg the universe background in \texttt{Space}.
One can expect many more cases.

\tightsubsection{Potential gains}
\label{sec:moti:improve}

Finally, we show that the temporal variability of quality sensitivity could be leveraged by ABR
algorithms to optimize QoE and save bandwidth, by {\em aligning quality adaptation
with dynamic user sensitivity}.
%\eg by avoiding low quality during a period of high quality sensitivity.
% If we could move the quality incidents from the places that has high impact on QoE to a lower
% place, the QoE could be improved significantly.
% As Figure~\ref{fig:ubiqui_biggest_gap} shows, we could improve up to 58.3\% QoE potentially.

We show the potential gains using an idealistic but clean experiment.
We create two simple ABR algorithms whose only difference is the QoE model they explicitly optimize:
one algorithm optimizes KSQI, the most accurate QoE model from Figure~\ref{fig:moti:accurcy} that is
{\em unaware} of dynamic quality sensitivity, and the other optimizes our eventual QoE model
from~\S\ref{sec:qoe_model}, which {\em is} aware of dynamic quality sensitivity.
Both algorithms take as input an entire throughput trace and the same video chunks encoded in the
same available bitrate levels (we assume 4-second chunks).
They then decide a bitrate-to-chunk assignment that maximizes their respective QoE model. (We assume
throughput is not affected by bitrate selections.) Note that these ABR algorithms are idealistic
because they have access to the entire throughput trace in advance, and hence know the future
throughput variability. However, this allows us to eliminate the confounding factor of throughput
prediction.
We pick one of the throughput traces (results are similar with other throughput traces) and rescale
it to $\{20, 40, \dots, 100\}\%$ to emulate different average network throughput.

\begin{figure}[t]
	\centering
	\includegraphics[width=0.3\textwidth]{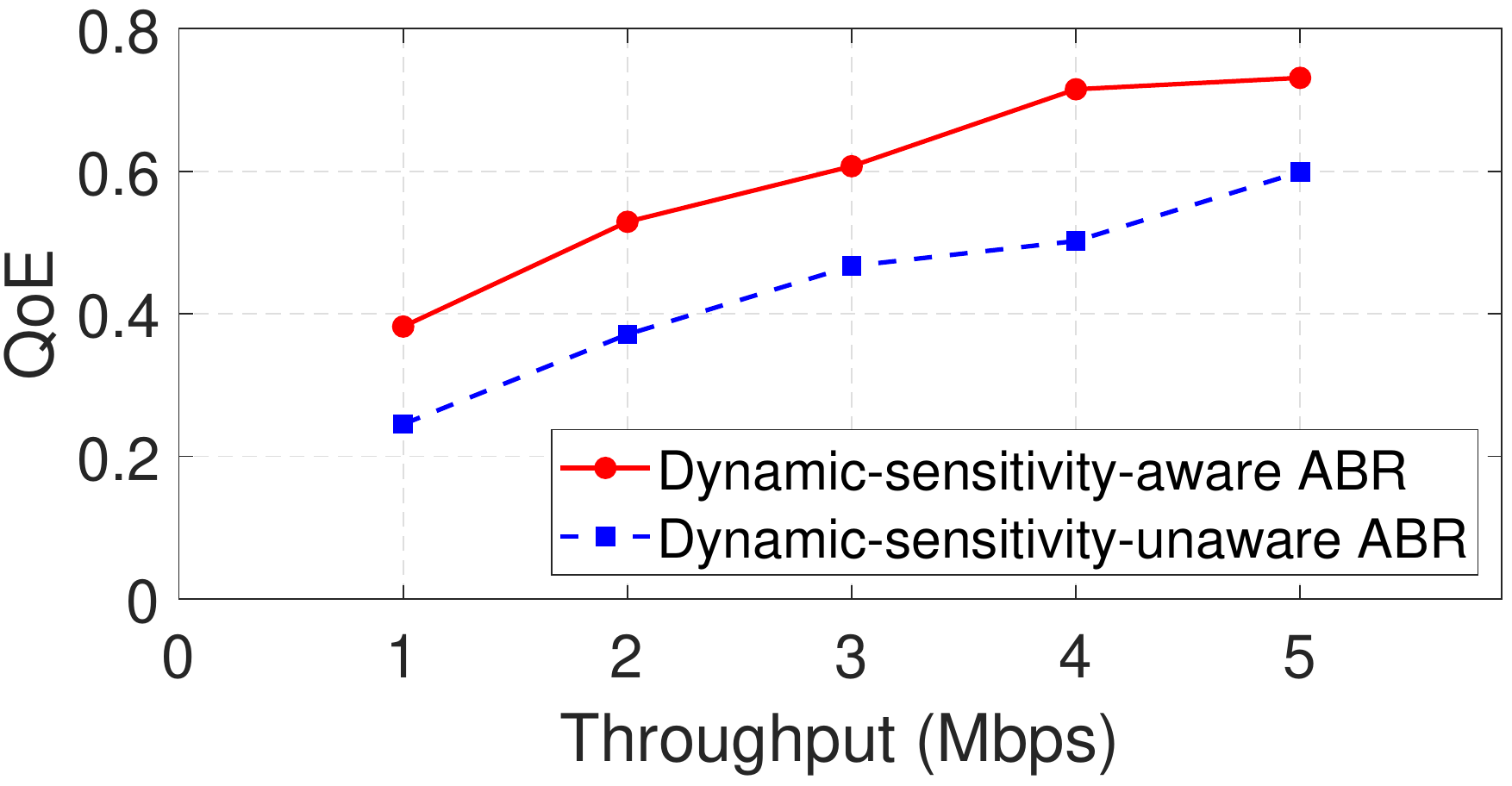}
	\tightcaption{Being aware of dynamic quality sensitivity can significantly improve QoE and save
	bandwidth.}
	\label{fig:moti:potgain}
\end{figure}

For each source video, we create the rendered video as if it were streamed by each ABR algorithm
(with rebufferings, bitrate switches, etc.). We use Amazon MTurk as before (see 
\S\ref{sec:qoe_model}) to assess the true QoE of the rendered video.
Figure~\ref{fig:moti:potgain} reports the average QoE of the two ABR algorithms across 16 source
videos and different average bandwidths.
We can see that being aware of dynamic quality sensitivity could improve QoE by 22-52\% while using
the same bandwidth, or save 39-49\% bandwidth while achieving the same QoE.
%save 36.2\% if we consider the temporal variability of the quality sensitivity.

\tightsubsection{Summary of findings}
\label{sec:moti:sum}

Our findings can be summarized as follows:
\begin{packeditemize}
    \item Existing QoE models have substantial room for improving their accuracy.
    In our dataset, recent models predict QoE with up to 35\% errors, which could mislead the
    design and choice of ABR algorithms.
    % 17.4-35.9\%, which causes them to rank the video sequences in wrong orders for 18.2-23.9\%.
    \item A common source of error in these QoE models is that they fail to capture
    content-induced variability in quality sensitivity. Quality sensitivity varies by 42\% on
    average in our videos and up to 121\% for some videos. 
    %\sid{Xu: can yo ucheck these numbers?
    %We should ensure they are consistent with the earlier text in this section.}\xu{have fixed
    %that, aligned with fig 3}
    % The temporal variability of quality sensitivity is content-dependent.
    % This variability is significant: for out of our source videos, the variance of MOS could be up to 0.152 if we put a quality incident at different positions.
    \item By making QoE models aware of the temporal variability of quality sensitivity, we could
    potentially improve QoE by up to 52\% while using the same bandwidth, or save up to 49\%
    bandwidth while achieving the same QoE.
\end{packeditemize}

%% file: qoemodel.tex
\tightsection{\name overview}
\label{sec:overview}

\begin{figure}[t]
    \centering
    \includegraphics[width=0.47\textwidth]{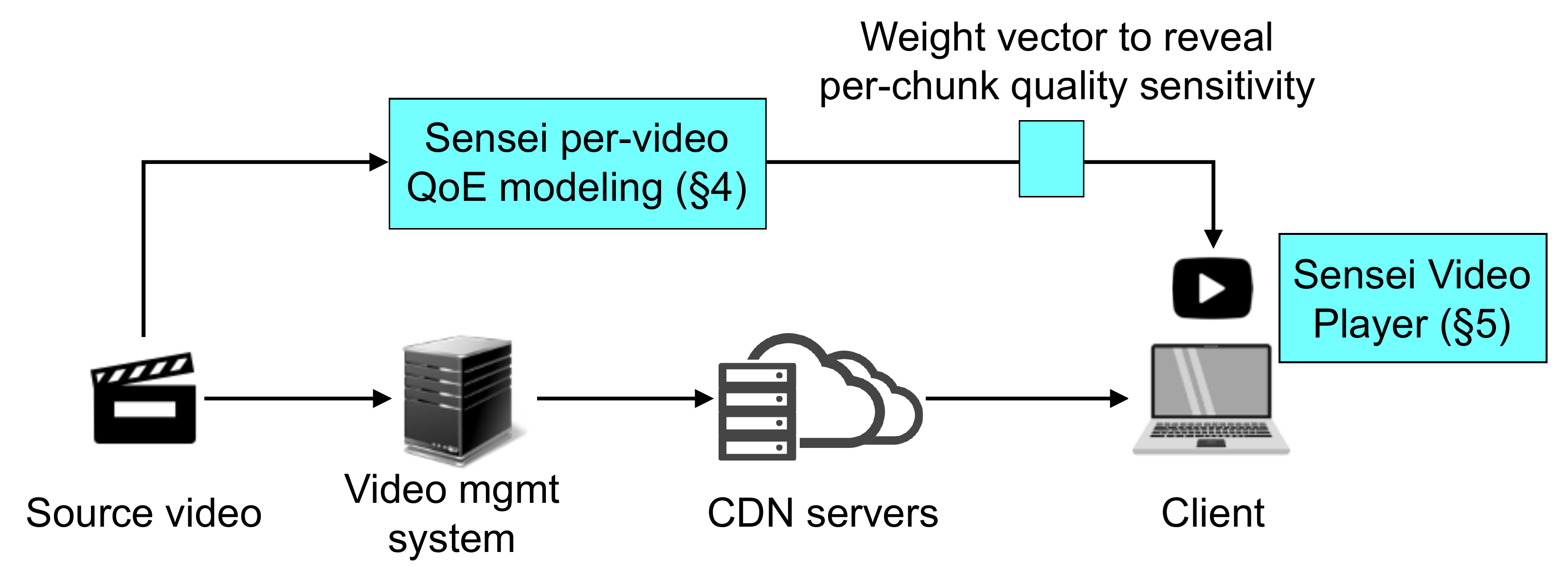}
    \tightcaption{Overview of \name.
    % \sid{This should be ``per-video'', not ``pre''}
    }
    \label{fig:system_overview}
\end{figure}

So far we have shown that true quality sensitivity is a key missing piece 
in today's QoE models that could significantly enhance the performance of ABR algorithms.
To unleash this potential in practice, we present \name, a video streaming system with two
main components (Figure~\ref{fig:system_overview}):
% a QoE profiler and a custom ABR algorithm.
% \name addresses two key challenges.

\begin{packeditemize}
\item {\em Scalable QoE modeling (\S\ref{sec:qoe_model}):}
As we observed in \S\ref{sec:moti:missing}, current approaches to QoE modeling are unable
to capture the complex relationship between video content (\eg key moments in
storylines) and user sensitivity to various video quality incidents.
Instead, we advocate for directly asking human viewers to rate the quality of a streamed video.
This reveals the true user sensitivity to various quality incidents, rather than inferring
it indirectly through heuristics.
Since quality sensitivity is influenced by each video's content, this user study must
scale out many videos. \name leverages crowdsourcing to automate the QoE profiling process per
video while maintaining reliable user rating quality.
\item {\em Sensitivity-aware ABR (\S\ref{sec:abr}):}
Video players today are designed to greedily maximize quality (high bitrate without
rebuffering) of each chunk.
This tenet is ill-suited to our goal of aligning quality adaptation with dynamic
quality sensitivity---quality should be optimized in proportion to the quality sensitivity of the
content. To achieve this, \name refactors the control layer of video players to
enable new adaptation actions that ``borrow'' resources from low-sensitivity chunks and give them to
high-sensitivity chunks.
\end{packeditemize}

\mypara{Key abstraction of per-chunk weights}
The enabler behind \name is the abstraction of video chunk-level weights that describe the
inherent quality sensitivity of different parts of a video.
It is inspired by the property of quality sensitivity that it is inherent
to video content and largely agnostic to the type of quality incident 
(\S\ref{sec:moti:missing}). At a high level, this abstraction has two practical benefits.
First, it allows us to reuse existing QoE models but reweight the quality of different chunks
based on their true quality sensitivity. This drastically cuts the cost and delay of QoE
crowdsourcing. Second, by using the sensitivity weights as input, the same \name ABR algorithm can
be used to optimize QoE for any new video.

\mypara{Cost-benefit analysis}
Compared to existing video streaming systems, \name essentially trades per-video profiling overhead
for improved QoE-bandwidth tradeoffs.
We argue that this is a favorable tradeoff, because the additional cost of profiling is
negligible compared to the overhead large content providers pay for the content and its
distribution.
For instance, the bid for the copyright of each popular TV episode (or making such an episode) can
cost on the order of \$10 billion~\cite{nflxlicense}, and the cost to distribute these videos to users can cost on the order of thousands for each minute of video (\eg at \$0.05/GB~\cite{cdn-price}, streaming a 1-minute video at 2.8Mbps to 1M users costs \$1K).
Compared to this, all of \name's overhead amounts to just \$30 per 1-minute worth of
video (\S\ref{sec:eval}). 
% \sid{Junchen: maybe we should omit the
% profiling time delay as this is a one-time cost? How do we amortize it over the many times the video
% will be (re)viewed?} 
In return, \name on average achieves 15.1\% higher real user ratings (which can
increase user engagement and revenue) or 32\% bandwidth savings (which reduces content
distribution cost). %These improvements are on par with those of complex video players.
These benefits apply for all users who watch the video, and possibly each
viewing of it (\eg if ads are shown).
% \sid{Junchen: check if this point about per-view ad benefit is valid}

Note that this cost-benefit analysis does not apply to videos with low viewership,
as \name's benefits may not outweight the costs, or to live videos, which have stringent delay
requirements.  
%That said, we believe \name is a reasonable starting point.
\name is a good fit for popular videos that contribute majority viewership (\eg 10\% of
YouTube videos account for 79\% viewership~\cite{youtube-distribution}).%\sid{We should mention Netflix too if possible}
%In theory, live videos could also use \name, by using user engagement (\eg percentage of users quit
%watching the video) as the user rating in QoE profiling.
%Such ratings are noisy, but video streaming industry has always used this
%measure~\cite{conviva,akamai,??}.

% \footnote{This cost-benefit analysis does not apply to videos with low viewership
% (\name's benefit may not compensate the cost) or live videos (which have stringent delay requirements).
% That said, we believe \name is a reasonable starting point.
% \name improves popular videos which contribute majority viewership (\eg \fillme\% YouTube videos
% account for \fillme\% viewership~\cite{??}).
% In theory, live videos could also use \name, by using user engagement (\eg percentage of users quit
% watching the video) as the user rating in QoE profiling.
% Such ratings are noisy, but video streaming industry has always used this
% measure~\cite{conviva,akamai,??}.
% }

% \jc{
% \begin{itemize}
% \item challenges
% \item crowdsourcing
% \item new abr with new actions
% \item key insight: chunk-level weights as a way to elevate user sensitivity as a first-class citizen
% \end{itemize}
% }

\tightsection{Profiling dynamic quality sensitivity at scale}
\label{sec:qoe_model}

%the quality sensitivity for each chunk; and 

This section builds an accurate and cost-efficient QoE model using two key ideas:
crowdsourcing to scalably profile the true quality sensitivity of each new video
(\ref{sec:qoe_model:crowd}), and chunk-level reweighting to reduce the cost of this profiling
(\ref{sec:qoe_model:our_model}).

\tightsubsection{Scaling out via crowdsourcing}
\label{sec:qoe_model:crowd}

% While some previous QoE methods do model how video content affects quality sensitivity, they derive this information by heuristics based on pixels and motions frames. 
% As we observed in \S\ref{sec:moti:missing}, these methods are necessarily insufficient to capture the complex interactions between dynamism of video content (\eg key moments in storylines) and subjective sensitivity to various video quality incidents.

% \mypara{Eliciting direct user ratings}
Instead of relying on complex, indirect heuristics, \name advocates for directly eliciting quality
ratings from human viewers to reveal their quality sensitivity to various quality incidents.
% to rate the quality of a streamed video session.
% The unique advantage is that user ratings reveal the {\em true} subjective sensitivity to various
% quality incidents, rather than indirectly inferring through heuristics.
% Now, as quality sensitivity is closely tied to the specific content at different part of a video,
% so the
However, the user ratings must be elicited {\em per video} and the sheer scale of this feedback can
be prohibitive! To put it into perspective, QoE models are usually built on user ratings from just a
handful of source videos (15-20\cite{bampis2018towards,duanmu2016quality}), but to get enough user ratings, a lab environment (or survey
platform) must be set up to recruit participants and have them watch at least two orders of
magnitude more video than the source videos (\ie 2000$\times$ the total length of source
videos\footnote{For instance, in the WaterlooSQOE-III dataset~\cite{duanmu2016quality}, each video is streamed over 13
throughput traces with 6 ABR algorithms, and each rendered video is then rated by 30
users.}).
This does not scale if we repeat the process per video.

\myparaq{Why crowdsourcing}
\name leverages commercial crowdsourcing platforms like Amazon MTurk~\cite{mturk} to automate the
user studies and scale them out to handle more videos.
Our rationale behind this design choice is two-fold:
\begin{packeditemize}
\item First, crowdsourcing reduces the overhead of participant recruitment and survey
dissemination (to about 78 minutes) and provides a large pool of participants. This allows
for repeated experiments to help control for human-related statistical noise.
\item Second, although the cost of crowdsourcing experiments grows linearly with videos, 
crowdsourcing platforms offer predictable and elastic pricing (more participants can be added
on-demand), allowing content providers to decide whether and
how to initiate a profiling given their budgets.
\end{packeditemize}

\begin{figure}[t]
    \centering
    \includegraphics[width=0.5\textwidth]{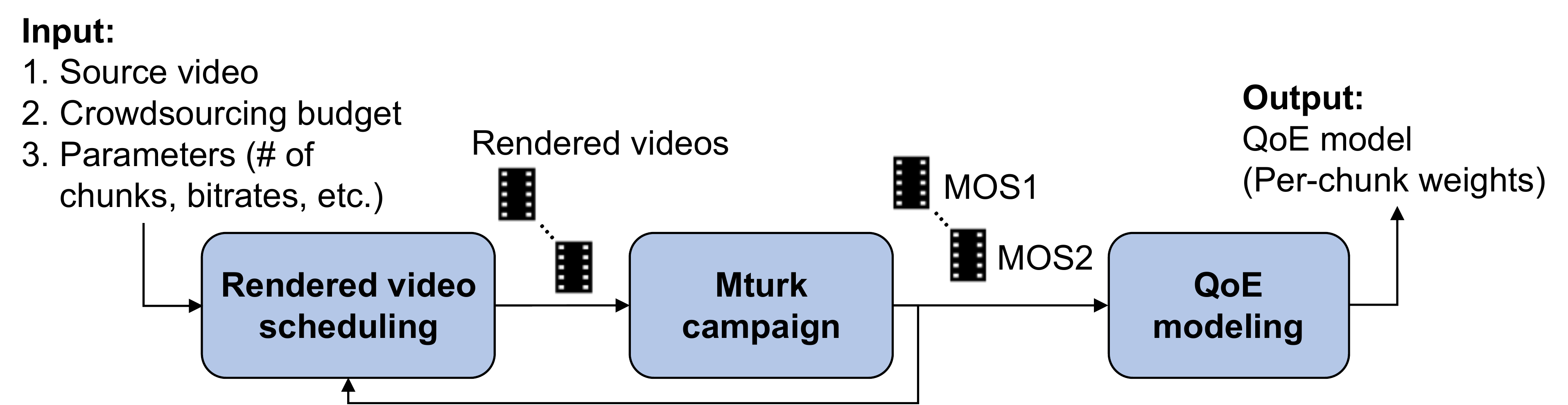}
    \tightcaption{Workflow of profiling dynamic quality sensitivity using crowdsourcing platforms. The
    arrow back to scheduler means the crowdsourced ratings can also be used to suggest more rendered
    videos to iteratively refine the QoE modeling. %\sid{Change to "QoE modeling"}
    }
    \label{fig:alg_overview}
\end{figure}

\begin{figure}[t]
      \begin{subfigure}[t]{0.44\linewidth}
         \includegraphics[width=1.0\linewidth]{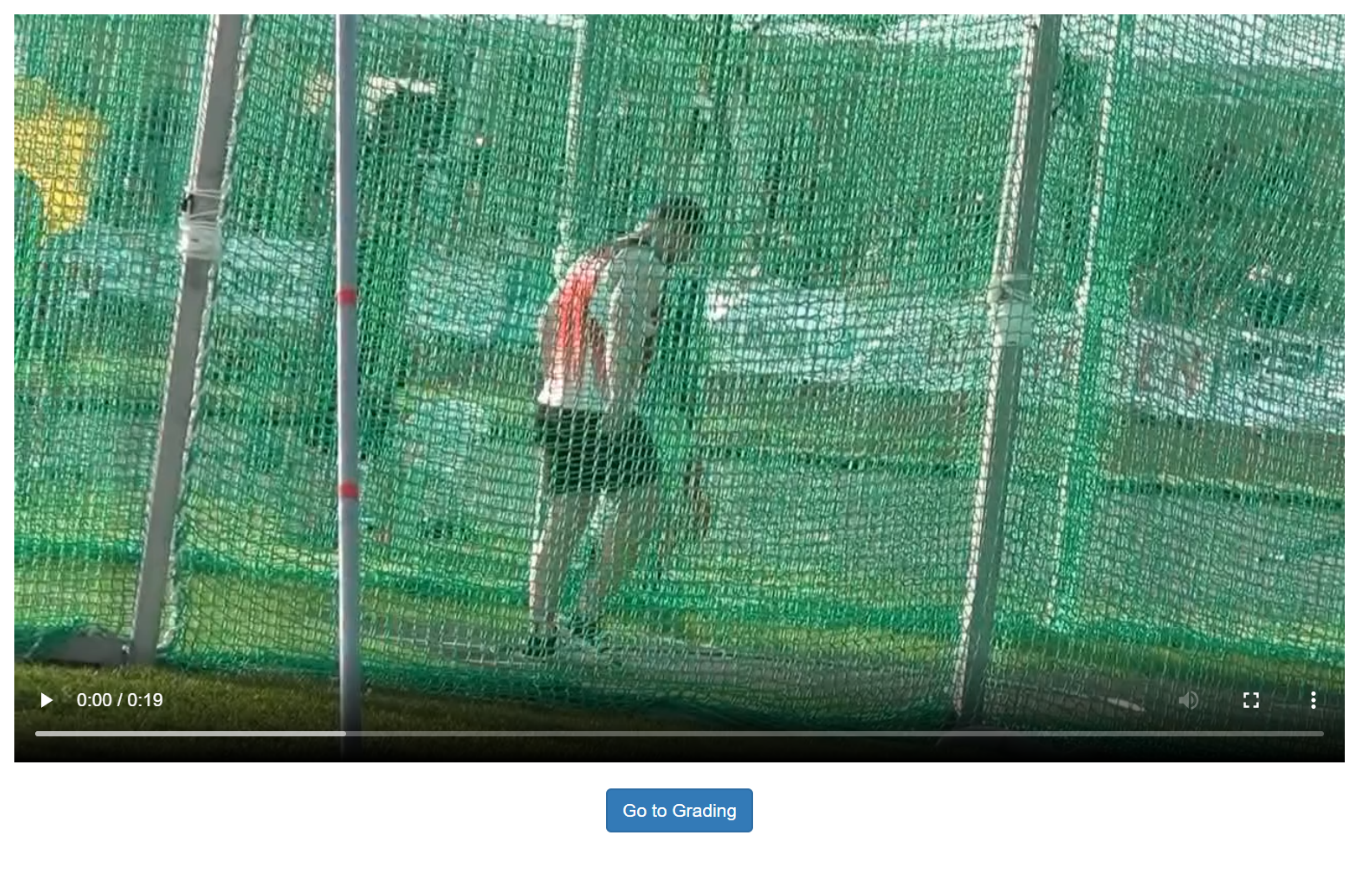}
         \caption{Video player page}
       \end{subfigure}
       \begin{subfigure}[t]{0.55\linewidth}
         \includegraphics[width=1.0\linewidth]{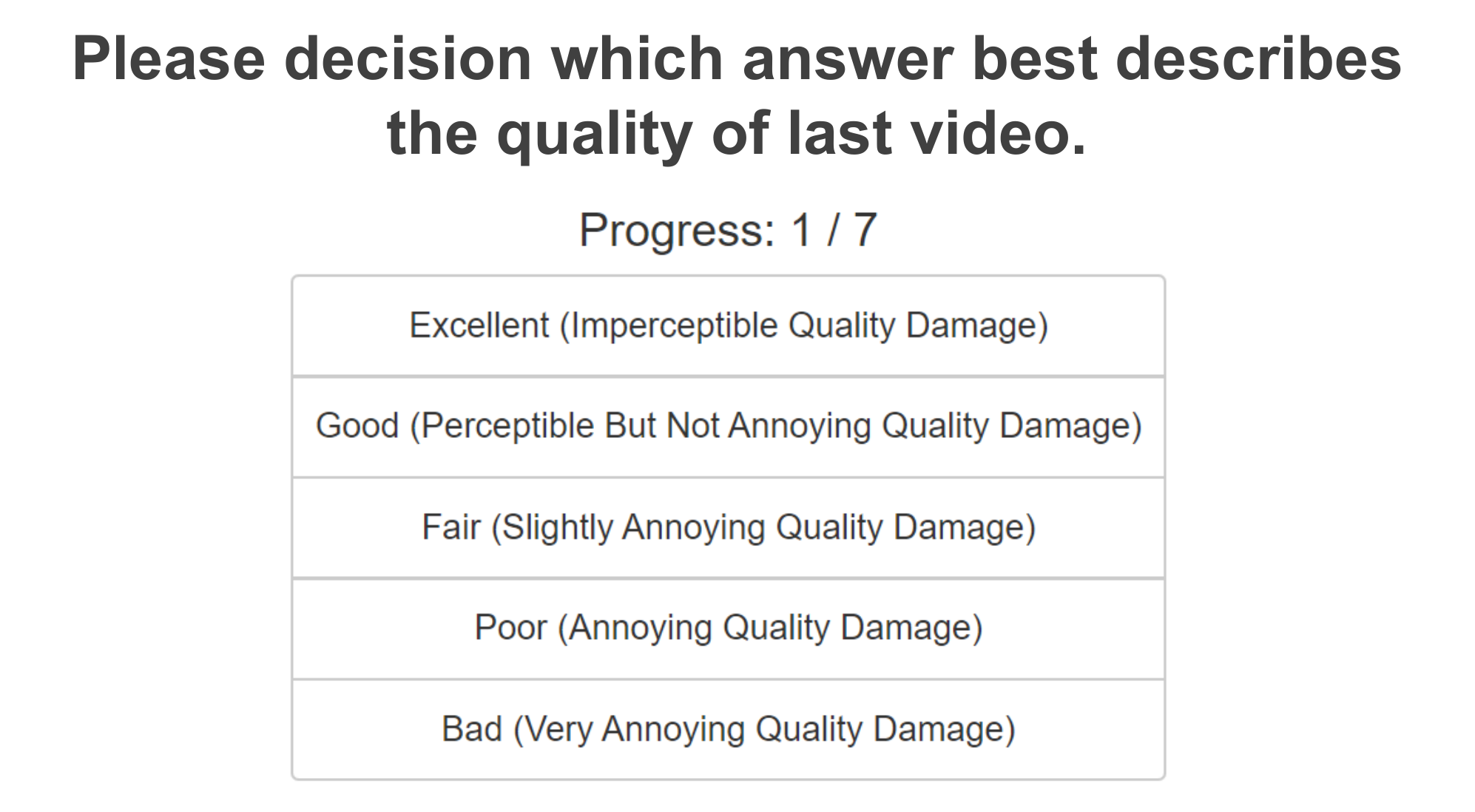}
        \caption{Rating page}
       \end{subfigure}
    \tightcaption{Screenshots of our QoE survey interface. 
    In each survey, a participant is asked to rate $\NumVideosPerSurvey$ rendered videos; after watching each rendered video, a participant is asked to rate its quality on the scale of 1 (worst) to 5 (best).}
    \label{fig:survey_screenshots}
\end{figure}

\mypara{Survey methodology}
Figure~\ref{fig:alg_overview} shows the workflow of QoE profiling in \name. 
At a high level, \name takes a source video and a monetary budget as input and returns a QoE model that incorporates dynamic quality sensitivity (customized for this video) as output.
\begin{packeditemize}
\item {\em Rendered video scheduling:} 
We first generate a set of {\em rendered videos} from the source video. 
Each rendered video is created by injecting a carefully selected low-quality incident (rebuffering event, bitrate drop, etc.) at a certain point in the video (\S\ref{sec:qoe_model:scheduling}).

\item {\em MTurk campaign:}
The rendered videos are then published on the MTurk platform and we specify how many Turkers, or  {\em participants}, we would like to recruit for this {\em campaign}.
Anytime a participant signs up, they will automatically start a {\em survey} during which they will be asked to watch $\NumVideosPerSurvey$ rendered videos and, after each video, rate its QoE on a scale of 1 to 5 (\ie Likert Scale). 
Figure~\ref{fig:survey_screenshots} shows screenshots of the video player and rating page. 
Once enough participants have signed up, the campaign automatically terminates.

\item {\em QoE modeling:} Finally, we use the mean opinion score (MOS) of each rendered video as its QoE and run a simple regression to derive the chunk-level weights. These weights are then incorporated into an existing QoE model to generate the final QoE model for this video (\S\ref{sec:qoe_model:our_model}).
\end{packeditemize}

%All ratings (as well as how long each participant spends to watch each test video) are logged for post processing (see \S\ref{sec:impl} for details).
%Notice that the number of test videos shown to each participant is separate from the number of test videos in each campaign. 

%\mypara{Sanity check of crowdsourced rating quality}
\mypara{Quality control of user ratings}
We take several principled measures to prevent and filter out spurious user ratings. We overview them here but provide more details in \S\ref{sec:sanitization}.
%\mypara{Sanity check of crowdsourced quality} 
The rendered videos are created with low-quality incidents as part of the rendering process. To minimize the influence of other quality incidents that may occur (\eg due to poor connections or browser issues), we ask participants to confirm the quality incident they observed after each video and eliminate inconsistent responses.
Importantly, we {\em randomize} the order in which the $\NumVideosPerSurvey$ rendered videos are shown to different participants. This eliminates biases due to viewing order and which videos were previously watched.
%the participants must rate quality after watching each video closely and should be fully aware the ratings need to reflect their feeling about the quality of the video.
Additional measures we take include limiting the number/length of videos watched per participant (to prevent fatigue), presenting the goals and rejection criteria of the survey upfront, calibrating rating ability using a "reference" video whose QoE is known, and logging whether a participant has watched all videos in full length (\S\ref{sec:impl}).

%\sid{I think we should mention the techniques Xu used to control for quality that were here before
%(hard rules, control questions, etc.). We can add details in the implementation section, but this
%is definitely part of the methodology, especially for an audience that doesn't often carry out user
%studies and will likely question the reliability of our user responses.}
%\jc{sid, there was a paragraph for that before but i thought the section was too long, but i agree it should be here. see the paragraph above this}

As a sanity check, we run an MTurk campaign with three 12-second video clips from a public dataset~\cite{duanmu2016quality}, which used an in-lab survey.
We compare the QoE measured from MTurk participants with the QoE recorded in the dataset and find that they highly agree with each other: after normalizing them to the same range, the relative difference between the MTurk rating and in-lab rating on the same video is less than 3\%.
That said, we acknowledge that our MTurk survey methodology is not perfect and user ratings are always susceptible to human factors, but these issues affect all QoE measurement studies.

\tightsubsection{Cutting cost via chunk-level reweighting}
\label{sec:qoe_model:our_model}

While crowdsourcing helps us scale profiling to more videos, profiling an individual video can still be prohibitively expensive.
Consider a strawman solution that build a QoE model for each chunk. If the quality sensitivity of a chunk depends on the quality incident as well as the quality of other chunks, such a model could require $\bigO(\NumChunks^2\cdot\NumParams)$ parameters, where $\NumChunks$ is the number of chunks and $\NumParams$ is the number of parameters in the QoE model. Eliminating the dependence on other chunks reduces this to $\bigO(\NumChunks\cdot\NumParams)$, but this may still be prohibitive---\eg KSQI has tens of parameters. Fortunately, we can leverage the insights from \S\ref{sec:moti:missing} to vastly reduce this complexity.
%To model how quality sensitivity varies across chunks (recall that quality switches happen at chunk boundaries, so variance within a chunk can be ignored), consider a strawman solution that builds a QoE model for each chunk. 
% As a strawman of modeling QoE under different quality sensitivity, let us consider building a separate QoE model for each part of a video. 
% We assume quality sensitivity can change at the boundaries chunk (this aligns with the quality adaptation later), 
%This strawman would require $O(\NumChunks\cdot\NumParams)$ parameters, where $\NumChunks$ is the number of chunks and $\NumParams$ is the number of parameters in a QoE model. If the quality of other chunks influences the quality of this chunk, this complexity could be as bad as $O(\NumChunks^2\NumParams)$.
%This number can easily explode; \eg KSQI has \fillme parameters.
% in each instance of QoE model. 
%With more parameters, it would require more QoE feedback to infer the parameters.
% This number grows quickly as the $\NumParams$ can be on the order of 10s (if not 100s); \eg KSQI has \fillme parameters.

%A strawman method built on the prior work is that we train the QoE model based on real user ratings of the video sequences which is an enumeration of all combinations of the quality of the video chunks.
%While it is functionally sufficient, the cost can be prohibitive.
%For example, for a source video with $10$ chunks, if the numbers of possible bitrate and rebuffering time levels are $5$ and $3$ respectively, we need to collect the ratings of $15^{10}$ video sequences, which is intractable.

\mypara{Encoding quality sensitivity in per-chunk weights}
To cut the profiling cost and delay without hurting profiling accuracy, \name exploits the fact that quality sensitivity is inherent to the video content and largely agnostic to the type of quality incident (\S\ref{sec:moti:missing}).  This allows us to assume a single weight per chunk that encodes its quality sensitivity, reducing the number of model parameters to $O(\NumChunks)$!

%Our goal is to create a QoE model that does not only depend on the pixel-level induced visual quality and quality incidents but also the dynamic of quality sensitivity, \ie how sensitive the users are to the quality of chunks.
With this insight, \name can incorporate the per-chunk weights into an existing QoE model. If the QoE model is {\em additive}, \ie overall QoE is the average of the QoE estimates of individual chunks, \name can directly reweight the chunks by their quality sensitivity. Though not all QoE models are additive (\eg LSTM-QoE), many state-of-the-art models including KSQI and its variants~\cite{yin2015control,mao2017neural} can be written as 
%\footnote{Some QoE models uses an exponential~\cite{??} or logarithmic function~\cite{??} that are not additive but can be made so.}
\begin{align}
\QoE = \sum\nolimits_{i=1}^{\NumChunks} \Quality_i, \label{eq:oldqoe_overall}
\end{align}
where $\Quality_i$ is the estimated QoE of the $i^{\textrm{th}}$ chunk. For KSQI, this takes into account the impact of visual quality, rebuffering, and quality switches.
Note that $\Quality_i$ is the contribution of the $i^{\textrm{th}}$ chunk inferred by the model which, in theory, could incorporate information about other chunks too. 
\name reweights the QoE model as follows:
\begin{align}
\QoE &= \sum\nolimits_{i=1}^{\NumChunks} \Weight_i\Quality_i, \label{eq:newqoe_overall}
\end{align}
where $\Weight_i$ is the weight of the $i^{\textrm{th}}$ chunk, reflecting how much more sensitive users are to quality incidents in this chunk compared to other chunks.
%Note that when the weights are the same (quality sensitivity does not vary), Equation~\ref{eq:newqoe_overall} will be reduced to Equation~\ref{eq:oldqoe_overall}.

\mypara{Weight inference}
Given any $\NumVideos$ rendered videos, if $\QoE_j$ is the QoE (MOS) of the $j^{\textrm{th}}$ rendered video and $\Quality_{i,j}$ is the estimated QoE of the $i^{\textrm{th}}$ chunk of the $j^{\textrm{th}}$ rendered video, then we can write  $\NumVideos$ equations, $\QoE_j=\sum_{i=1}^{\NumChunks} \Weight_i\Quality_{i,j}$ for $j=1,\dots,\NumVideos$. We can then infer the $\Weight_i$ using a linear regression.
% \sid{The example below doesn't really say anything new and doesn't showcase the regression in action. We could consider adding an example that actually shows the regressed values, but space and time are a problem now.}
%To use the example in Figure~\ref{fig:soccer_content}, if the rebufferings happens at the beginning of each chunk (in total 5 chunks), then the weight of the $4^{\textrm{th}}$ chunk should be the highest (since the quality incident that happened to the $4^{\textrm{th}}$ chunk has the highest impact).

In the remainder of the paper, we assume that KSQI reweighted by Equation~\ref{eq:newqoe_overall} is the QoE model of \name.

\tightsubsection{Crowdsourcing scheduler}
\label{sec:qoe_model:scheduling}

Having reduced our parameters to the per-chunk weights, we now turn our attention to reliably estimating these weights by asking users to rate as few rendered videos as possible.

\mypara{Two-step scheduling}
For a given source video, \name's rendered video scheduler uses a two-step process to decide which rendered videos to publish and how many participants to elicit ratings from.
\begin{packeditemize}
\item First, \name creates a set of $\NumChunks$ rendered videos, each with a single 1-second rebuffering event at a different chunk ($\NumChunks$ is the number of chunks).
It then publishes these videos and asks $\NumTurkers_1$ participants to rate each video.
The total rendered video duration is $\bigO(\NumChunks\cdot\NumTurkers_1)$.
Once the videos are rated, we infer the per-chunk weights as described above. 
\item Second, we pick $\NumChunks' \ll N$ chunks whose inferred weights are $\ThreshDiff$-high or low (\eg 6 \% higher or lower than the average weight).
We then repeat the first step with two differences: 
(1) low-quality incidents are added only to these chunks, and 
(2) the quality incidents include $\NumBitrates$ bitrates (below the highest bitrate) and $\NumBufferings$ rebuffering events (1,2,\ldots seconds). We publish the rendered videos and ask $\NumTurkers_2$ participants to rate them, for a total video duration of $\bigO(\NumChunks'\cdot\NumBitrates\cdot\NumBufferings\cdot\NumTurkers_2)$.
\end{packeditemize}
We microbenchmark the key parameters of the scheduler (Figure~\ref{fig:cost_accuracy}) and empirically select the values that balance accuracy and cost:
$\NumBitrates$=2 incidents, $\NumBufferings$=1 rebuffering incidents, and $\NumTurkers_1$=10 and $\NumTurkers_2$=5 participants per rendered  video.

\mypara{Intuition}
The rationale behind the two-step scheduler is as follows. 
With a small $\NumTurkers_1$ (\eg 10), the estimated weights will have nontrivial variance. But these estimates are already indicative enough to identify a subset of chunks whose quality sensitivity is very high or low, so we can {\em focus} the second iteration on these chunks.
In general, for an ABR algorithm to improve QoE-bandwidth tradeoffs, it is more important to identify which chunks have very high/low quality sensitivity than to precisely estimate the quality sensitivity of each chunk.

% \mypara{Analysis on MTurk cost}
Cutting the total rendered video duration and the number of participants directly impacts the cost and delay of the QoE profiling step.
Cost is proportional to the total length of videos watched by all participants, because each participant is paid by a predetermined hourly rate. 
{\em Delay} is proportional to the number of participants, because it takes longer for participants to sign up asynchronously (on the order of tens of minutes to get 100 participants) than it takes to complete the survey, which can happen in parallel. See Appendix~\ref{sec:sanitization} for further discussion.
% \jc{there is a tradeoff between hourly rate and delay, but we haven't explore it yet}
We can ignore the delay of familiarizing themselves with the system and providing the ratings, which account for a small fraction of the time spent by each participant.

%% file: abr.tex
\tightsection{\name's ABR logic}
\label{sec:abr}

% Next, we present the ABR logic in \name. 
The key difference between \name's ABR logic and traditional ABR logic is that it aligns quality adaptation with the temporal variability of quality sensitivity.
We first show how \name modifies a traditional ABR framework (\ref{sec:abr:formu}).
Since these changes are external to the core ABR logic, existing ABR algorithms only require marginal changes to benefit from \name (\S\ref{sec:abr:alg}). 
% \name ABR can be built on existing ABR algorithms with minor changes and leverage their advantageous throughput prediction or adaptation logic 

\tightsubsection{Enabling new adaptation actions}
\label{sec:abr:formu}

\begin{figure}[t]
    \centering
    % \begin{subfigure}[t]{0.48\linewidth}
    %     \includegraphics[width=1.0\linewidth]{graphs/current_abr.pdf}
    %     \caption{Canonical ABR framework}
    % \end{subfigure}
    %  \hfill
    % \begin{subfigure}[t]{0.48\linewidth}
    % \includegraphics[width=1.0\linewidth]{graphs/our_abr.pdf}
    %     \caption{Nature1}
    %   \end{subfigure}
    % \includegraphics[width=1.0\linewidth]{graphs/abr_contrast.pdf}
    \includegraphics[width=0.97\linewidth]{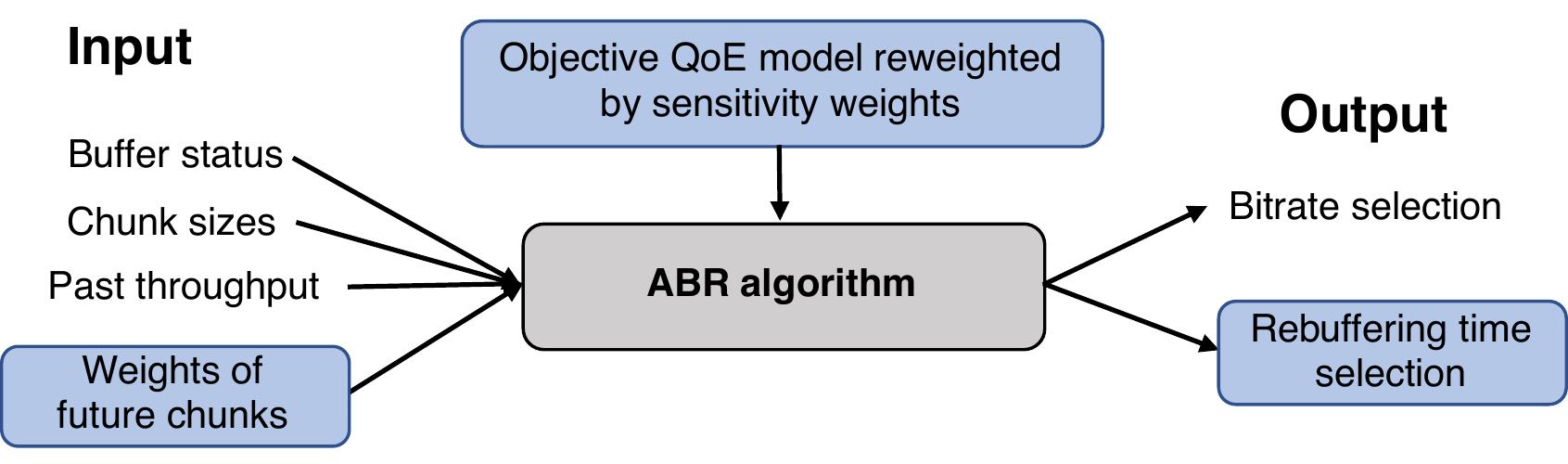}
    \tightcaption{ABR framework of \name. The differences with traditional ABR framework are highlighted.
    % \xu{sensei!!}
    % \jc{add change of objective function}
    }
    \label{fig:abr_framework}
\end{figure}

% Instead of reinventing video players from scratch, 
\name takes a pragmatic approach by working within the framework of existing players. It proposes specific changes to their input and output, as highlighted in Figure~\ref{fig:abr_framework}.
% , so that the ABR algorithms themselves only need marginal changes leverage the benefits of \name.
% This calls for several changes to existing ABR framework. 
% contrasts the \name ABR framework with typical ABR interface.
% \begin{packeditemize}

\mypara{Input} Besides the current buffer length, next chunk sizes, and history of throughput measurements, \name's ABR algorithm takes as input the sensitivity weights of the next $h$ chunks. A larger $h$ allows us to look farther into the horizon for opportunities to trade current quality for future quality, or vice versa, depending on the quality sensitivity of chunks. In practice, we are also constrained by the reliability of our bandwidth prediction for future chunks. We pick $h=5$ since we observe that QoE gains flatten beyond a horizon of 4 chunks. 
%so the player can be aware of the dynamic sensitivity. %, which indicates which future chunks have relatively higher quality sensitivity.
% \name's input has the weights of the future chunks, which helps \name make decisions favor the quality sensitive chunks.

\mypara{Output} 
\name's ABR algorithm selects the bitrate for future chunks as well as when the next rebuffering event should occur.\footnote{In practice, \name only makes adaptation decisions for the next chunk. This is a practical choice to shrink the action space, but it is not fundamental. Note that since the player invokes the ABR algorithm after each chunk is downloaded, \name can still plan adaptations for multiple chunks in the future even if it only acts on one chunk at a time.}
% This adds a new action space that the player can now shift a rebuffering event to an earlier point to avoid rebuffering at later high sensitive chunks.
In contrast, traditional players only initiate rebuffering events when the buffer is empty. 
% before any unplayed chunk, while the canonical ABR algorithms invoke a rebuffering event only when the player buffer drains out.

\mypara{Objective QoE model}
If the ABR algorithm explicitly optimizes an additive QoE model, \name can modify its objective as  described in \S\ref{sec:qoe_model:our_model}. 
While \name can be applied to many recent ABR algorithms (\eg~\cite{mao2017neural, yan2020learning, yin2015control}), some ABR algorithms (\eg BBA) do not have explicit QoE objectives and thus cannot be optimized by \name as is.
% we acknowledge that they cannot be optimized directly. 
% Though \name makes minimum assumption about the internals of the base ABR logic, it does require the base ABR algorithm to use an explicit QoE objective function. 
% For instance, \fillme \jc{bba?}

% \jc{why not projecting actions into future}

\begin{figure}[t]
    \centering
    \includegraphics[width=0.95\linewidth]{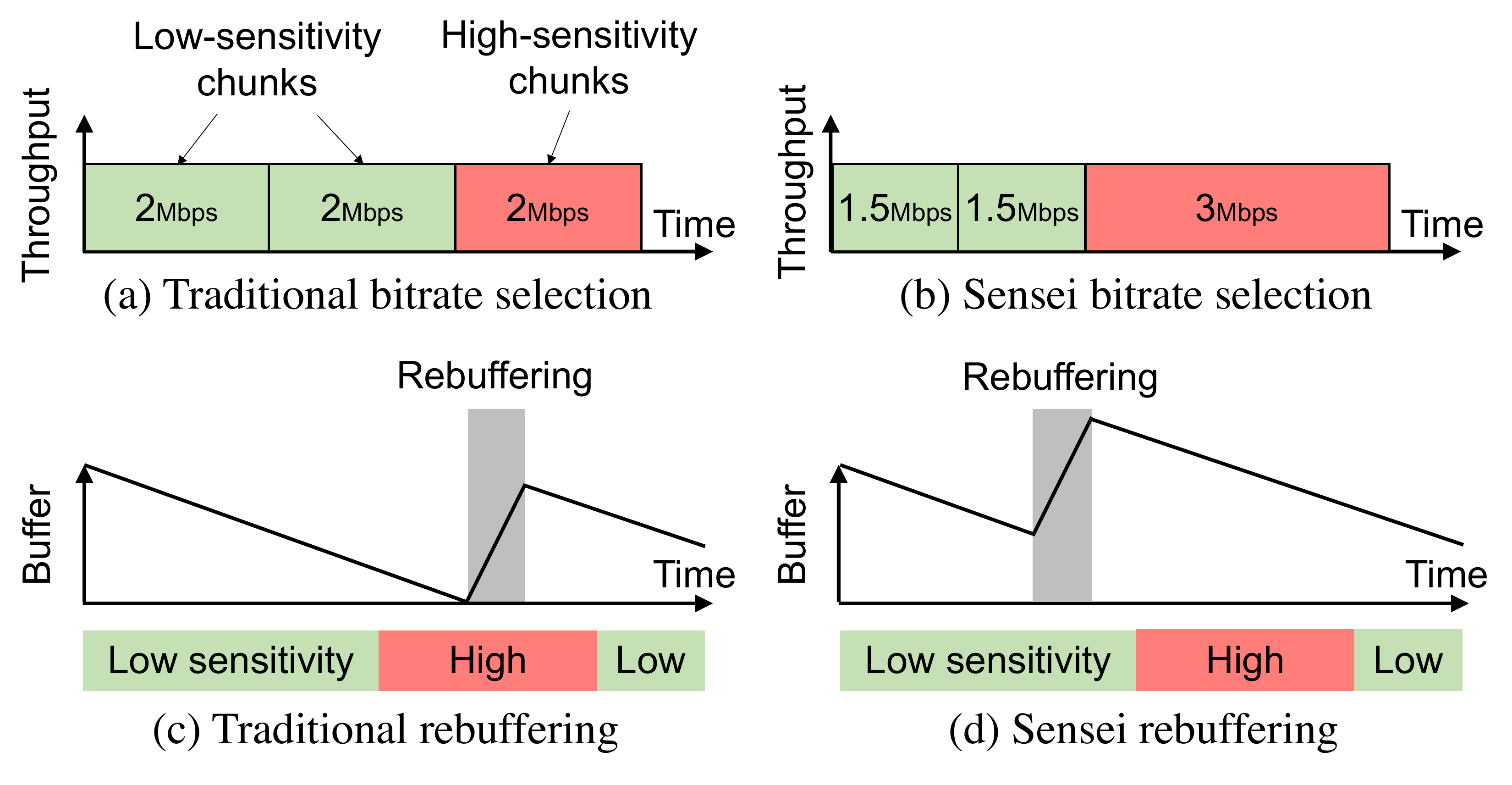}
    \vspace{-0.3cm}
    \tightcaption{Illustrative examples of \name vs traditional ABR logic: how \name improves quality (a vs. b) or avoids bad quality (c vs. d) at high-sensitivity chunks.}
    \label{fig:abr_examples}
\end{figure}

In theory, these changes are sufficient to enable at least the following optimizations, which traditional ABR algorithms are unlikely to explicitly do.
(1) Lowering the current bitrate so that it can raise the bitrate in the next few chunks, if they have higher quality sensitivity (Figure~\ref{fig:abr_examples}(a) and (b));
(2) Raising the current bitrate slightly over the sustainable level if quality sensitivity is expected to decrease in the next few chunks; and
(3) Initiating a short rebuffering event now in order to ensure smoother playback for the next few chunks, if they have higher quality sensitivity (Figure~\ref{fig:abr_examples}(c) and (d)).

% \jc{all these relies on accurate bandwidth prediction}

\tightsubsection{Refactoring current ABR algorithms}
\label{sec:abr:alg}

We apply \name to two ABR algorithms: Pensieve~\cite{mao2017neural}, based on deep reinforcement learning, and Fugu~\cite{yan2020learning}, a more traditional algorithm based on bandwidth prediction.
% For fairness, we use KSQI as the QoE model for the Pensieve and Fugu and the base QoE model of the \name-modified variations.
% We make sure this change will strictly improve the quality of Pensieve and Fugu, since the QoE models used in their original work are special cases of KSQI.

% \name can be implemented by making minor modifications to some current ABR algorithms.
% Here. we pick two recent proposals as \name's base ABR: Fugu~\cite{yan2020learning} and Pensieve~\cite{mao2017neural}.
% As we will see, it does not assume certain internal structure of the base ABR algorithms (Fugu is based on throughput prediction and Pensieve is a blackbox neural net).
% \jc{mention that we move all baselines to use KSQI for their QoE models are simplifications of KSQI}

\mypara{Applying \name to Pensieve}
\name leverages the flexibility of deep neural networks (DNNs) and augments Pensieve's input, output and QoE objective---its states, actions, and reward, in the terminology of reinforcement learning---as described in \S\ref{sec:abr:formu}. It then retrains the DNN model in the same way as Pensieve;
% modifies its input and output and retrain the DNN model, in almost the same way of Pensieve.
we call this variation \name-Pensieve. \name-Pensive makes two minor changes to reduce the action space (which now includes rebuffering).
% In particular, \name-Pensieve adds the sensitivity weights of the next $h$ chunks (by default $h=5$).
% The action output at the state of chunk $i$ includes the bitrate {\em and} the rebuffering time of each chunk $j$ ($j=i,\dots,i+h-1$).
% As \name-Pensieve has a larger action space, we observe that this leads to high performance variance.
First, we restrict possible rebuffering times to three levels (\{0,1,2\} seconds) that can only happen at chunk boundaries.
% \name-Pensieve the selection bitrate and rebuffering time decisions separately. 
Second, instead of choosing among combinations of bitrates and rebuffering, \name-Pensieve either selects a bitrate or initiates a rebuffering event at the next chunk. 
%If it is the former, it will use that bitrate without initiating rebuffering.
If it chooses the latter, \name-Pensieve will increment the buffer state by the chosen rebuffering time and rerun the ABR algorithm immediately.
% The reward function

% We will describe the necessary changes to the rule-based (\eg Fugu \cite{yan2020learning}) and non-rule-based (\eg Pensieve \cite{mao2017neural}) ABR algorithms.

\mypara{Applying \name to Fugu}
% To understand the modifications needed to make Fugu an \name ABR, 
Let us first explain how Fugu works.
% The aim of the rule-based ABR algorithms is to maximize the expected QoE of future $H$ chunks, given the number of look-ahead chunks $H$  and the network throughput prediction.
% The rule-based ABR algorithm usually select the bitrate for the chunks based on the explicit network throughput.
% We will take Fugu as an example and show how to apply dynamic quality sensitivity.
% At a high level, after downloading the $i^{\textrm{th}}$ chunk, Fugu takes as input throughput prediction for the next $h$ chunks (in the form of a distribution $\textrm{Pr}(R_j=r)$, where $R_j$ is the throughput of $i^{\textrm{th}}$ chunk, $i<j\leq i+h$, and $r$ is one of the discretized throughput levels).
% It then uses a dynamic programming logic to decide what bitrate should be used to maximize
At a high level, before downloading the $i^{\textrm{th}}$ chunk, Fugu considers the throughput prediction for the next $h$ chunks.
% (in the form of a predicted throughput distribution for each of the future chunks).
For any throughput variation $\gamma$ (with predicted probability $p(\gamma)$) and bitrate selection $B=(b_i,\dots,b_{i+h-1})$, where $b_j$ is the bitrate of the $j^{\textrm{th}}$ chunk, it simulates when each of the next $h$ chunks will be downloaded and estimates the rebuffering time $t_j(B,\gamma)$ of the $j^{\textrm{th}}$ chunk (which could be zero). 
%with the assumption that rebuffering starts when the buffer is empty until a chunk is downloaded. 
It then picks the bitrate vector $(b_i,\dots,b_{i+h-1})$ that maximizes the expected total quality over the next $h$ chunks and possible throughput variations.
\begin{align}
\sum_{\gamma}p(\gamma)\sum_{j=i}^{i+h-1}\Quality(b_j,t_j(B,\gamma)) \label{eq:fugu-1}
\end{align}
Here $\Quality(b,t)$ estimates the quality of a chunk with the bitrate $b$ and rebuffering time $t$ using a simplified model of KSQI. 
% (\eg replacing logarithmic penalty of rebuffering time with a linear penalty).

Now, the \name variation of Fugu, which we call \name-Fugu, uses Fugu's throughput prediction and the sensitivity weights $w_j$ of the next $h$ chunks. %($\Weight_j$ is the weight of the $j^{\textrm{th}}$ chunk). 
% For any throughput variation $\gamma$ and bitrate selection $B=(b_i,\dots,b_{i+h-1})$, \name-Fugu also enumerate the feasible rebuffering time $t_j$ at of the $j^{\textrm{th}}$ chunk. 
% Rebuffering time $t_j$ at the $j^{\textrm{th}}$ chunk is feasible as long as the buffer length is never negative.
\name-Fugu  picks the bitrate vector $B=(b_i,\dots,b_{i+h-1})$ and rebuffering time vector $T=(t_i,\dots,t_{i+h-1})$, where $t_j$ is the rebuffering time of the $j^{\textrm{th}}$ chunk, that maximizes the expected total quality over the next $h$ chunks and possible throughput variations.\footnote{We omit further details that are orthogonal to \name's modifications
%, such as how the $\Quality$ function models the penalty of bitrate switches and how to make the optimization tractable via dynamic programming. 
}
\begin{align}
\sum_{\gamma}p(\gamma)\sum_{j=i}^{i+h-1}\Weight_j\Quality(b_j,t_j) \label{eq:fugu-2}
\end{align}
Here, the rebuffering times must be feasible, \ie the buffer length can never be negative.

% A comparison between objective~\ref{eq:fugu-1} and~\ref{eq:fugu-2} reveals two changes: 
In short, \name-Pensieve and \name-Fugu add an extra action (rebuffering time per chunk), and their objective function reweights the contribution of each chunk's quality using the sensitivity weights provided by our QoE model.

%% file: system.tex
\tightsection{Implementation}
\label{sec:impl}

% We have implemented crowdsourcing-based QoE modeling 
% % (including the automation of MTurk tests) 
% and the integration of \name ABR with a canonical DASH video player.
% Here we highlight key implementation details.

% \tightsubsection{Automation of MTurk tests}
% \label{sec:impl:turker}

\mypara{Automation of MTurk tests}
We implement the pipeline shown in Figure~\ref{fig:alg_overview} in an (almost) fully automated manner. 
It uses a combination of Python (for logic) and Javascript (for the video server).
Given a source video, it first uses FFmpeg to create the rendered videos by adding specific low-quality incidents.
It then uploads the rendered videos to a video server, from which Turkers (participants) will later download the video.
After that, it generates a unique link for this campaign and posts it on the MTurk website (this is the only step that needs manual intervention). 
Turkers can join the test by clicking the link, which redirects them to our video server.
Once a Turker completes a survey (\ie having rated all assigned videos), our server logs it 
% (and the turker ID and how long a turker spends on each video, etc) 
and notifies us.
When enough ratings have been collected, the server uses a script to train the quality sensitivity weights, as described in \S\ref{sec:qoe_model:our_model}.
Appendix~\ref{sec:lessons} describes useful lessons we learned from MTurk experiments.

% \jc{details to be left to implementation are: how much to pay each turker, how to streamlining the publication process? any MTurker selection criterion? any feedback postfiltering rules? too long video?}

% \tightsubsection{Video player integration}

\mypara{Video player integration}
We implement \name on DASH.js~\cite{dashjs}, an open source video player from which many commercial players are developed.
% \mypara{Augmenting manifest file}
% Before streaming a video, the DASH player will download a manifest file of the video.
We augment the DASH manifest file with per-chunk sensitivity weights (by adding a new XML field under \texttt{Representation}) and change the manifest file parser \texttt{ManifestLoader} to parse the weights of the chunks.
% In the DASH systems, the player should load the video's manifest file before playing this video.
% The manifest file (\emph{a.k.a.,} MPEG-DASH file) contains the URL, resolution, sizes of each video chunk.
% The media-related information of each chunk is stored under the field \texttt{Representation}.
% We add a subfield \texttt{weight} under \texttt{Representation} to denote the quality sensitive of the chunk.
% Accordingly, in Dash.js, we need to modify the class \texttt{ManifestLoader} to parse the weights of the chunks.
% \mypara{Controlled rebuffering}
% We customize DASH.js to enable adaptation actions of \name. 
Compared to other ABR algorithms, a unique challenge faced by \name is to actively initiate rebuffering when the buffer is not empty.
We use Media Source Extensions (MSE)~\cite{mseapi} (an API that allows browsers to change player states) to delay a downloaded chunk that is in the browser buffer from being loaded into the player buffer. 
To initiate a short rebuffering,
% If we want to the player initiate a short rebuffering while the browser buffer still has content (\ie the buffer is not empty yet), 
\name sets a callback to trigger \texttt{SourceBufferSink} function (which loads a chunk from the browser buffer into the player buffer) after a controlled delay.

% appending the downloaded chunks to the player buffer.

% by using Media Source Extensions (MSE) API~\cite{mseapi} that is widely supported by the popular web browsers, \eg Chrome, Firefox, IE and Safari.

% \name enables the player to add rebuffering events even the buffer does not drain out.
% To support this, we manually delay appending the downloaded chunks to the player buffer.
% If we want to add an rebuffering event to the player, we will set a timer for \texttt{SourceBufferSink} that is an MSE API for appending the downloaded chunk to the player buffer.
% We also change the timing of downloading the chunks.
% Previously, a chunk's download event will be triggered only after its previous chunk has been put into the player buffer.
% Now, we trigger this event after the completion of the previous chunk and do not have to wait for putting the previous chunk into the player.

% \jc{
% \begin{itemize}
% \item need a way to sanity check user engagement. 
% \item mturk reputation. rejection not exceed (10\%, not over 30\%)
% \item msster turker good reputation
% \item around 15 rating (different environment/platforms). how many more users
% \item user devices
% \end{itemize}
% }

%% file: eval_xu_modified.tex
\tightsection{Evaluation}
\label{sec:eval}

Our evaluation of \name shows several key findings:
\begin{packeditemize}
    \item Compared to recent proposals, \name can improve QoE by 7.7-52.5\% without using more bandwidth or can save 12.1-50.3\% bandwidth while achieving the same QoE.
    Note that these improvements are on par with those of recent ABR algorithms that use complex ABR logic.
    \item The performance gains of \name come at a cost of \$31.4 for a one-minute long
    video, which is marginal compared to the investments made by content providers.
    \item \name can improve QoE prediction accuracy by 11.8-37.1\% over state-of-the-art QoE models.
    \item \name's ABR algorithm consistently outperforms baseline ABR algorithms even when
    bandwidth fluctuates.
\end{packeditemize}

\begin{table}[t]
\footnotesize
\begin{tabular}{llll}\hline
\textbf{Name} & \textbf{Genre} & \textbf{Length} & \textbf{Source dataset} \\ \hline\hline 
(a) Basket1 & Sports & 3:40 & LIVE-MOBILE \\ \hline
(b) Soccer1 & Sports & 3:20 & LIVE-NFLIX-II \\ \hline
(c) Basket2 & Sports & 3:40 & YouTube-UGC \\ \hline
(d) Soccer2 & Sports & 3:40 & YouTube-UGC \\ \hline
(e) Discus & Sports & 3:40 & YouTube-UGC \\ \hline
(f) Wrestling & Sports & 3:40 & YouTube-UGC \\ \hline
(g) Motor & Sports & 3:40 & YouTube-UGC \\ \hline
(h) Tank & Gaming & 3:40 & YouTube-UGC \\\hline 
(i) FPS1 & Gaming & 3:40 & YouTube-UGC \\\hline
(j) FPS2 & Gaming & 3:40 & YouTube-UGC \\\hline
(k) Mountain & Nature & 1:24 & LIVE-MOBILE \\\hline
(l) Animal & Nature & 3:40 & YouTube-UGC \\\hline
(m) Space & Nature & 3:40 & YouTube-UGC \\\hline
(o) Girl & Animation & 3:40 & YouTube-UGC \\\hline
(n) Lava & Animation & 3:40 & LIVE-NFLIX-II \\\hline
(p) BigBuckBunny & Animation & 9:56 & WaterlooSQOE-III \\\hline
\end{tabular}
\tightcaption{Summary of the test video set. (See appendix for the detailed descriptions.)}
\label{tab:summary}
\end{table}

\tightsubsection{Experimental Setup}

\mypara{Test video set and throughput traces} 
Table~\ref{tab:summary} summarizes the genres, lengths and source datasets of our test video set.
The videos are selected from four popular datasets. LIVE-MOBILE~\cite{ghadiyaram2017subjective}, LIVE-NFLX-II~\cite{bampis2018towards} and WaterlooSQOE-III~\cite{duanmu2016quality} are professional-grade datasets sometimes used to train QoE models in the literature. We complement these sources with videos from a user-generated dataset (YouTube-UGC \cite{wang2019youtube}).
The videos are randomly selected to cover four video genres.
To create an adaptive video streaming setup, we chop videos into 4-second chunks and encode each chunk with H.264/MPEG-4 AVC \cite{wiegand2003overview} in five bitrate levels: $\{300,750,1200,1850,2850\}$Kbps (which correspond to $\{240, 360, 480, 720, 1080\}$p on YouTube).
We randomly select 10 throughput traces from two public datasets, FCC~\cite{federal2016raw} and 3G/HSDPA~\cite{riiser2013commute}. 
We restrict our selection to those whose average throughput is between 0.2Mbps and 6Mbps, so that the ABR algorithms will make non-trivial bitrate selection decisions. \sid{This example actually sounds ``trivial'' given how simple it is. I would either cut everything after ``\eg always'' or pick a different example.}
% (We train Pensieve with 1000 network traces from \cite{mao2017neural}.)

\begin{figure}[t]
    \centering
    \begin{subfigure}[t]{0.48\linewidth}
         \includegraphics[width=0.95\linewidth]{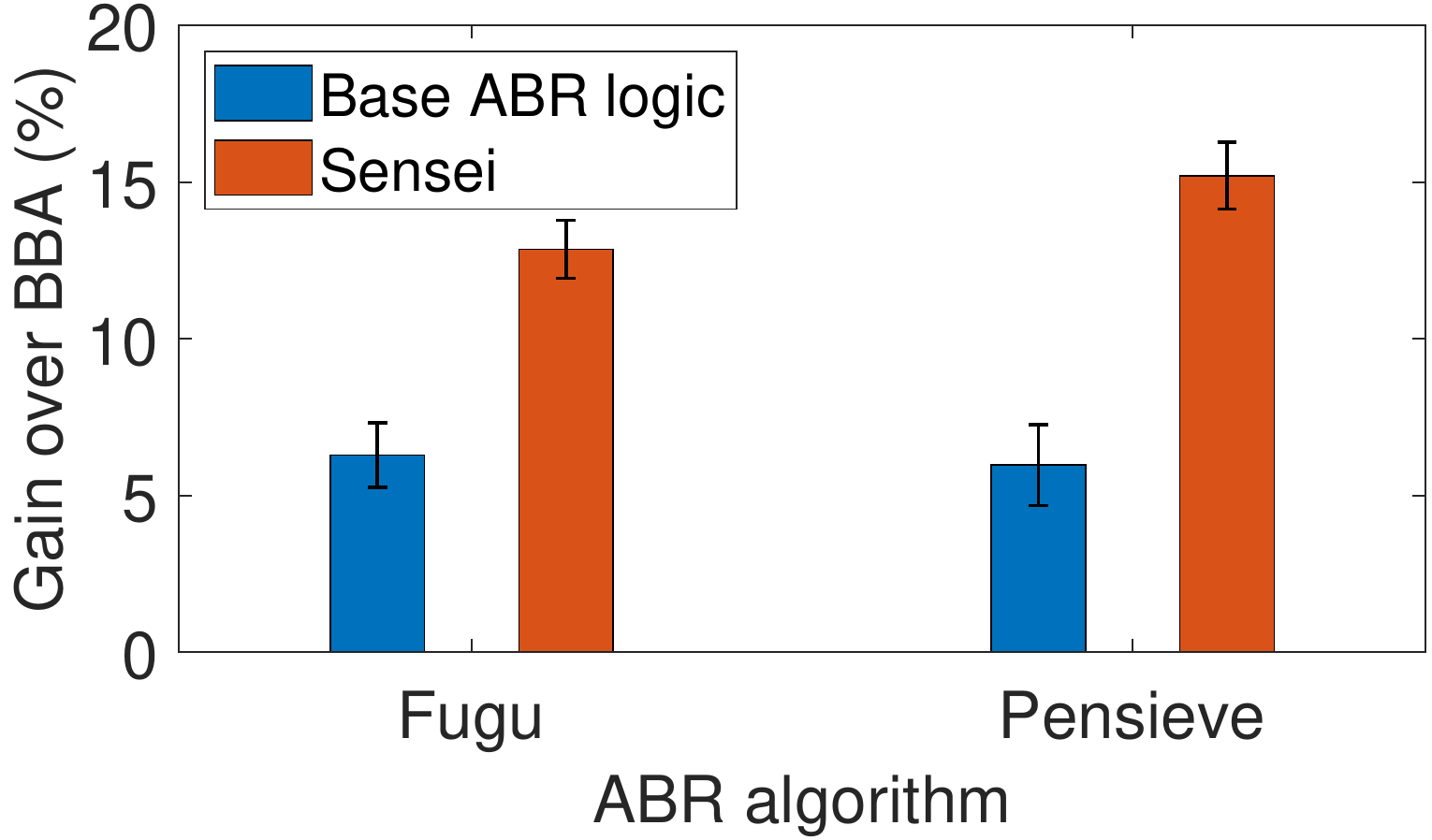}
         \caption{Impact of base ABR logic}
         \label{fig:base-abr}
       \end{subfigure}
       \hfill
         \begin{subfigure}[t]{0.48\linewidth}
         \includegraphics[width=1.0\linewidth]{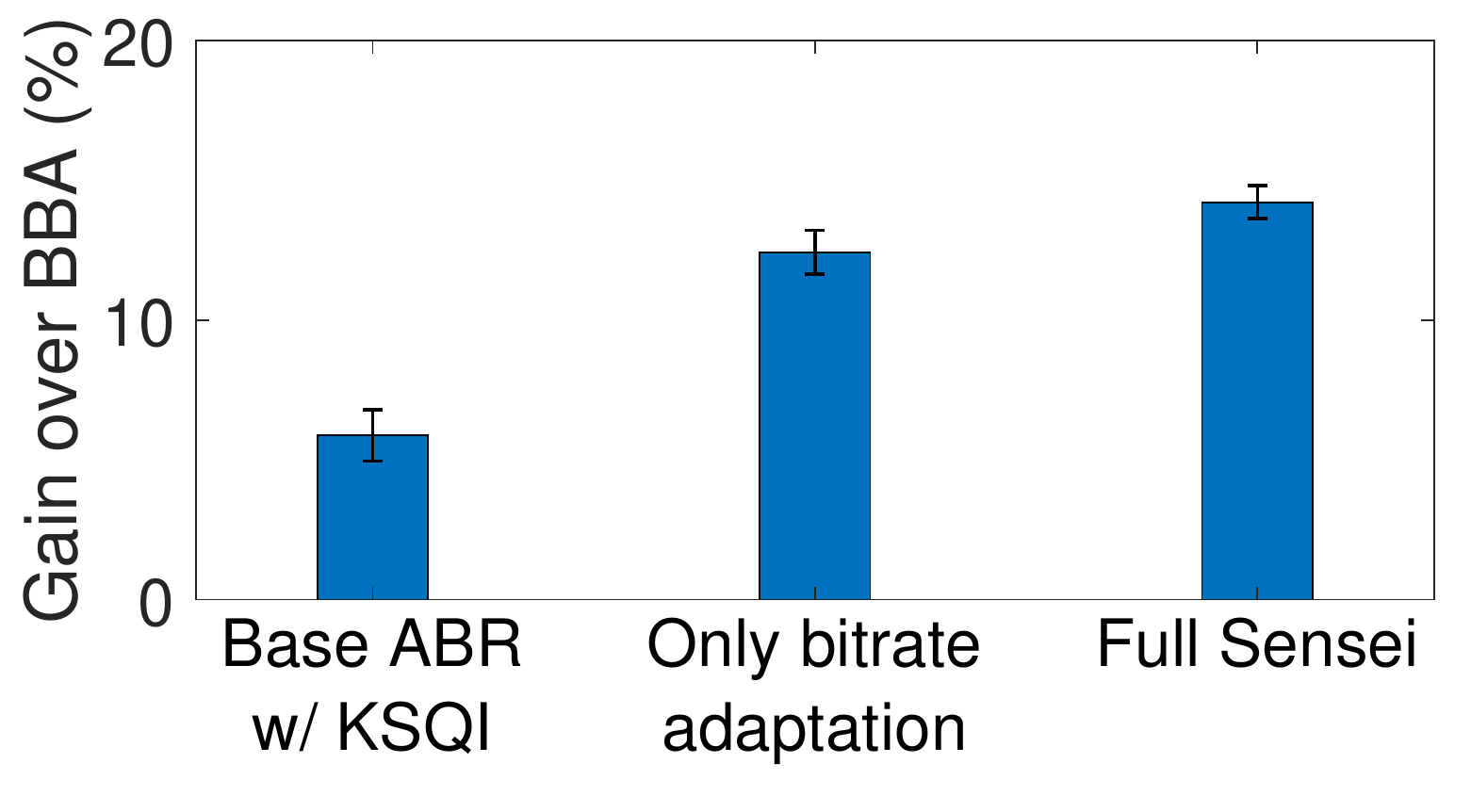}
         \caption{\name QoE breakdown}
         \label{fig:breakdown}
       \end{subfigure}
    \tightcaption{Understanding \name's improvements.}
    \label{fig:understanding}
\end{figure}

\mypara{Baselines}
We compare \name with the following ABR algorithms: Buffer-based adaptation (BBA)~\cite{huang2014buffer}, Fugu~\cite{yan2020learning}, and Pensieve~\cite{mao2017neural}. 
We keep their default settings (\eg same DNN architecture and training network traces for Pensieve, etc).
For fairness, we use KSQI as the QoE model for Pensieve, Fugu, and the base QoE model of the \name variants.
This modification strictly improves the quality of Pensieve and Fugu, because the QoE models used in their original implementation are special cases of KSQI.
Also, we restrict the intentionally chosen rebuffering time levels to be the same as those selected in the video sequences rated by crowdsourcing workers, \ie \{0, 1, 2\} seconds rebuffering time.\sid{I don't understand what ``while other...'' means}
We use \name-Pensieve (\ie the application of \name to Penseive) as \name, but we confirm that the improvements of \name-Fugu are on par with \name-Pensieve (Figure~\ref{fig:base-abr}).

\mypara{Performance metrics}
We compare ABR algorithms using three metrics.
We evaluate their QoE (normalized to [0,1]) using the same source video and throughput trace, and report the {\bf QoE gain} of one ABR algorithm ($\QoE_1$) over another ($\QoE_2$), \ie $(\QoE_1-\QoE_2)/\QoE_2$.
We scale down a given throughput trace by different ratios and, given a target QoE, calculate the {\bf bandwidth savings} by determining the minimum throughput under which each ABR algorithm achieves the target QoE.
We measure the {\bf crowdsourcing cost} paid to MTurk to get enough ratings to profile a 1-minute video. Other than \name, this cost is zero.
We also evaluate the performance of the our model by accuracy prediction in Pearson’s Coefficient (PLCC) and the rank correlation in Spearman’s Coefficient (SRCC).

\tightsubsection{End-to-end evaluation}

\begin{figure*}[t]
    \centering
    \begin{subfigure}[t]{0.31\linewidth}
         \includegraphics[width=1.0\linewidth]{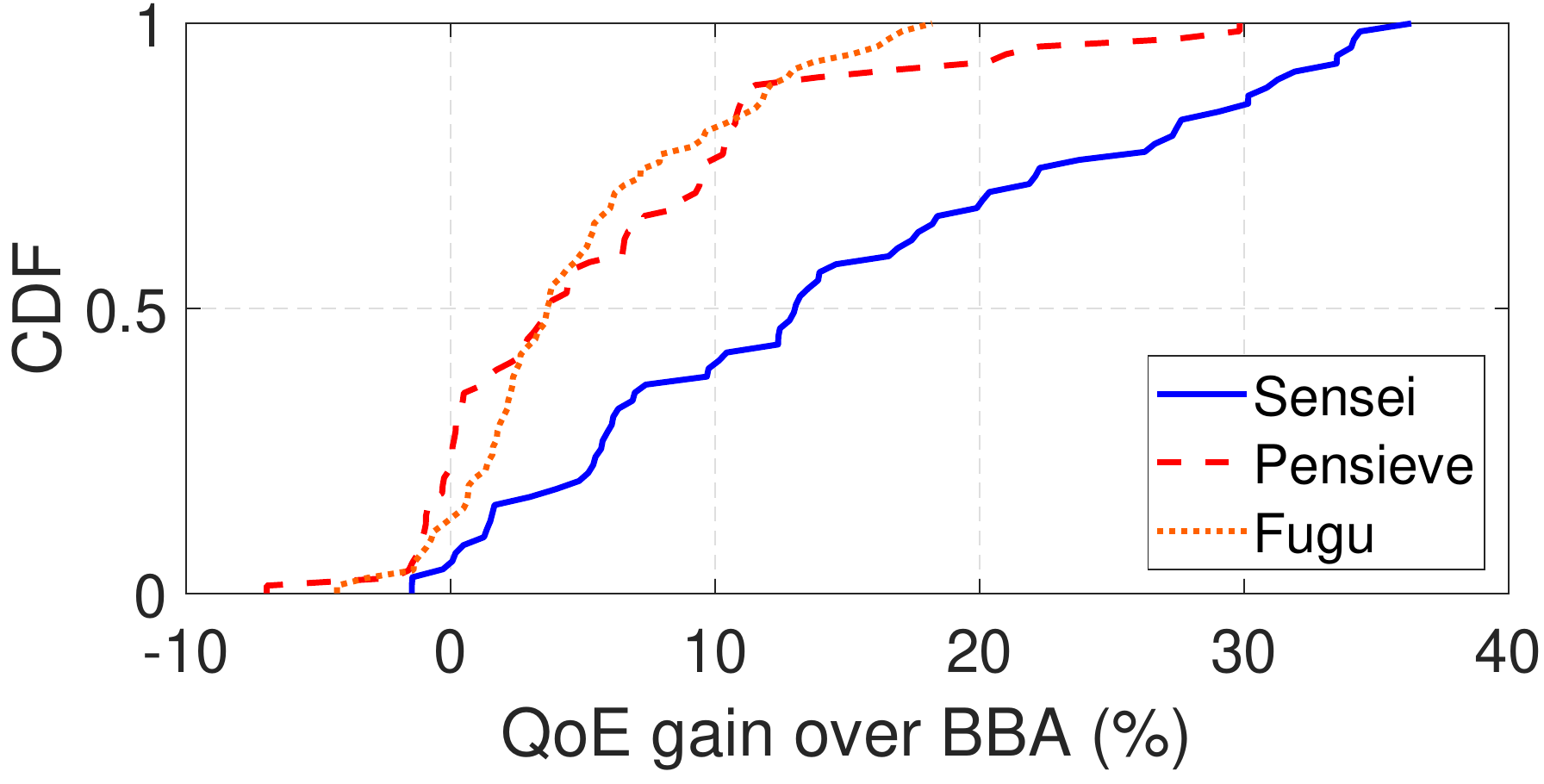}
         \caption{Distribution of QoE gains over BBA}
         \label{fig:overall_qoe_trace}
       \end{subfigure}
       \hfill
         \begin{subfigure}[t]{0.30\linewidth}
         \includegraphics[width=1.0\linewidth]{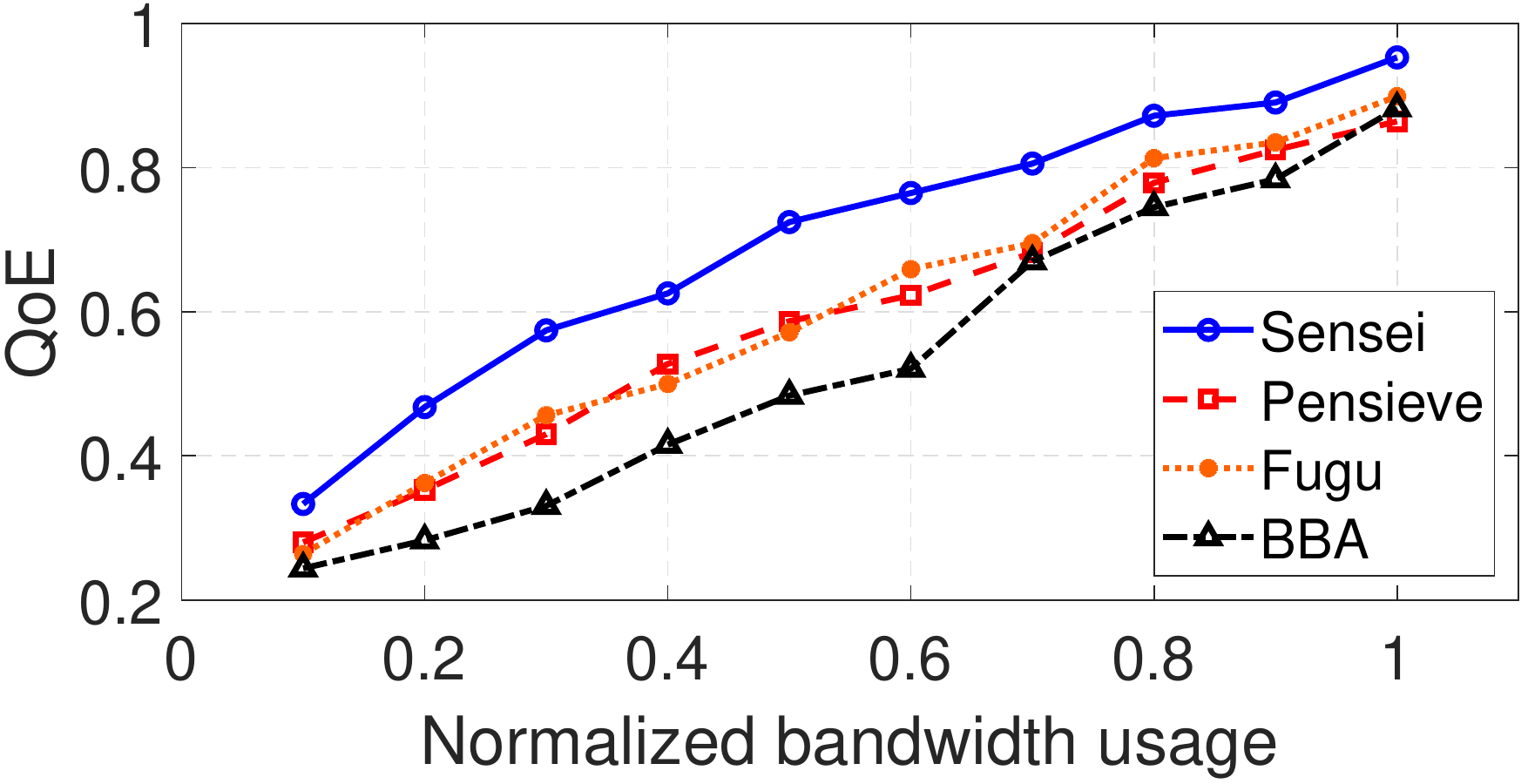}
         \caption{QoE vs. bandwidth usage}
         \label{fig:overall_performance_qoe_throughput}
       \end{subfigure}
       \hfill
         \begin{subfigure}[t]{0.31\linewidth}
         \includegraphics[width=1.0\linewidth]{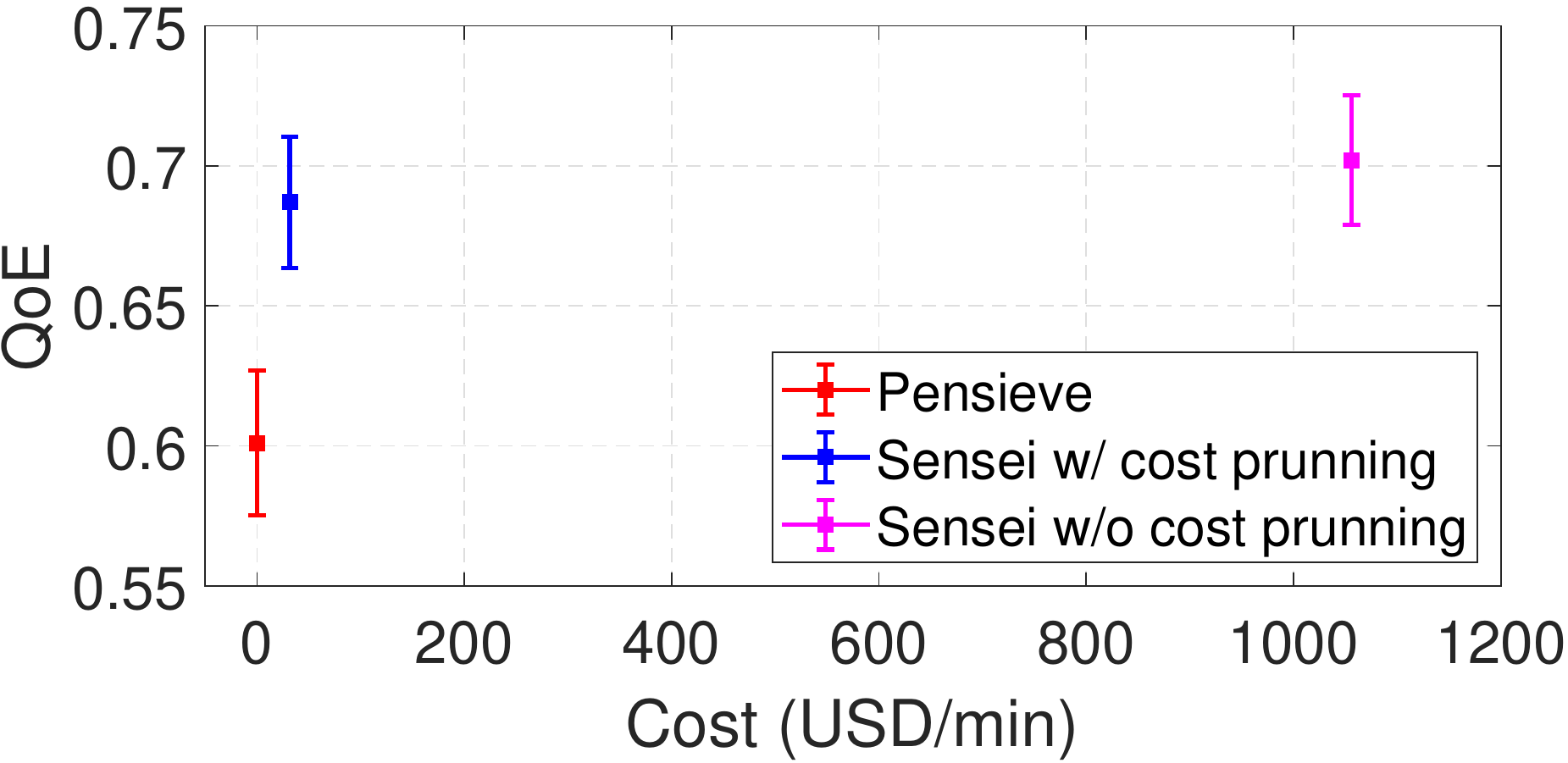}
         \caption{USD per min vs. QoE \sid{Maybe say ``USD per min'' or ``USD per minute'', otherwise sounds like ``minimum''. Also, ``Saving'' sounds too generic. Maybe say ``w/ cost pruning'' if it fits?}}
         \label{fig:qoe_cost_overall}
       \end{subfigure}
    \tightcaption{End-to-end performance of \name compared with three baselines, across all videos.
    }
    \label{fig:e2e}
\end{figure*}

\mypara{Overall QoE gains}
Figure~\ref{fig:overall_qoe_trace} shows the distributions of QoE gains of \name, Pensieve, and Fugu over BBA, across all combinations of the 16 source videos and 10 network traces.
Compared to BBA, \name has at least 14.4\% QoE gain for half of the trace-video combinations, whereas Pensieve's and Fugu's median QoE gains are around 5.7\%.
The tail improvement of \name is greater: \name's QoE gain at the 80th percentile is 5.9\%, whereas Pensieve's and Fugu's are 0.1\% and 1.3\% respectively.

\mypara{Bandwidth savings} 
Figure~\ref{fig:overall_performance_qoe_throughput} shows the average QoE of different ABR algorithms across the source videos, under one throughput trace scaled down by different ratios (x-axis). 
We confirm the results are consistent across different throughput traces.
We see that when setting a target QoE of 0.8, the bandwidth savings of \name is about 27.9\% compared to Pensieve and Fugu, and 32.1\% compared to BBA.

\begin{figure}[t]
    \centering
    \includegraphics[width = 0.43\textwidth]{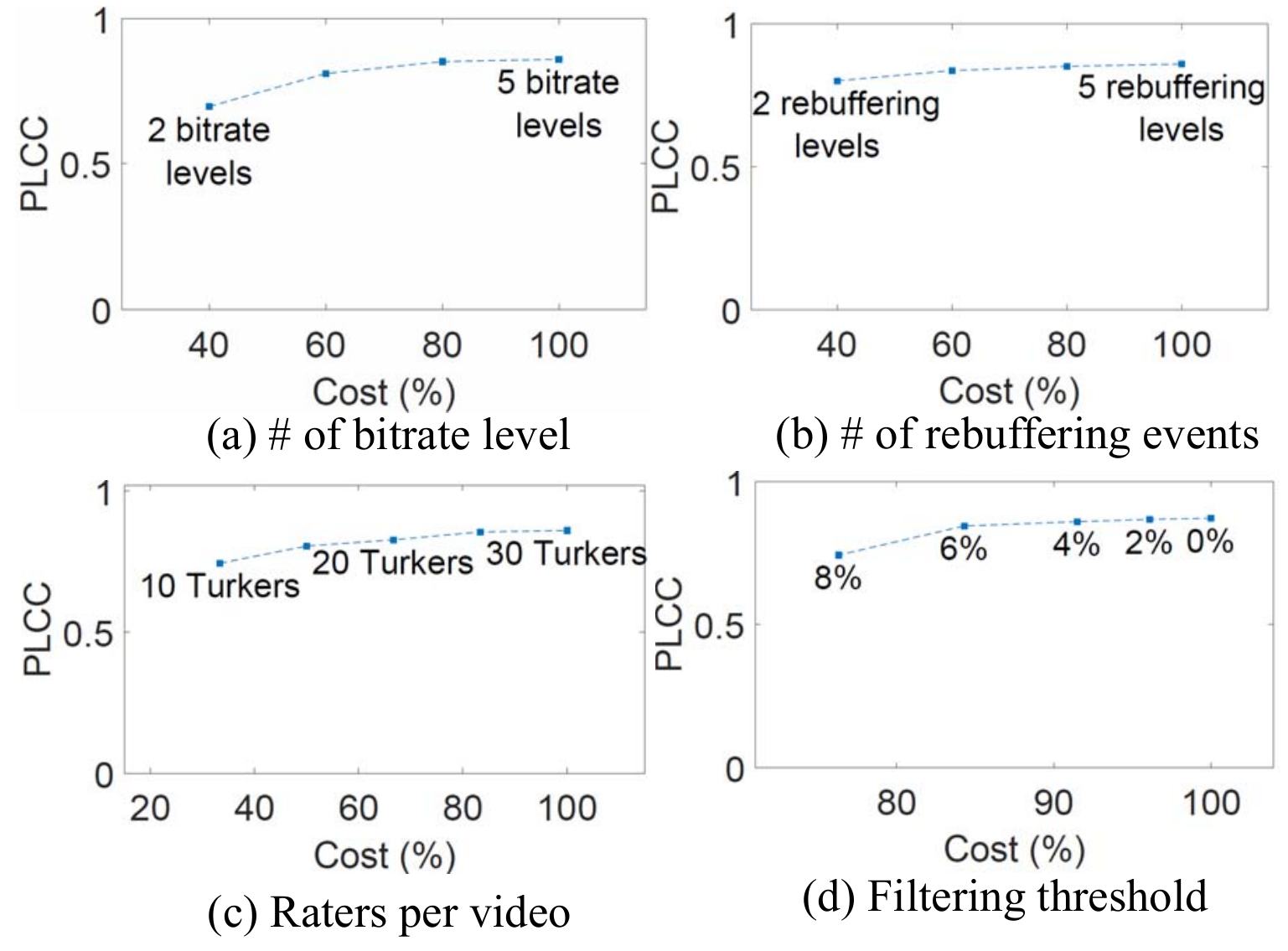}
    \tightcaption{QoE model accuracy changes with cost.}
    \label{fig:cost_accuracy}
\end{figure}

\mypara{QoE vs. crowdsourcing cost}
\name's QoE gains and bandwidth savings come at a cost.
Figure~\ref{fig:qoe_cost_overall} shows the crowdsourcing cost and resulting QoE of \name relative to Penseive, both with and without the cost-pruning optimization (which is evaluated separately in Figure~\ref{fig:cost_accuracy}).
Compared to enumerating all combinations of the quality incidents, we see that costs can be pruned by 96.7\% with only a 3.1\% degradation in QoE, and \name is still 14.7\% better on average than its base ABR logic (Pensieve with KSQI).
This cost amounts so $\sim$\$31 per 1-minute video, which is negligible compared to the other expenses incurred by content providers (see discussion in \S\ref{sec:overview}).

\begin{figure}[t]
    \centering
    \includegraphics[width=0.33\textwidth]{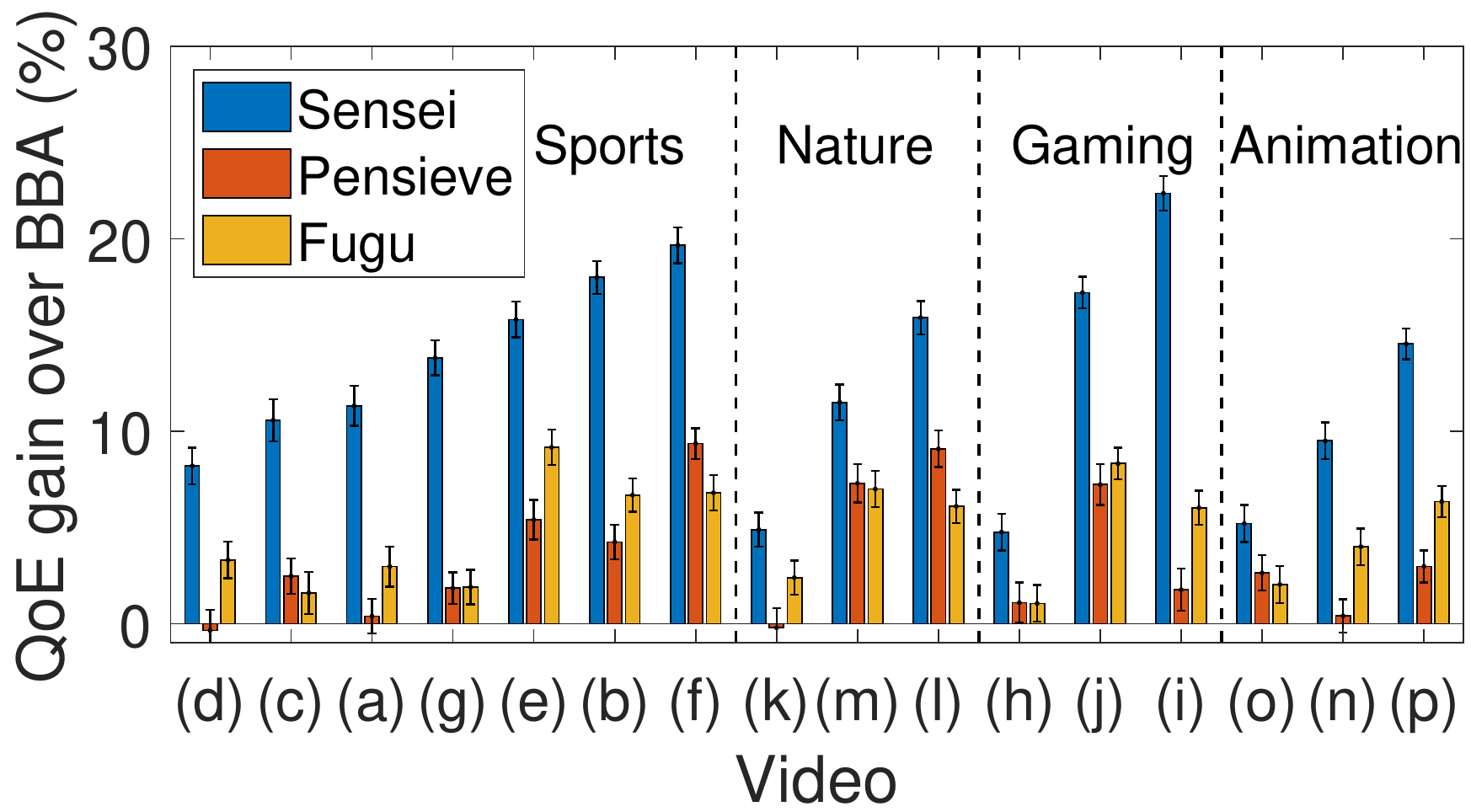}
    \tightcaption{QoE gains over BBA for each source video, grouped by genre.}
    \label{fig:overall_performance_videos}
\end{figure}

\begin{figure}[t]
    \centering
    \includegraphics[width=0.33\textwidth]{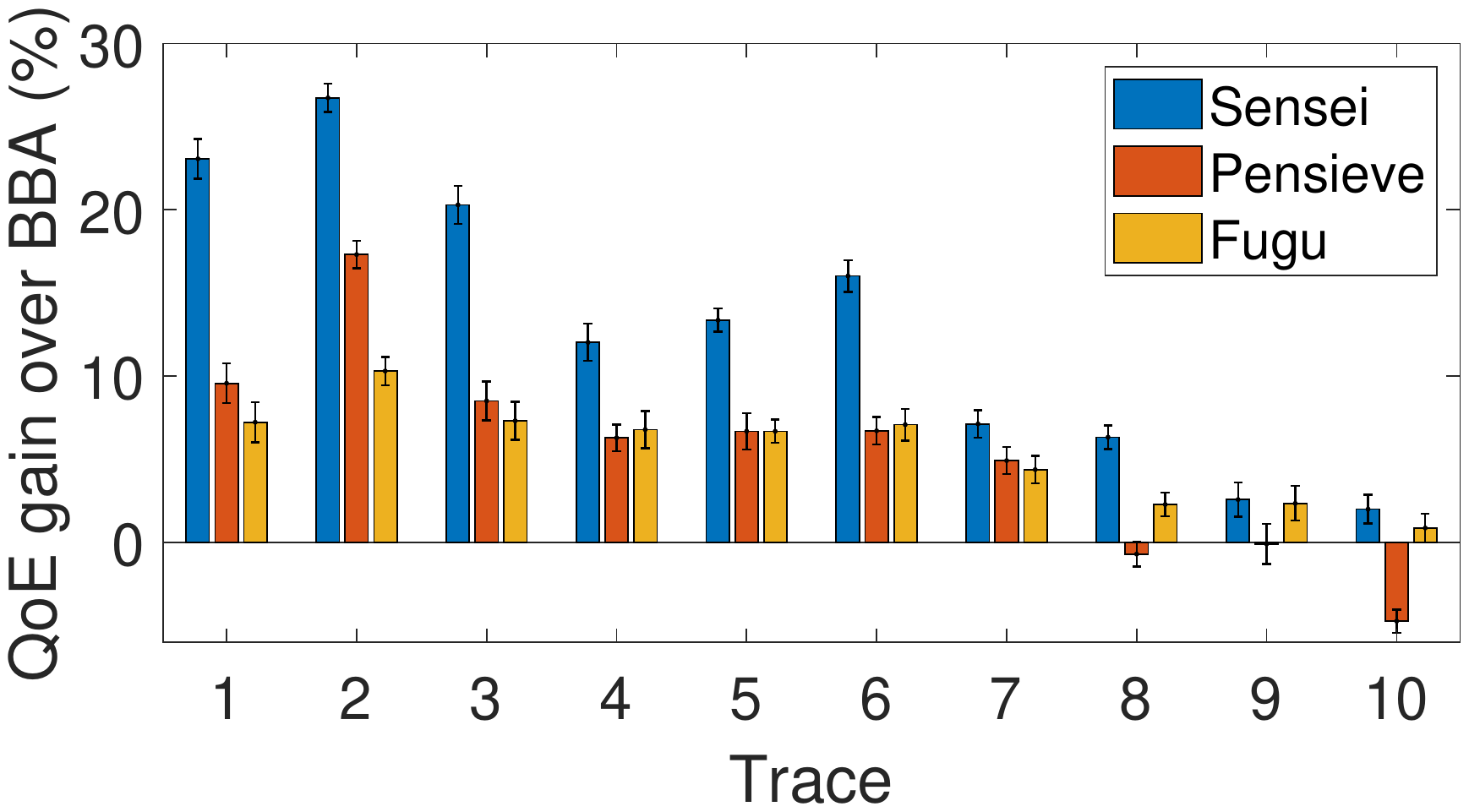}
    \tightcaption{QoE gains over BBA for each throughput trace (ordered by increasing average throughput).}
    \label{fig:overall_performance_traces}
\end{figure}

\mypara{Improvements by video and trace}
Figure~\ref{fig:overall_performance_videos} shows the QoE gains for each video.
Each bar shows the average QoE gain on the same video across all throughput traces.
We can see there is a significant variability in the QoE gains across videos even within the same genre, which shows that the dynamic quality sensitivity of a video is hardly determined by its content genre. 
Figure~\ref{fig:overall_performance_traces} shows the QoE gains for each network trace.
Each bar shows the average QoE gain on the same trace across all videos.
Overall, \name yields more improvement when the average throughput is lower (towards the left). 
This confirms our intuition that, by leveraging dynamic quality sensitivity, \name is better at maintaining high QoE even when the network is under stress.


\mypara{Implications} 
The video streaming literature has largely focused on developing new ABR algorithms (BBA vs. Pensieve vs. Fugu).
However, the fact that \name's gains over Pensieve (its base ABR logic) are similar to Pensieve's gains over BBA suggests that there is significant untapped potential in making an existing ABR algorithm aware of dynamic quality sensitivity!

\tightsubsection{\name's QoE model}
We now microbenchmark \name's QoE model (\S\ref{sec:qoe_model}). 
For each source video in our dataset, we divide it into multiple 4-second chunks and randomly select each chunk's bitrate from $\{$300, 750, 1200, 1850, 2850$\}$Kbps, and also select a rebuffering time (the start-up delay for the first chunk) from $\{0, 1, 2\}$ seconds.\sid{Isn't the reburring time chosen from $\{0,1,2\}$ seconds?}
We obtain the ground truth QoE of video sequences using our MTurk survey methodology (\S\ref{sec:qoe_model:crowd}).
We compare \name's QoE model with KSQI~\cite{duanmu2019knowledge}, LSTM-QoE~\cite{eswara2019streaming}, and P.1203~\cite{robitza2017modular} (introduced in \S\ref{sec:moti:modelling}).
We randomly split the 640 rendered videos generated by the combination of ABR algorithms, source videos, and network traces into a train set (400) and a test set (240), which are used to train and test these QoE models.

\begin{figure}[t]
      \begin{subfigure}[t]{0.27\linewidth}
         \includegraphics[width=1.0\linewidth]{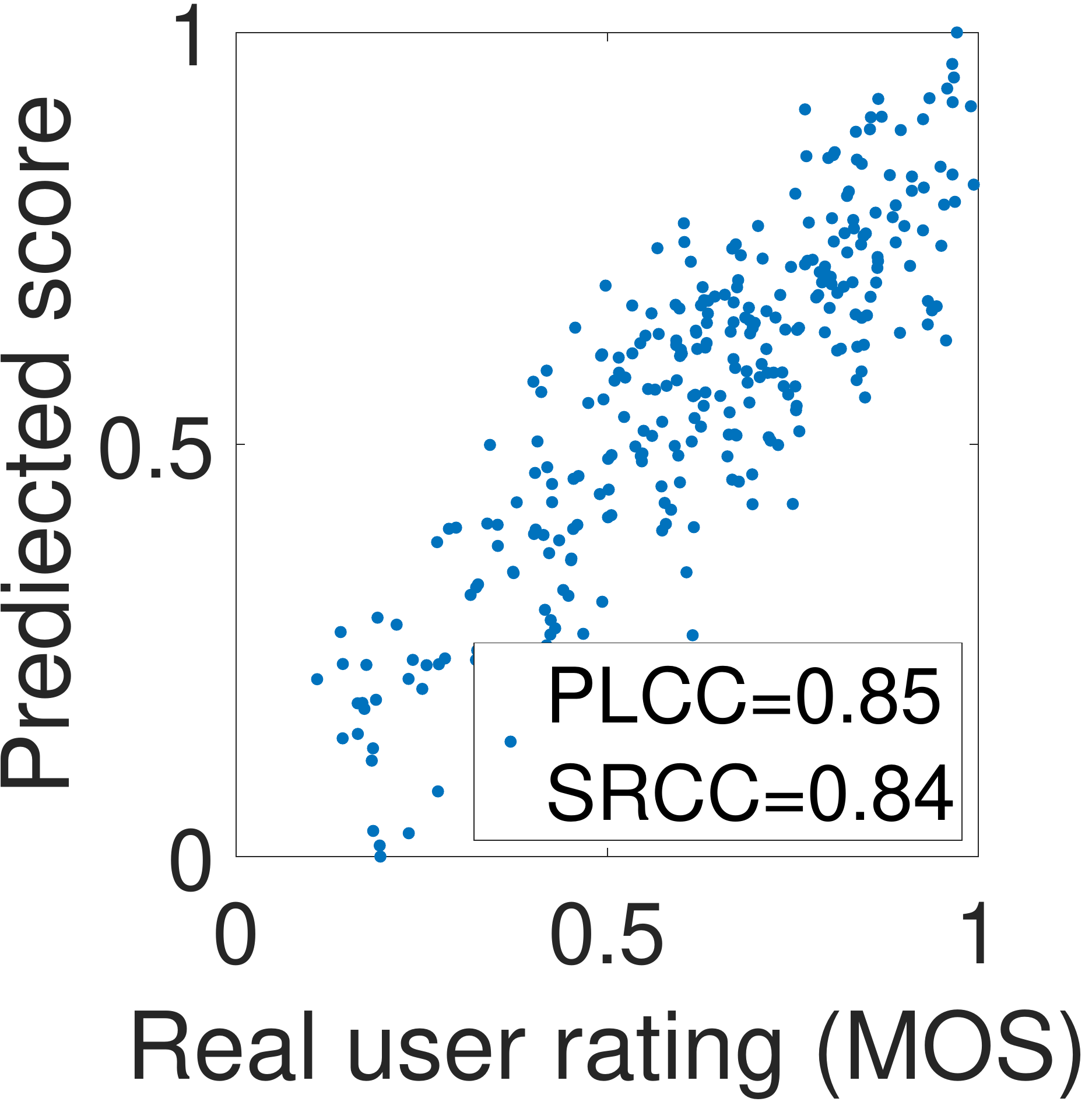}
         \caption{\name}
       \end{subfigure}
       \hfill
       \begin{subfigure}[t]{0.23\linewidth}
         \includegraphics[width=1.0\linewidth]{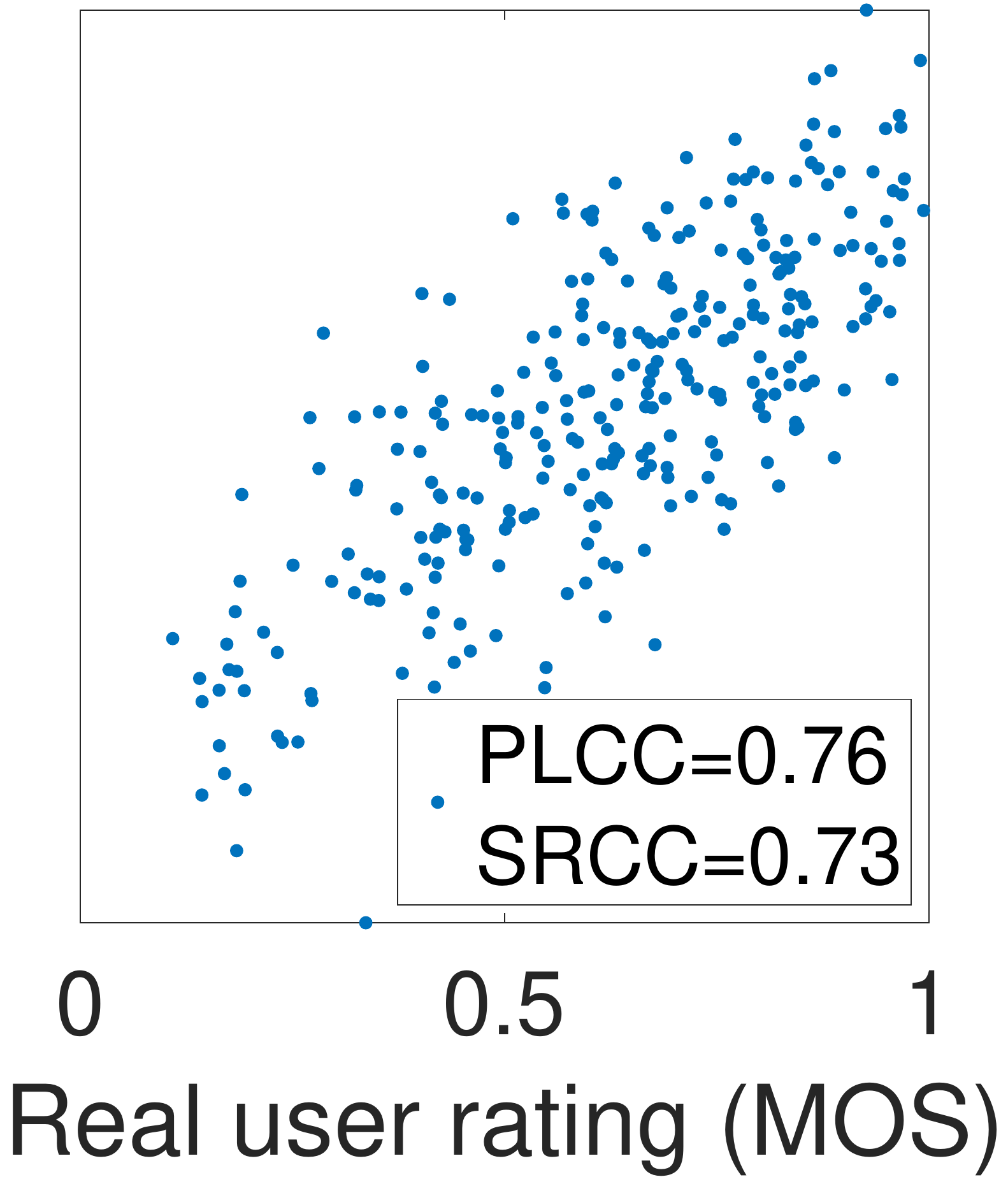}
        \caption{KSQI}
       \end{subfigure}
        \begin{subfigure}[t]{0.225\linewidth}
         \includegraphics[width=1.0\linewidth]{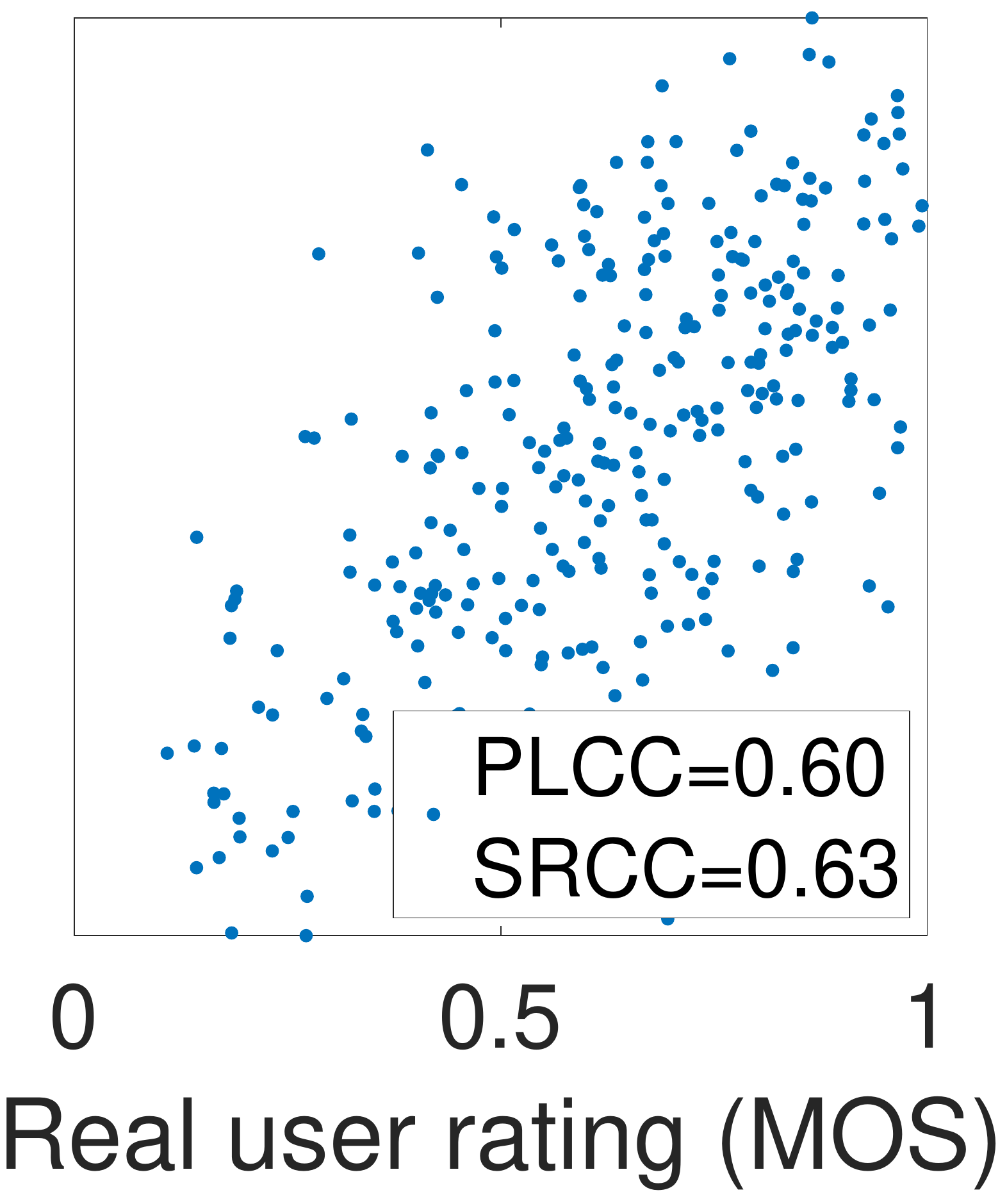}
            \caption{\hspace{-0.1cm}LSTM-QoE}
       \end{subfigure}
       \hfill
       \begin{subfigure}[t]{0.225\linewidth}
         \includegraphics[width=1.0\linewidth]{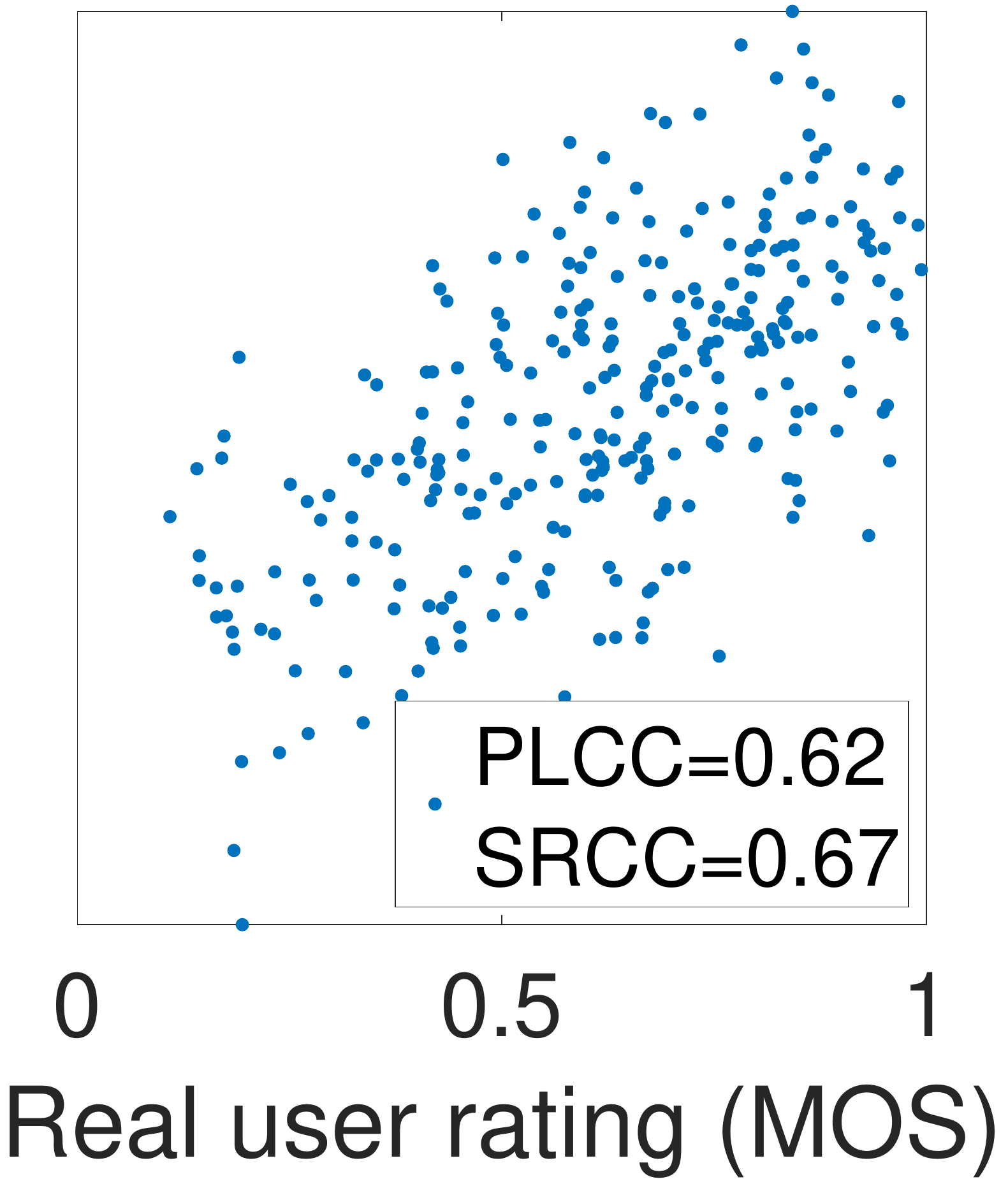}
            \caption{P.1203}
       \end{subfigure}
    \tightcaption{QoE prediction accuracy of \name and baseline QoE models.}
    \label{fig:model_accurate}
\end{figure}

\mypara{QoE prediction accuracy}
Figure~\ref{fig:model_accurate} shows the QoE scores estimated by \name's QoE model and the baseline models.
Each dot is a generated rendered video.
We can see that \name's QoE model is visually more concentrated along the diagonal (indicating high correlation with the ground truth) than any of the baselines.
\sid{Shouldn't we define these in the ``Performance metrics paragraph?''}\xu{already added}
On average, \name is over 0.85 in PLCC and 0.84 in SRCC, whereas the baselines are below 0.76 in PLCC and 0.73 in SRCC.

\mypara{MTurk cost pruning}
Figure~\ref{fig:cost_accuracy} evaluates our MTurk cost-pruning optimization and its impact on QoE prediction accuracy (measuring using PLCC).
We examine four parameters (introduced in \S\ref{sec:qoe_model:scheduling}): the number of bitrate levels ($\NumBitrates$) and the number of rebuffering lengths ($\NumBufferings$) that affect a rendered video, the number of ratings per rendered video across the two steps ($\NumTurkers = \NumTurkers_1+\NumTurkers_2$) \sid{People will wonder how $M$ is broken down in our experiments?}, and the difference threshold $\ThreshDiff$ used to pick which chunks to investigate in the second step.
As the figure shows, by lowering each parameter from its highest value, we can greatly reduce the cost while incurring less than 3\% drop in accuracy.
This holds if we reduce the number of bitrates from 5 to 3, the number of rebuffering times from 5 to 2, the number of ratings per video from 30 to 10, or raise the difference threshold from 0\% to 6\%.
We conclude that MTurk costs can be dramatically reduced by setting these parameters at the desired ``sweet spot'' between cost savings and QoE prediction accuracy.

\tightsubsection{\name's ABR logic}

Next, we microbenchmark \name's ABR logic (\S\ref{sec:abr}).
To scale our experiment out, we use real videos and throughput traces but use the QoE predicted by \name (instead of real user ratings) to evaluate QoE. 
This modification yields the same QoE estimation on average compared to real user ratings under the same setting (see Figure~\ref{fig:model_accurate} (a)).\sid{I don't see how the figure shows this point though.}

\begin{figure}[t]
    \centering
    \includegraphics[width=0.23\textwidth]{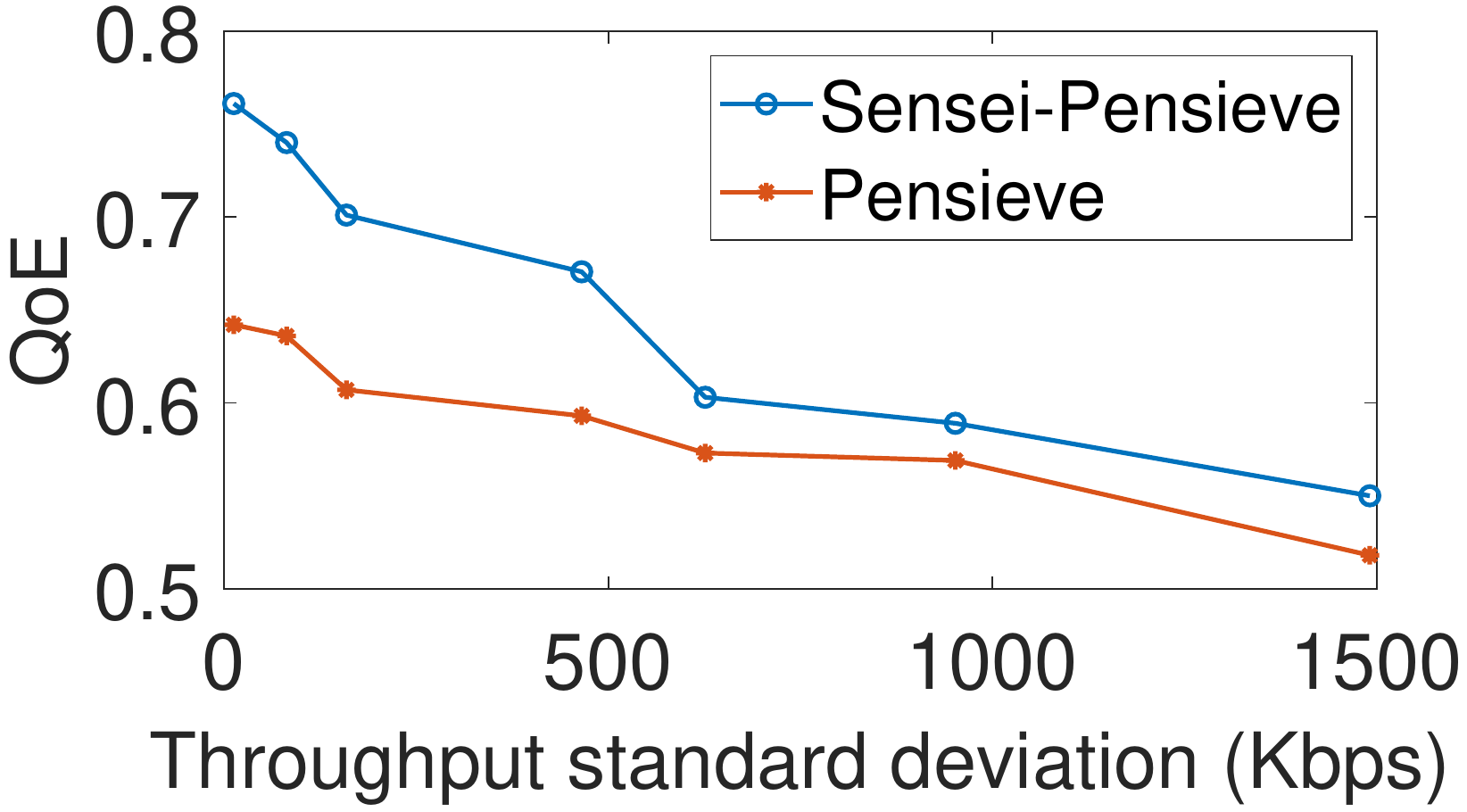}        \includegraphics[width=0.23\textwidth]{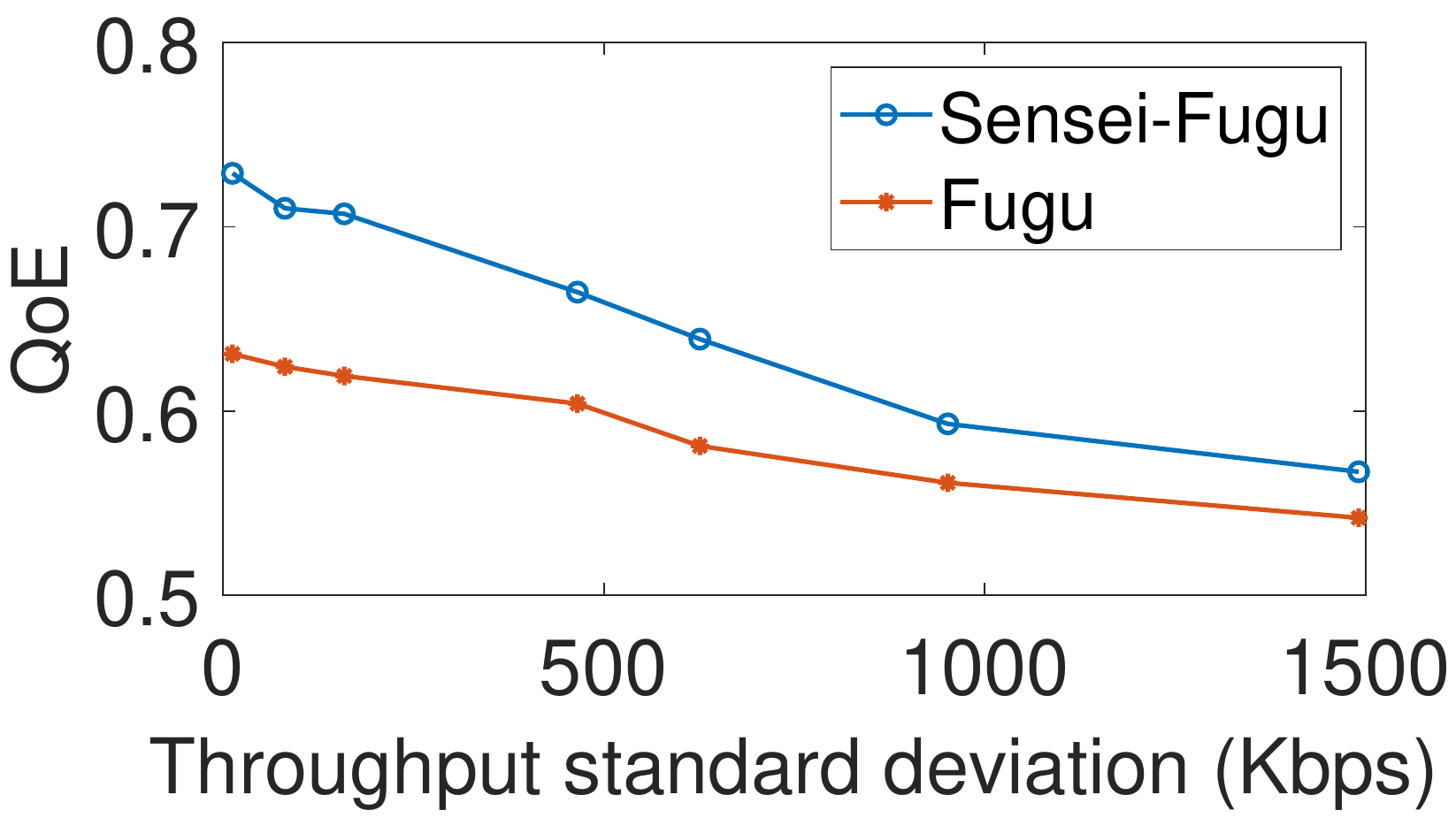}
    \tightcaption{QoE under increasing bandwidth variance.\sid{Should we label them as
    Sensei-Pensieve and Sensei-Fugu here?}}
    \label{fig:bw-var}
\end{figure}

\mypara{Impact of bandwidth variance}
Figure~\ref{fig:bw-var} shows the performance of \name under different levels of throughput variance.
We pick one throughput trace and increase its throughput variance by adding a Gaussian noise with zero mean.\sid{Should we specify the parameters?} 
We confirm the results are similar for other throughput traces.
The graph begins at the variance of the original throughput trace; as variance increases, \name's QoE degrades, but it still maintains a significant gain over its base ABR logic. 
\name is robust because it relies on the ability to predict how likely low quality is to occur on {\em only} high quality-sensitivity chunks, not all future chunks. 
So as long as the average throughput between now and the next high quality-sensitivity chunk is stable, \name will work well.

\mypara{Understanding \name's improvements}
Figure~\ref{fig:base-abr} shows that \name achieves comparable improvement when either Pensieve or Fugu are the base ABR logic. 
This suggests that \name's effectiveness does not depend on the design of the base ABR logic.
Figure~\ref{fig:breakdown} separates the two sources of \name's improvement: (1) making the QoE objective aware of dynamic quality sensitivity, shown by the gap between the 1$^{\textrm{st}}$ and 2$^{\textrm{nd}}$ bars, and (2) additional new adaptation actions (\eg initiating rebuffering proactively), shown by the gap between the 2$^{\textrm{nd}}$ and 3$^{\textrm{rd}}$ bars.
We see that both sources improve QoE, but changing the QoE objective has a greater contribution. 
So even if a content provider cannot control rebuffering, it can still benefit significantly from \name by incorporating dynamic user sensitivity.

\mypara{Systems overhead}
We confirm empirically that compared to a video player without \name, the runtime overhead of \name is less than 1\% in both CPU cycles and RAM usage.

%% file: relatedwork.tex
\tightsection{Related Work}
\label{sec:relatedwork}

% We briefly survey the most related work on video QoE model, ABR algorithm and user study platform.

\mypara{Video QoE modeling}
Video QoE modeling takes two general forms: visual quality assessment (VQA) and adaptation quality assessment (AQA). \sid{A bit odd that the second form doesn't have an acronym}
% The former assesses visual appearance under various video encoding.
Classic VQA (\eg~\cite{gonzalez2004digital,wang2004image,rehman2015display,soundararajan2012video}) focuses on the user's perception of pixel value changes due to encoding.
% and structural information in a frame. 
Recent work uses more data-driven models, including SVM~\cite{vmaf} and deep learning models (\eg~\cite{kim2018deep}).\sid{From a quick scan of the paepr it doesn't seem like they use RL}
%tries to links the video pixels to user ratings by deep reinforcement learning.
% computes the pixel difference between the source video and the test video, and
%  and STRRED \cite{soundararajan2012video} use structural information of frames.
% Recently, QoE modeling for adaptive video streaming has attracted much attention from industry and academia.
% Due to the intractable number of the factors that affect QoE, the QoE models can be categorized by their observation space.
% The video quality assessment (VQA) models evaluates the pixel-level visual quality of the test videos.
% To estimate the visual quality, PSNR \cite{gonzalez2004digital} computes the pixel difference between the source video and the test video, and
% SSIM \cite{wang2004image, rehman2015display} and STRRED \cite{soundararajan2012video} use structural information of frames.
% Recent VQA models exploit the user feedback in term of ratings to estimate the user-perceived quality.
% VMAF \cite{vmaf} uses a support vector machin (SVM) based regression, while DeepVQA \cite{kim2018deep} tries to links the video pixels to user ratings by deep reinforcement learning.
AQA considers streaming-related incidents (join time, quality switches, rebuffering, etc.) which can influence user experience and engagement (\eg~\cite{dobrian2011understanding,balachandran2013developing,krishnan2013video}).
% Other than pixel-level visual quality, advanced QoE models also take the quality incidents into account.
Various models have been developed to capture their effect~\cite{bampis2017continuous,bampis2017augmented, duanmu2016quality,duanmu2019knowledge,de2013model,chen2014modeling,eswara2019streaming} as well as users' attention over space (\eg\cite{ozcinar2019visual,nguyen2018your,xu2020introduction}) and 
time (\eg\cite{ghadiyaram2018learning,duanmu2018quality,gao2019content}).
\name is complementary to these efforts: they apply the same heuristics to all videos, but \name customizes itself for each video (in a cost-efficient way) to capture the true user sensitivity to video quality.

% Those models weight the QoE influnce of the quality incidents and the visual quality quality by different regression models: nonlinear autoregressive exogenous (NARX) model \cite{bampis2017continuous, bampis2017augmented}, quadratic programming \cite{duanmu2016quality, duanmu2019knowledge}, linear regression \cite{de2013model}, Hammerstein-Wiener model \cite{chen2014modeling}, long short-term memory network \cite{eswara2019streaming} etc.

\mypara{Adaptive bitrate (ABR) algorithms}
% Recent years have witnessed the prevalence of DASH systems where the ABR algorithms is the core part.
ABR algorithms can be grouped into buffer-based (\eg~\cite{jiang2012improving, li2014probe}) and rate-based algorithms (\eg~\cite{huang2014buffer,spiteri2016bola,spiteri2019theory}).
% They both adapt bitrate though rate-based ABR uses throughput measurement as input and buffer-based ABR uses buffer occupancy as input.
% The rate-based algorithms (\eg~\cite{jiang2012improving, li2014probe}) tries to match the bitrate to the network throughput, while the buffer-based algorithms (\eg~\cite{huang2014buffer,spiteri2016bola,spiteri2019theory}) uses buffer occupancy as a more reliable control signal to adapt bitrate. 
%the bitrate version of the next chunks.
More recent ABR algorithms combine the two approaches to optimize for explicit QoE objectives, via control theory~\cite{yin2015control}, machine learning-based throughput prediction~\cite{sun2016cs2p,yan2020learning}, or deep reinforcement learning~\cite{mao2017neural}.
Key parameters of the ABR logic can be customized to the network conditions or device~\cite{akhtar2018oboe}.
% both buffer occupancy and future network throughput into account. MPC \cite{yin2015control} uses a control-theory-based algorithm to optimize the QoE of the chunks in a horizon. CS2P's \cite{sun2016cs2p} uses the MPC's ABR algorithm but a better network predictor that regards the network throughput as discreted states and will notify the systems when the throughput state changes. Puffer uses a machine-learning method to predict throughput as a discrete probability distribution, by which Puffer optimizes the expected QoE of the future chunks by MPC controller.
% Pensieve takes the buffer occupancy and the past throughput as the input, and use a deep reinforcement learning model to select the bitrate for the next chunks.
% Oboe~\cite{akhtar2018oboe} claims that the performance of the ABR algorithms highly depend on the parameters, and Oboe designs an algorithm for finding the best set of the parameters for the existing ABR algorithsm.
Though \name reuses existing ABR algorithms, 
% changes the input/output information and QoE objective, 
its contribution is to identify a minimal set of changes (such as adaptation actions they never would have taken) needed for these algorithms to fully leverage dynamic quality sensitivity.

\mypara{QoE research using crowdsourcing} 
Some prior studies explore the use of crowdsourcing platforms to obtain direct QoE feedback.
% These efforts have so far been focused on providing more data to build better QoE models or compare application-level protocols (\eg ABR).
% Recent research on QoE heavily relies on the subjective evaluation collected from real end users.
% The low cost and flexible of crowdsourcing has attracted much attention from  industry and academia.
% The crowdsourcing platforms, \eg Amazon MTurk\cite{mturk} and Microworkers\cite{microworkers}, provide the most direct and unfiltered access to the participants, which enables a fast deployment of a large-scale user study.
% Upon those platform, a number of frameworks are built for multimedia user study.
Some of these works (\eg~\cite{hossfeld2011quantification,chen2010quadrant,rainer2013web}) provide methodologies for using commercial crowdsourcing platforms, \eg, Amazon MTurk~\cite{mturk} and Microworkers~\cite{microworkers}, to systematically model user perception using objective quality metrics. \sid{Make MTurk and Microworkers parenthetical?}
Other works build crowdsourcing platforms themselves for similar purposes (\eg~\cite{varvello2016eyeorg,yan2020learning}). 
Compared to these efforts, \name faces a unique challenge of scaling crowdsourcing to a {\em per-video} basis. % in order to identify dynamism of quality sensitivity on a new video. 
While our cost pruning techniques are conventional, we enable them through the insight that per-chunk sensitivity weights provide minimal yet sufficient information to embrace dynamic quality sensitivity.

% to customize QoE modeling to any new video.

% WESP~\cite{rainer2013web} provides APIs to configure each component of the user study, \eg pre-questionnaire, and control questions.
% Its APIs also allows us to design the survey on the common UI, \eg single stimulus, double
% stimulus, pair comparison or continuous quality evaluation.
% Quadrant of Euphoria (QE) \cite{chen2010quadrant} uses pair comparison instead of a MOS test.
% In QE, the QoE of a test multimedia content is estimated by the probability that this content has larger QoE than another random content.
% Eyeorg \cite{varvello2016eyeorg} studies relationship between user-perceived web loading time and QoE. Like QE, Eyeorg shows the web page loading event side by side to increase the realibility of the user ratings.
% To control the page loading time, Eyeorg shows the videos that record the page loading events with different latency to the raters.
% YoutubeQoE \cite{hossfeld2011quantification} is a platform that can collect the user feedback for the videos.
% It collects the user ratings (1-5) to quantify the QoE of the test videos.
% To ensure the reliability of the user ratings, YouTubeQoE uses the control questions, \eg what content the video is about.

%% file: appendix.tex
\appendix

\vspace{0.1cm}
% \noindent Appendices are supporting material that has not been peer reviewed.

\section{Dataset}

\begin{figure*}[t]
    \centering
    \includegraphics[width=0.9\linewidth]{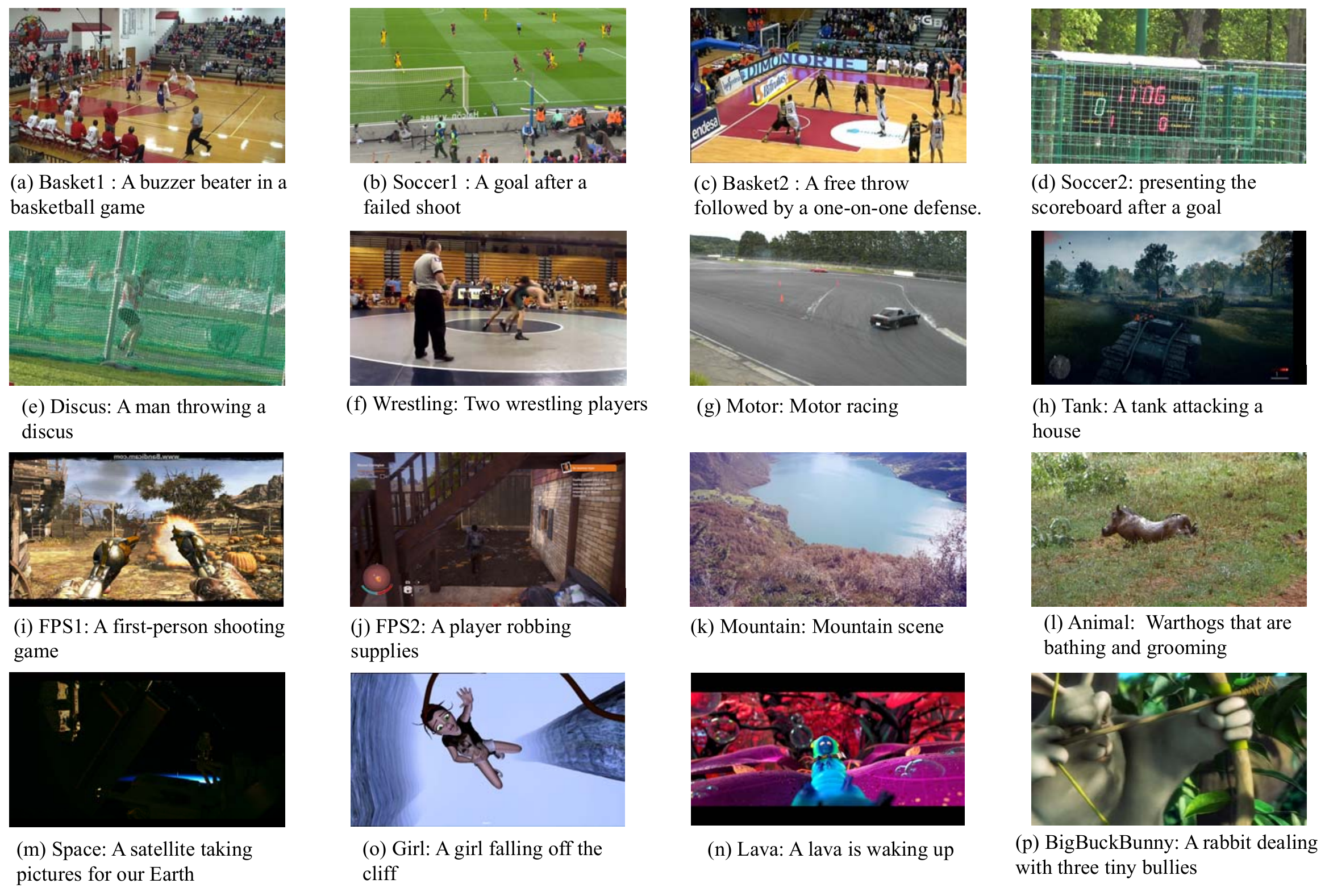}
    \caption{Summary of source videos in our dataset. They span four genres: sports (a - g), gaming (h - j), natural (k - m), and animation (n - p). 
    They are compiled randomly from public QoE datasets: LIVE-MOBILE~\cite{ghadiyaram2017subjective}(a,k), LIVE-NFLX-II~\cite{bampis2018towards} (b, n), YouTube-UGC~\cite{wang2019youtube} (c - j and l - o); WaterlooSQOE-III~\cite{duanmu2016quality} (p). }
    \label{fig:dataset_screenshot}
\end{figure*}

Figure~\ref{fig:dataset_screenshot} provides screenshots and descriptions of the 16 source videos used in our dataset.

\section{MTurk rating sanitization}
\label{sec:sanitization}

As mentioned in \S\ref{sec:qoe_model:crowd}, the user ratings can be noisy and biased for multiple reasons. 
To get reliable quality ratings, we take the following steps.
\begin{packeditemize}
\item {\em Rejection criteria:} 
There is always one {\em reference} video as one (at a random position) of the rendered videos in each survey. The reference video has the highest quality (highest bitrate and no rebuffering) and if a Turker gives any other rendered video a higher rating than the reference video, we will reject the scores of this Turker. 
We will reject a Turker if one of the rendered videos is viewed for less than its length (\ie the video is not fully watched). 
The rejection criteria help to ensure that the Turkers give quality-induced ratings after watching each video in full.
When a Turker's data is rejected, the Turker will not receive payment.\footnote{Of course, this criteria is highlighted on the first page of the survey and we have received disputes from Turkers but not often.}
\item {\em Randomizing viewing orders:} 
Every time a Turker joins the survey, we randomize the order in which rendered videos are shown. 
This allows us to perform postanalysis to check (and confirm) that the position of a video within a survey has little impact on its rating.
\item {\em Biases by Turker population:}
We also check that about 43.8\% (or 67.3\%) ratings from Turkers who participate our survey only once (or at most twice). This confirms that the Turker pool is large enough avoid small population bias. 
This corroborate with the sanity-check result in \S\ref{sec:qoe_model:crowd} that on average MTurk quality ratings are strongly correlated with in-lab survey results.
\end{packeditemize}

To prevent people from gaming the system by sitting on a job for too long, we pay each participant a fixed amount equal to the fixed hourly rate times the estimated amount of time needed for the survey (which is proportional to the total length of the videos per participant). 
We use an hourly rate of \$10 (roughly matching the minimum wage standard in our state), though we have not explored the impact of raising/lowering this wage.
Of course, participants will not get paid if they only watch part of the videos. %(see other rejection criteria in \S\ref{sec:impl:turker}).

\section{Lessons from MTurk Experiments}
\label{sec:lessons}
We conclude with key lessons learned from running MTurk experiments and focus on the difference between in-lab and crowdource-based QoE studies.

\mypara{How many Turkers are needed}
An essential tradeoff between crowdsourcing and in-lab study (\eg~\cite{duanmu2016quality, bampis2018towards}) is crowdsourcing can easily find more participants for us but their work is less reliable.
So in chooseing between in-lab and crowdsourcing, a key parameter is how many Turkers are needed to get a reliable signal. 
We did a head-to-head comparison with WaterlooSQOE-III \cite{duanmu2016quality} and found that as far as QoE assessment is concerned, we need 17\% more Turkers to reduce the variance of QoE rating down to the same level if the same video is rated in an in-lab study.

\mypara{Reputation matters}
MTurk platform (and other crowdsourcing tools) dramatically cuts the cost of making our survey visible to potential Turkers, but if Turkers sign up slowly this can slow down the whole process.
The speed at which they sign up depends largely on the reputation of the publisher (us) and in general, Turkers are more willing to sign up if the publisher historically has a low rejection rate. 
This means we must be very clear upfront about the studies rejection criteria.
% Unlike in-lab studies, each MTurk survey will recruit Turkers from scratch. 

Reputation is crucial to get quality feedback.
We follow a common practice (\eg~\cite{lovett2018data}) and restrict ourselves to {\em master Turkers}, a class of {\em reliable} Turkers who have at least participated in 1000 surveys and whose overall rejection rate is lower than 1\%. 
It also corroborates our experience: rejection rate from these Turkers over $4\times$ lower than normal Turkers, though their hourly rate is generally higher than normal Turkers. 

\begin{figure*}[t]
      \begin{subfigure}[t]{0.24\linewidth}
         \includegraphics[width=1.0\linewidth]{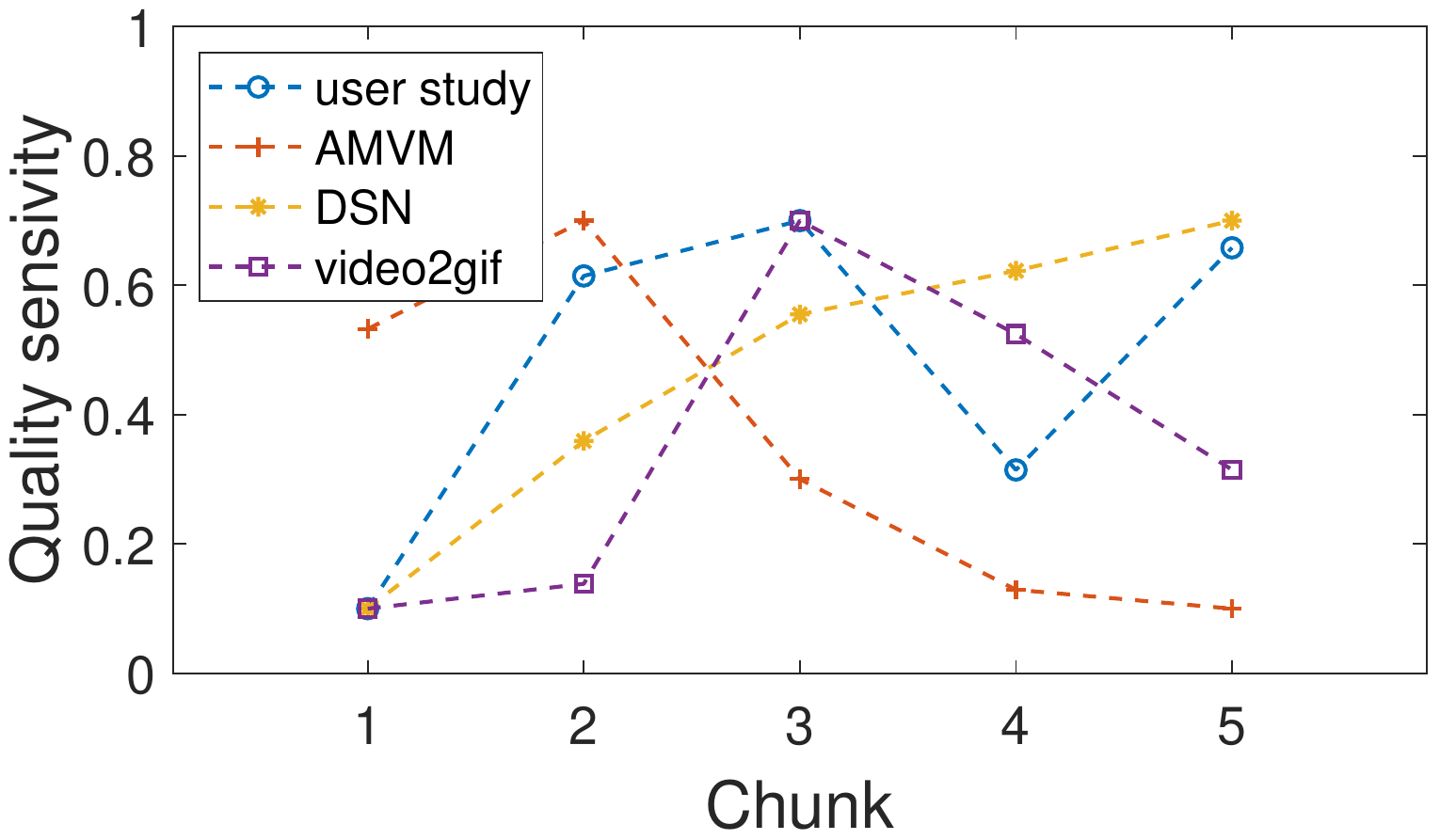}
        %  \caption{Lava}
        \caption{Lava}
      \end{subfigure}
      \hfill
      \begin{subfigure}[t]{0.24\linewidth}
         \includegraphics[width=1.0\linewidth]{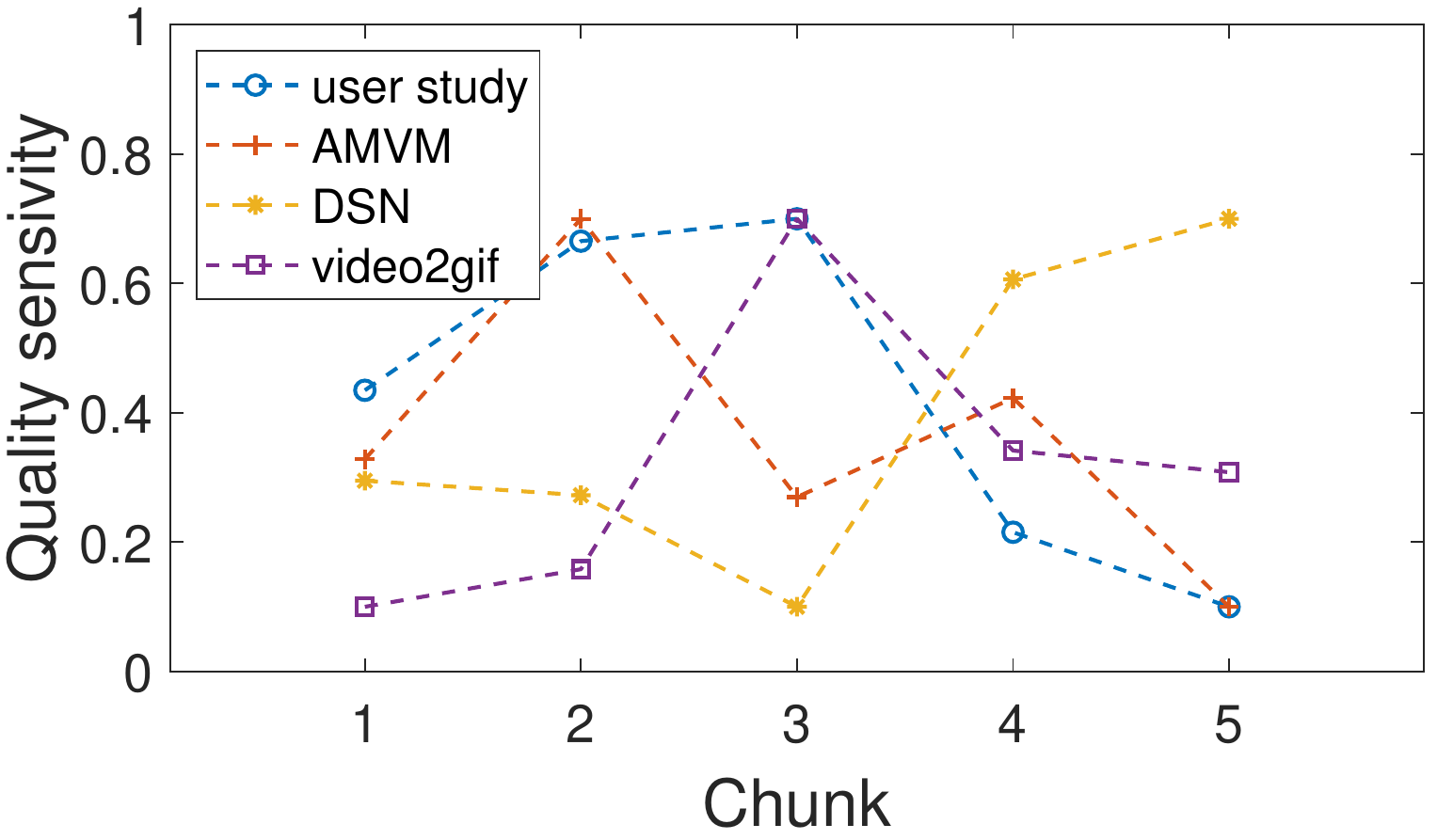}
        % \caption{Tank}
        \caption{Tank}
      \end{subfigure}
      \hfill
      \begin{subfigure}[t]{0.24\linewidth}
            \includegraphics[width=1.0\linewidth]{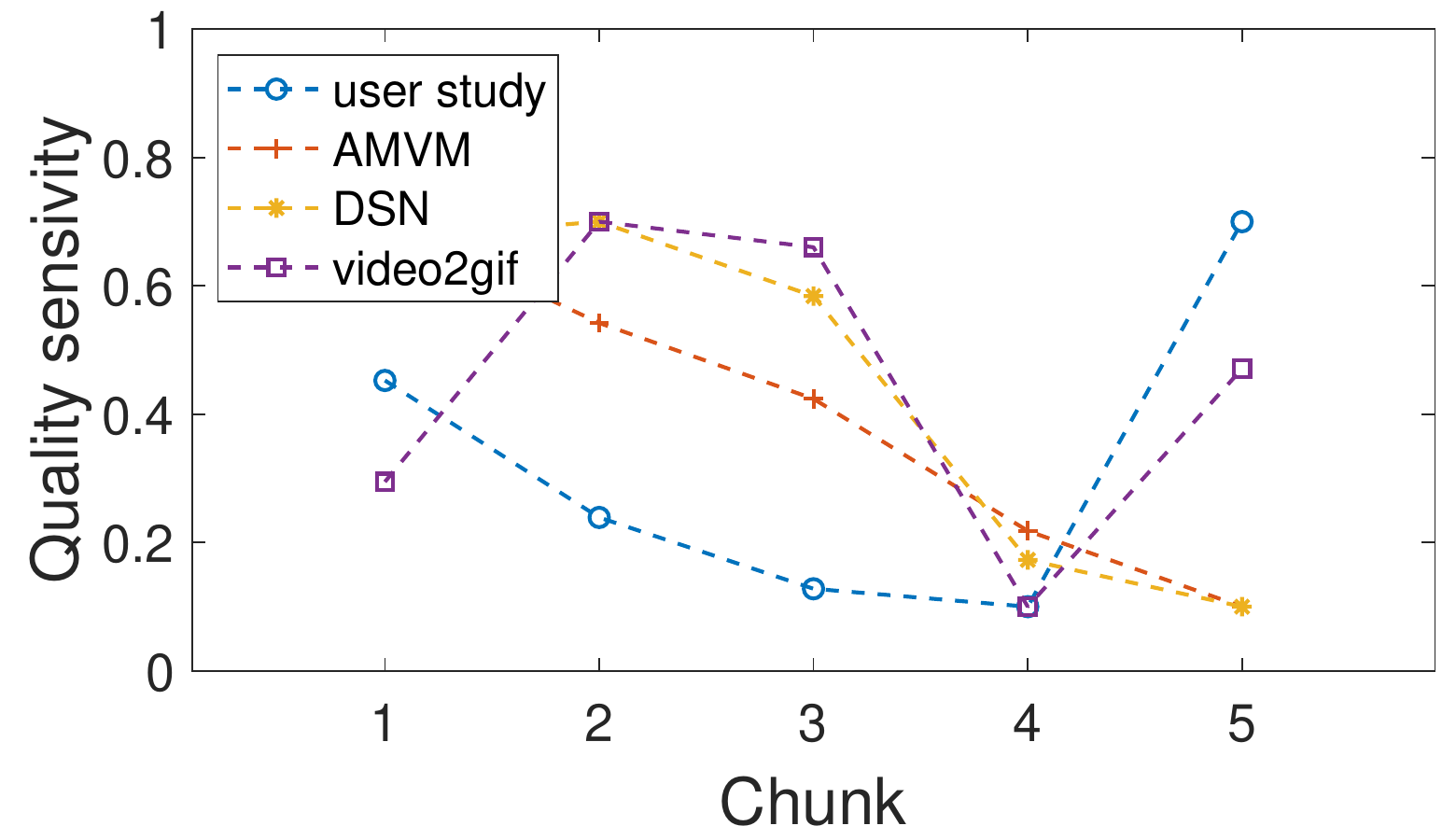}
            \caption{Animal}
      \end{subfigure}
      \hfill
        \begin{subfigure}[t]{0.24\linewidth}
            \includegraphics[width=1.0\linewidth]{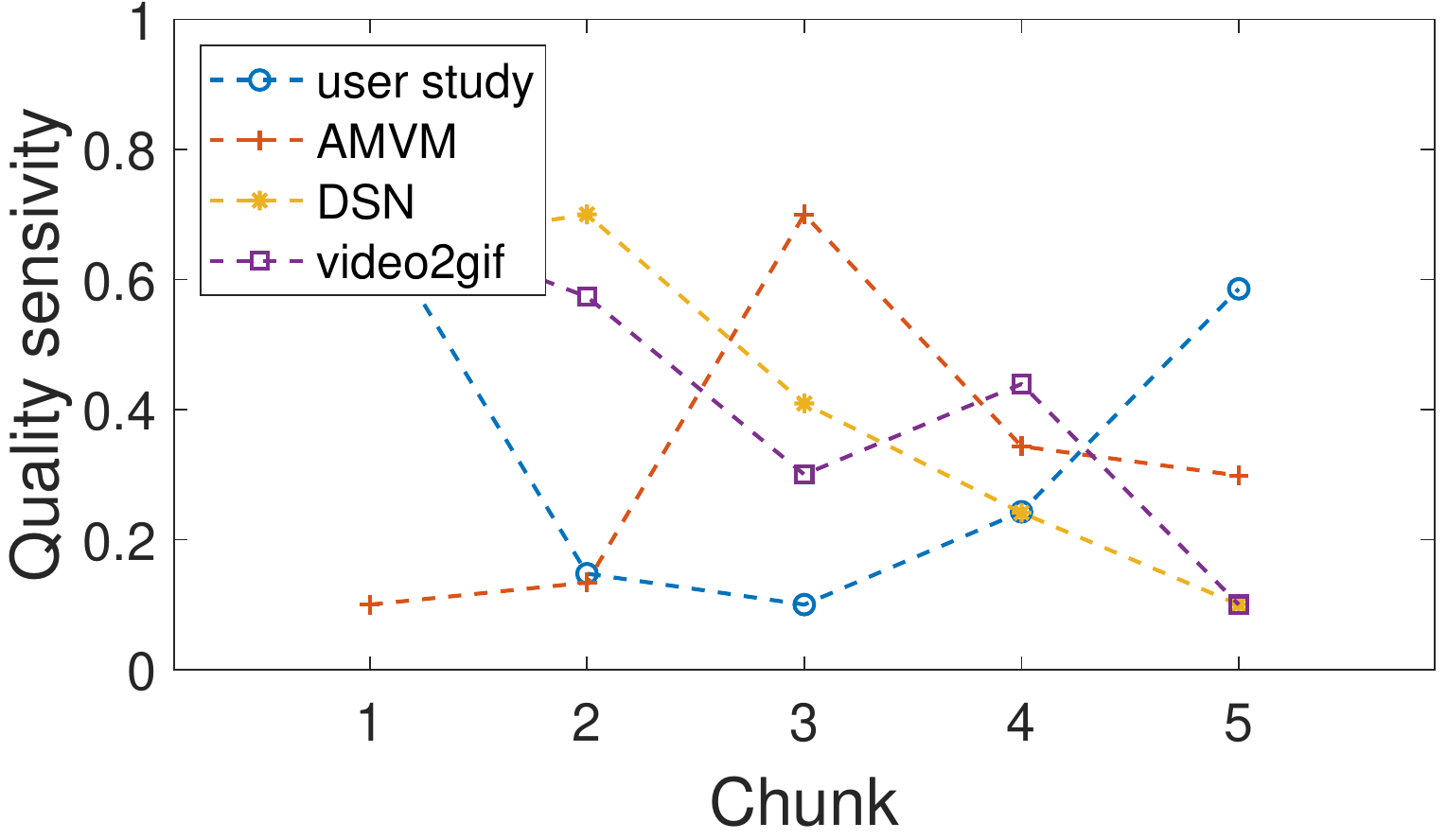}
            \caption{Soccer2}
      \end{subfigure}
    \caption{Quality sensitivity estimation by CV-based models.}
    \label{fig:cvtech_fail}
\end{figure*}

\section{Why not use vision-based techniques?}
\label{sec:cv}

A stronger alternative for inferring the per-chunk quality sensitivity, which goes beyond the visual quality metrics discussed in \S\ref{sec:moti:missing}, are computer vision (CV) models such as those for video highlights detection and video summarization.
We tested three recent computer vision models, AMVM~\cite{liu2015deriving}, DSN~\cite{zhou2018deep} and Video2GIF~\cite{gygli2016video2gif}, but their results do not correlate well with the quality sensitivity weights inferred by \name.
However, as we show in Figure~\ref{fig:cvtech_fail}, the trend of the quality sensitivity of the chunks is not aligned with that predicted from CV-based models.
% Here, the quality sensitivity of a chunk is calculated by 1 minus the QoE of the video sequence with a 1-second rebuffering event inserted to the front of chunk.
We speculate two reasons for this.
First, these models tend to highlight video segments that are information-rich (\eg lots of objects), but information richness does not necessarily imply higher quality sensitivity. For example in the soccer video in Figure~\ref{fig:soccer_content}, the period leading up to the goal is the most quality sensitive, but the CV models believe the scenes showing the audience (a lot of people) are the most important.
% that the user would pay more attention, and for each chunk, the degree of user attention is not exactly aligned its sensitivity. 
% Thus, when the model is training, the reward or loss function designed for their own purposes would lead the models to another way.
Second, as we observed in \S\ref{sec:moti:missing}, quality sensitivity can be influenced by different factors in different videos, so learning-based methods may not be a good fit because they are biased towards the data they were trained on.